\documentclass[12pt]{article}

\usepackage[dvipdfmx]{graphicx}
\usepackage{amsfonts,amsmath,amssymb,epsf}
\usepackage{amsmath,braket}
\usepackage{graphicx,color}
\usepackage[top=30truemm,bottom=30truemm,left=25truemm,right=25truemm]{geometry}
\usepackage{latexsym}
\usepackage{here}

\newcommand{\de}{\partial}
\newcommand{\be}{\begin{equation}}
\newcommand{\ba}{\begin{eqnarray}}
\newcommand{\ea}{\end{eqnarray}}
\newcommand{\ee}{\end{equation}}

\newcommand{\ca}{\mathcal}
\newcommand{\lr}{\leftrightarrow}
\newcommand{\f}{\frac}
\newcommand{\s}{\sqrt}

\newcommand{\ti}{\tilde}
\newcommand{\ap}{\alpha}

\newcommand{\ddd}{\cdot\cdot\cdot}
\newcommand{\no}{\nonumber \\}
\newcommand{\la}{\langle}
\newcommand{\lb}{\rangle}
\newcommand{\bea}{\begin{eqnarray}}
\newcommand{\eea}{\end{eqnarray}}
\newcommand{\bes}{\begin{equation*}}
\newcommand{\beas}{\begin{eqnarray*}}
\newcommand{\eeas}{\end{eqnarray*}}
\newcommand{\bas}{\begin{array*}}
\newcommand{\eas}{\end{array*}}
\newcommand{\ees}{\end{equation*}}

\newcommand{\ep}{\epsilon}

\newcommand{\ov}{\overline}

 \def\ep{{\epsilon}}

\def\tr{{\text{tr}}}

 \def\cm{{\checkmark}}

\newcommand{\ex}[1]{\mathrm{e}^{#1}}

\newcommand{\pa}[1]{\left(#1 \right)}

\newcommand{\BR}[1]{\Biggl[#1 \Biggr]}

\newcommand{\abs}[1]{\left|#1\right|}

\newcommand{\ar}[1]{\xrightarrow[#1]{}}

\newcommand{\dg}[1]{#1^{\dagger}}
\newcommand{\fr}{\frac}

\begin{document}

\begin{titlepage}
\thispagestyle{empty}

\begin{flushright}
YITP-19-123
\\
IPMU19-0182
\\
\end{flushright}

\bigskip

\begin{center}
\noindent{{\large \textbf{Looking at Shadows of Entanglement Wedges}}}\\
\vspace{1cm}
Yuya Kusuki$^{a}$, Yuki Suzuki$^{b}$,  Tadashi Takayanagi$^{a,c}$ and Koji Umemoto$^{a}$
\vspace{0.5cm}\\

{\it $^a$Center for Gravitational Physics,\\
Yukawa Institute for Theoretical Physics,
Kyoto University, \\
Kitashirakawa Oiwakecho, Sakyo-ku, Kyoto 606-8502, Japan}\\

{\it $^b$ Faculty of Science, Kyoto University,\\
Kitashirakawa Oiwakecho, Sakyo-ku, Kyoto 606-8502, Japan} \\

{\it $^{c}$Kavli Institute for the Physics and Mathematics
 of the Universe (WPI),\\
University of Tokyo, Kashiwa, Chiba 277-8582, Japan}

\end{center}

\begin{abstract}
We present a new method of deriving shapes of entanglement wedges directly from CFT calculations.
We point out that a reduced density matrix in holographic CFTs possesses a sharp wedge structure such that inside the wedge 
we can distinguish two local excitations, while outside we cannot. We can determine this wedge, which we call a CFT wedge, by computing 
a distinguishability measure.  We find that CFT wedges defined by the fidelity or Bures distance as a distinguishability measure, 
coincide perfectly with shadows of entanglement wedges in AdS/CFT. We confirm this agreement between CFT wedges and entanglement wedges for 
two dimensional holographic CFTs where the subsystem is chosen to be an interval or double intervals, as well as  
higher dimensional CFTs with a round ball subsystem. On the other hand if we consider a free scalar CFT, we find that there are no sharp 
CFT wedges. This shows that sharp entanglement wedges emerge only for holographic CFTs owing to the large N factorization.
 We also generalize our analysis to a time-dependent example and to a holographic boundary conformal field theory (AdS/BCFT). 
Finally we study other distinguishability measures to define CFT wedges. We observe that some of measures lead to CFT wedges which
slightly deviate from the entanglement wedges in AdS/CFT and we give a heuristic explanation for this. This paper is an extended version of our earlier letter 
arXiv:1908.09939 and includes various new observations and examples.

\end{abstract}

\hspace{6cm}
{\it Dedicated to the memory of Tohru Eguchi}

\end{titlepage}

\newpage

\tableofcontents

\newpage

\section{Introduction}

The AdS/CFT has provided a key framework to explore quantum gravity aspects of string theory \cite{Ma}.
The principle of AdS/CFT relates  quantum gravity  in an anti de-Sitter (AdS) spacetime equivalently to 
a conformal field theory which lives on the boundary of AdS. The basic rule of the correspondence is 
given by the bulk-boundary correspondence \cite{GKP,W}, which says the gravity partition function is 
equal to the CFT partition function. 

To better understand the AdS/CFT, it is useful to decompose the correspondence into subregions.
Namely, we would like to understand
which subregion in AdS is dual to a given region $A$ in a CFT. 
The answer to this question has been argued to be the entanglement wedge 
$M_A$ \cite{EW1,EW2,EW3}, 
the region surrounded by 
the subsystem $A$ and the extremal surface $\Gamma_A$ whose area gives the 
holographic entanglement entropy \cite{RT,HRT,Review}.  
Here we consider a static spacetime and assume a 
restriction on the canonical time slice.  In more general time-dependent spacetimes, the genuine 
entanglement wedge is given by the domain of dependence of $M_A$. 

In this correspondence, called entanglement wedge reconstruction, the bulk reduced density matrix on the entanglement 
wedge $\rho^{bulk}_{M_A}$ is equivalent to the CFT reduced density matrix $\rho_A$. So far this subregion-subregion duality has been explained
by combining several known facts: the gravity dual of bulk local field operator (called HKLL map \cite{HKLL}
and its generalization \cite{Faulkner:2017vdd}),
the formula of quantum corrections to holographic entanglement entropy \cite{FLM,JLMS} and
the conjectured connection between the AdS/CFT and quantum error correcting codes \cite{ADH,Dong:2016eik}.
However, since this explanation highly relies on the dual AdS geometry and its dynamics from the beginning, 
it is not clear how the entanglement wedge geometry naturally emerges from a CFT itself. 

Recently, a new approach to entanglement wedges was reported briefly in \cite{STU}, where purely CFT analysis reveals the structure of 
entanglement wedge for the first time.  In the present paper, which is a full paper accompanying with  the letter \cite{STU}, we would like to 
provide not only detailed explanations but also more evidences for this construction with various new examples. This includes a precise derivation of entanglement wedge from Bures metric when the subsystem $A$ consists of  double intervals. Moreover, we give purely CFT derivations of  
the entanglement wedges in a time-dependent setup and in AdS/BCFT \cite{AdSBCFT}.
Though most of our examples are two dimensional conformal field theories (2d CFTs), in a later part of this paper, we will analyze the higher dimensional CFTs and derive the entanglement wedges from CFTs.

In our analysis, it is important to remember that only a special class of CFTs, called holographic CFTs, can have
classical gravity duals which are well approximated by general relativity. A holographic CFT is characterized by
a large central charge $c$ (or large rank of gauge group $N$) and 
very strong interactions. The latter property leads to a large spectrum gap \cite{He,Hartman2013a}. Thus
we expect that the entanglement wedge geometry is available only when we employ holographic CFTs.
Indeed, our new framework will explain how entanglement
wedges emerge from holographic CFTs.

Consider a locally excited state in a 2d CFT, created by inserting a primary operator $O_\ap(w,\bar{w})$ on the vacuum.
The index $\ap$ distinguishes different primaries.
As the first example, we focus on a 2d CFT on an Euclidean complex plane R$^2$. We write the coordinates of this space by
$(w,\bar{w})$ or equally $(x,\tau)$ such that $w=x+i\tau$.
We choose a subsystem $A$ on the $x$-axis and define the reduced density matrix on $A$, tracing out its complement $B$:
\ba
\rho_A(w,\bar{w})={\cal N_\ap}\cdot \mbox{Tr}_B\left[O_\ap(w,\bar{w})|0\lb \la 0|O_\ap^\dagger(\bar{w},w)\right], \label{redb}
\ea
where ${\cal N_\ap}$ is a normalization factor to secure $\mbox{Tr}\rho_A=1$.
This state was first introduced in \cite{Nozaki} to study its entanglement entropy. Refer also to \cite{Alcaraz:2011tn} 
for calculations of entanglement entropy of primary states.

\begin{figure}
  \centering
  \includegraphics[width=8cm]{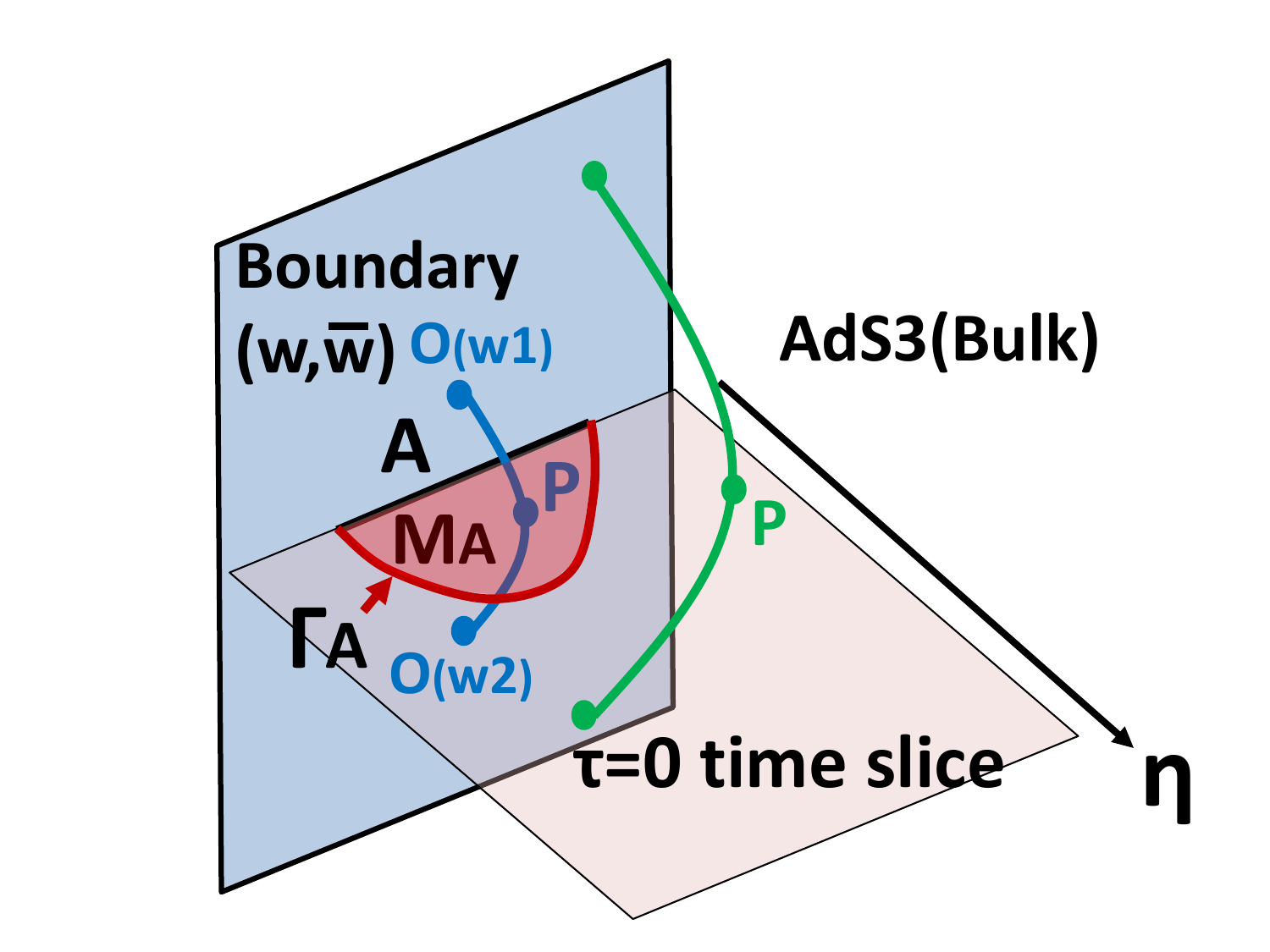}
  \caption{We sketched an entanglement wedge $M_A$ for an interval $A$ in AdS$_3/$CFT$_2$. We also show  
holographic computations of two point functions dual to geodesics. The blue (or green) geodesic does (or does not) intersect with $M_A$ at $P$.}
\label{fig:EW}
  \end{figure}

We would like to choose the (chiral and anti chiral) conformal dimension $h_\ap$ of the primary operator $O_\ap$ in the range:
\be
1\ll h_\ap \ll c.  \label{rangeh}
\ee
This assumption allows us to neglect its back reaction in the gravity dual and to approximate the two point function
$\la O(w_1,\bar{w}_1) O^\dagger(w_2,\bar{w}_2)\lb$ by the geodesic length in the gravity dual
between the two points  $(w_1,\bar{w}_1)$ and $(w_2,\bar{w}_2)$ on the boundary $\eta\to 0$ of the Poincare AdS$_3$
\be
ds^2=\eta^{-2}(d\eta^2+dwd\bar{w})=\eta^{-2} (d\eta^2+dx^2+d\tau^2), \label{po}
\ee
where we set the AdS radius one.
Thus, by projecting on the bulk time slice $\tau=0$, the state $\rho_A(w,\bar{w})$ is dual to a bulk excitation at a bulk point $P$, which is defined by the intersection between the time slice $\tau=0$ and the geodesic. This procedure is sketched in Fig.\ref{fig:EW}.

In this way, we can probe the bulk point by using the locally excited reduced density matrix (\ref{redb}).
If the entanglement wedge reconstruction is correct, then we should be able to distinguish $\rho_A(w,\bar{w})$ and $\rho_A(w',\bar{w}')$ when 
$w\neq w'$ if either of their bulk points $P$ and $P'$ is in the entaglement wedge. If both of them are outside, we should not be able to distinguish 
$\rho_A(w,\bar{w})$ and $\rho_A(w',\bar{w}')$. Remarkably this argument of distinguishability is based on purely CFT calculations and we can define a CFT counterpart of entanglement wedge from this analysis, which we call the CFT wedge. We can regard the CFT wedges are shadows of entanglement wedges when we 
interpret the geodesics in Euclidean spaces as light rays.
In other words, the entanglement wedge reconstruction argues that the CFT wedge coincides with the true entanglement wedge.
The main part of this paper is to confirm this expectation in various examples of AdS/CFT. 

This paper is organized as follows. In section two, we will give a brief review of distance (or distinguishability) measure of quantum states and introduce the concept of CFT wedges. In section three, we will analyze the geometry of CFT wedge from the measure $I(\rho,\rho')$ in the single interval case of 2d CFTs and confirm that this reproduces the entanglement wedges.
In section four, we will study the Bures information metric in the single interval case of 2d CFTs 
and confirm that this reproduces the entanglement wedges.
In section five, we will analyze how the time dependent excited states correctly probe the entanglement wedges in a simple example. In section six, we turn to the double interval example in 2d CFTs and we confirm that the Bures metric reproduces the entanglement wedges, while the measure $I(\rho,\rho')$ leads to a small deviation. In section seven, we will analyze the CFT wedges for global quantum quenches and theromfield double state, where the correct entanglement wedge is reproduced under a reasonable assumption. In section eight, we extend our calculations of CFT wedges in higher dimensional holographic CFTs and confirm that the Bures metric reproduces the correct entanglement wedges. In section nine,
we discuss other distinguishability measures, where we observe that CFT wedges for most of them fall into the two classes of the Bures metric and  $I(\rho,\rho')$. In section ten, we will discuss how we can reproduce the entanglement wedge if we employ the HKLL operators instead of local operators. In section eleven we will summarize our conclusions and discuss future problems. In appendix A, 
we willl give the detailed calculations of  $I(\rho,\rho')$ in the single interval. In appendix B, we will present the detailed analysis of the Bures metric in $c=1$ CFT. In appendix C, we discuss the Bures metric in a general time dependent case. In appendix D, we will list properties of various distinguishability measures.

\section{Distance Measure of Quantum States and CFT Wedges}

The main analysis in this paper is to study the distinguishability of reduced density matrices of the form (\ref{redb}). Therefore in this section we would like to summarize relevant measures of distances between 
two density matrices $\rho$ and $\rho'$. Refer to \cite{Hayashi} for a text book.  After these preparations, we will introduce the notion of CFT wedges
which are finally identified with shadows of entanglement wedges in AdS/CFT.

\subsection{Fidelity and Related Quantities}

First we would like to introduce quantities which provide analogues of inner product of two density matrices.
One of the best quantities is the fidelity $F(\rho,\rho')$ defined by
\ba
F(\rho,\rho')=\mbox{Tr}[\s{\s{\rho}\rho'\s{\rho}}].  \label{fidelity}
\ea
The fidelity is symmetric under an exchange of $\rho$ and $\rho'$ and 
takes values in the following range
\be
0\leq F(\rho,\rho')=F(\rho',\rho)\leq 1.  \label{propa}
\ee 
Moreover it satisfies 
\ba
&& F(\rho,\rho')=1\ \mbox{if and only if}\  \rho=\rho', \label{propb}\\
&& F(\rho,\rho')=0\ \mbox{if and only if}\ \rho\rho'=0.  \label{propc}
\ea
Therefore we can employ the fidelity to distinguish 
two quantum states. 

There are many other measures which satisfy the above basic properties (\ref{propa}), 
(\ref{propb}) and (\ref{propc}) (which are listed in App.\ref{ap:list}).
One of them is the Affinity $A(\rho,\rho')$ \cite{Luo2004}:
\ba
A(\rho,\rho')=\mbox{Tr}[\s{\rho}\s{\rho'}].  \label{aff}
\ea 
This quantity has upper and lower bounds in terms of the fidelity as 
\begin{equation}
 F^2(\rho, \rho') \leq A(\rho, \rho') \leq F(\rho, \rho').
\end{equation}

For the actual computations, taking a square root of a given density matrix 
is not always tractable. This motivates us to consider a quantity $I(\rho,\rho')$
\begin{equation}
I(\rho,\rho')\equiv \fr{\tr \rho \rho'}{\s{ \pa{\tr \rho^2} \pa{\tr \rho'^2}   }}.  \label{Irho}
\end{equation}
This quantity is called geometric mean (GM) fidelity, introduced in \cite{Wang2008}  
(see also \cite{Liu2012, Liang2018}) and satisfied the basic properties  (\ref{propa}), 
(\ref{propb}) and (\ref{propc}). This was employed to study non-equilibrium 
dynamics of quantum systems in \cite{CardyR}. 
We might be able to think that this quantity $I(\rho,\rho')$ is analogous to 2nd Renyi entropy, while 
the fidelity is analogous to von-Neumann entropy. Indeed the total power of $\rho$ 
and $\rho'$ is two in the former, while one in the latter.

It is also useful to evaluate these quantities when the states are pure, which are 
expresses as $\rho=|\phi\lb\la\phi|$ and $\rho'=|\phi'\lb\la\phi'|$. From the definitions
we obtain 
\ba
&& F(\rho,\rho')=|\la\phi|\phi'\lb|, \label{purea} \\
&& A(\rho,\rho')=|\la\phi|\phi'\lb|^2, \label{pureb} \\
&& I(\rho,\rho')=|\la\phi|\phi'\lb|^2. \label{purec} \\
\ea

\subsection{Distance Measures}

Now we would like to move onto distance measures between two quantum states 
$\rho$ and $\rho'$. First of all, the Bures distance is defined from the fidelity as follows
\ba
&& D_B(\rho,\rho')^2=2(1-F(\rho,\rho')). \label{disb} 
\ea
It is obvious that this quantity is symmetric
and this takes values in the range:
\be
0\leq D_B(\rho,\rho')=D_B(\rho',\rho)\leq 2.   \label{proppa}
\ee
In addition, this satisfies
\ba
D_B(\rho,\rho')=0\ \ \mbox{if and only if}\ \ \rho=\rho'.  \label{proppb}
\ea

There are three more important distance measures: the trace distance $D_{tr}(\rho,\rho')$  \cite{nielsen2002quantum}, 
relative entropy distance $D_R(\rho,\rho')$ and Hellinger distance $D_H(\rho,\rho')$, each given by
\ba
&& D_{tr}(\rho,\rho')=\frac{1}{2}|\rho-\rho'|_1=\frac{1}{2}\mbox{Tr}[\s{(\rho-\rho')^{2}}], \label{trdis} \\
&& D_R(\rho,\rho')^2=\mbox{Tr}[\rho(\log\rho-\log\rho')], \label{reladis}\\
&& D_H(\rho,\rho')^2=2(1-A(\rho,\rho')),  \label{heldis}\\
&& D_{I}(\rho,\rho')^2=2(1-I(\rho,\rho')). \label{gmdis}
\ea
The three of them, namely $D_{tr},D_H$ and $D_I$ satisfy the basic properties (\ref{proppa}) and (\ref{proppb}).
On the other hand, the relative entropy distance $D_R(\rho,\rho')$ is not symmetric and 
takes the values $0\leq D_R(\rho,\rho')<\infty$, though   (\ref{proppb}) holds.
Refer to \cite{Zhang2019,Zhang2019a} for computations in integrable 2d CFTs.
and to \cite{Zhang:2019kwu} for an application to locally excited states (see also \cite{Bhattacharyya:2019ifi}).

It is useful to note the following relations between these distances:
\ba
&& D_R(\rho,\rho')\geq 2D_{tr}(\rho,\rho')^2, \label{ineqra} \\
&& 1-F(\rho,\rho') \leq D_{tr}(\rho,\rho') \leq \s{1-F(\rho,\rho')^2}.    \label{ineqtra}
\ea

\subsection{Information Metrics and Quantum Cramer-Rao Theorem}

Furthermore, we can introduce so called the information metric when the density matrix is parameterized by continuous  valuables $\lambda^i$, denoted by $\rho(\lambda)$. For the Bures distance, this metric is defined as follows
\ba
D_B(\rho(\lambda+d\lambda),\rho(\lambda))=G_{Bij}d\lambda^i d\lambda^j+\ddd,  \label{infometr}
\ea 
where $d\lambda_i$ are infinitesimally small and $\ddd$ denotes the higher powers of $d\lambda^i$. This metric $G_{Bij}$ is called the Bures metric.   In the same way we can define another metric from the relative entropy distance $D_R$, called quantum Fisher metric $G_R$.  It is also possible to define the metrics $G_H$ and $G_I$ for the distance measures $D_H$ and $D_I$, respectively.

The quantum version of Cramer-Rao theorem \cite{Hel} (see also the text book \cite{Hayashi})  tells us that when we try to estimate the value of 
$\lambda_i$ from physical measurements, the errors of the estimated value is bounded by the inverse of the Bures metric $G_B$ as follows
\ba
\la \delta \lambda^i\delta\lambda^j \lb \geq (G^{-1}_B)^{ij}.  \label{CRaB}
\ea
In particular, when $G_{Bij}=0$, the uncertainty gets divergent and we cannot estimate the value of 
$\lambda_i$ at all.
This is simply because the density matrix does not depend on $\lambda_i$ and we cannot distinguish density matrices for various values of $\lambda_i$.

More precisely, the quantum Cramer-Rao theorem is stated as follows.
A physical measurement is described by the  POVM operator $M_\omega(\geq 0)$ such that $\sum_{\omega}M_{\omega}=I$, 
where $\omega$ corresponds to each value of the measurement. Tr$[\rho M_\omega]$ denotes the probability that the measured value
is given by $\omega$. We would like to estimate the value of $\lambda^i$ from the measured value $\omega$ 
following a arbitrary chosen function $\lambda^i\to \hat{\lambda}^i(\omega)$. We introduce an error in this process as
\ba
\la \delta \lambda^i\delta\lambda^j \lb\equiv \sum_{\omega}(\lambda_i-\hat{\lambda}^i(\omega))(\lambda_j-\hat{\lambda}^j(\omega))
\mbox{Tr}[\rho_\lambda M_\omega].
\ea
To be exact we actually consider $n$ copies of the system $\rho_\lambda^{\otimes n}$ and take the asymptotic limit 
\ba
\la \delta \lambda^i\delta\lambda^j \lb_n\equiv \sum_{\omega}(\lambda_i-\hat{\lambda}^i(\omega))(\lambda_j-\hat{\lambda}^j(\omega))
\mbox{Tr}[\rho_\lambda^{\otimes n} M^n_\omega].
\ea
The quantum Cramer-Rao Theorem \cite{Hel} argues the lower bound by the inverse of the Bures metric:
\ba
\lim_{n\to\infty} n\la \delta \lambda^i\delta\lambda^j \lb_n\geq (G^{-1}_B)^{ij}.
\ea

\subsection{Simple Example of Information Metric: Pure States in CFTs}

For pure states $\rho=|\phi\lb\la\phi|$ and $\rho'=|\phi'\lb\la\phi'|$, the distance measures look like

\ba
&& D_B(\rho,\rho')^2=2(1-|\la\phi|\phi'\lb|), \label{disbura}\\
&& D_H(\rho,\rho')^2=2(1-|\la\phi|\phi'\lb|^2). \label{dishela}
\ea
We omit the relative entropy distance because $D_R$ gets divergent when $|\phi\lb \neq |\phi'\lb$.

Consider locally excited states $|\phi(w,\bar{w})\lb=O_\ap(w,\bar{w})|0\lb$ in a 2d CFT. We simply find
\ba
|\la\phi(w)|\phi'(w')\lb|=\frac{|w-\bar{w}|^{2h}|w'-\bar{w}'|^{2h}}{|w-\bar{w}'|^{4h}}.
\ea 
This leads to the Bures metric 
\be
D_B^2\simeq \frac{h_\ap}{\tau^2}(d\tau^2+dx^2),  \label{pureburesc}
\ee
and the Hellinger metric
\be
D_H^2\simeq \frac{2h_\ap}{\tau^2}(d\tau^2+dx^2). \label{purehelinc}
\ee

Interestingly, the information metric is proportional to the two dimensional hyperbolic space $H_2$.
This looks like a time slice of the gravity dual i.e. the Poincare AdS$_3$ (\ref{po}). 
This coincidence is very natural because the distinguishability between two excitations
should increase when the corresponding bulk points are geometrically separated.  
This was already noted essentially in \cite{MNSTW}. However, this result is universal for any 2d CFTs as
the computation only involves two point functions. This implies the study of information metric of 
reduced density matrix $\rho_A$ has more opportunities to explore deep mechanisms of AdS/CFT,
which is the main motivation of this paper.

\subsection{CFT Wedges in Holographic CFTs}

Distinguishability measures for reduced density matrices (\ref{redb}) 
crucially depend on the nature of CFTs such as multi-point correlation functions, 
as opposed to those for pure states. The special properties of holographic CFTs allow 
us to introduce a CFT counterpart of entanglement wedge as we will explain in this
paper for various examples. We call this geometrical structure in holographic CFTs 
the CFT wedges, which we would like to introduce below.

Consider an information metric $G_{\#}$ (here $\#=B,I,$ etc. specifies the type of distance measure) for 
a reduced density matrix  $\rho_A$ of locally excited state given by (\ref{redb}), regarding 
the operator insertion point $X=(w,\bar{w})$ as the parameter
 $\lambda$ in (\ref{infometr}). The information metric 
has the components $G_{\# ij}$ with $i,j=w,\bar{w}$ and depends on the 
location $(w,\bar{w})$. Since the restriction to 2d CFTs is not necessary in this subsection, 
we have in mind holographic CFTs in any dimensions below.

In this setup, we introduce the geometrical structure in a CFT, which we call the CFT wedge $C^{(\#)}_A$ for the 
subsystem $A$ as follows:
\ba
\mbox{If}\ X\in C^{(\#)}_A,  \ \mbox{then}\  \ G_{\# ij}(X)>0,\no
\mbox{If}\ X\notin C^{(\#)}_A, \ \mbox{then}\  \ G_{\# ij}(X)\simeq 0.   \label{CFTw}
\ea

In the case of the Bures metric, we can equivalently write this in terms of Fidelity as follows:
\ba
\mbox{If}\ X=X'\in C^{(B)}_A , \ \mbox{then}\  \ F(\rho(X),\rho(X'))\simeq 1,  \no
\mbox{If}\ X\notin C^{(B)}_A \mbox{and}\ X'\notin C^{(B)}_A, \ \mbox{then}\  \ F(\rho(X),\rho(X'))\simeq 1, \no
\mbox{If otherwise},\ \ \mbox{then}\  \ F(\rho(X),\rho(X'))\simeq 0. \label{fidelityews}
\ea

Also for another distance measure $I(\rho,\rho')$ we can express the CFT wedge $C^{(I)}_A$ by
\ba
\mbox{If}\ X=X'\in C^{(I)}_A , \ \mbox{then}\  \ I(\rho(X),\rho(X'))\simeq 1,  \no
\mbox{If}\ X\notin C^{(I)}_A \mbox{and}\ X'\notin C^{(I)}_A, \ \mbox{then}\  \ I(\rho(X),\rho(X'))\simeq 1, \no
\mbox{If otherwise},\ \ \mbox{then}\  \ I(\rho(X),\rho(X'))\simeq 0. \label{igmews}
\ea

Note that the sharp geometrical structures (\ref{CFTw}), (\ref{fidelityews}) and (\ref{igmews}) only appear in holographic CFTs, where
we take the limit $h_\ap\gg 1$ as in (\ref{rangeh}). The non-vanishing information metric  in (\ref{CFTw}) scales as $O(h_\ap)$.
For generic CFTs, such as free field CFTs, we only find smeared behaviors, 
which prohibit us to define a CFT wedge, though qualitatively the behaviors of distance measures are often similar. In other words,
the sharp CFT wedges emerge only when we consider holographic CFTs.  

We also would like to stress that the CFT wedges can depend on the choice of distance measures. Indeed as we will see later, 
for generic setups, $C^{(B)}_A$ and $C^{(I)}_A$ can differ. In the end we would like to argue that the correct choice which probes the 
low energy states in AdS/CFT (i.e. the code subspace) will be the Bures metric. We will comment more on this point in the final part of 
this paper.

\section{Entanglement Wedge from $I(\rho,\rho')$ in the Single Interval Case}

We start with the simplest example, namely the CFT wedges $C^{(I)}_A$ (\ref{igmews}) 
for the measure $I(\rho,\rho')$  (\ref{Irho}) when $A$ is a single interval in a 2d CFT. 
Consider a 2d CFT on the flat space R$^2$, whose Euclidean time and space coordinate are denoted by 
$\tau$ and $x$. We employ a complex coordinate $(w,\bar{w})$ or equally a Cartesian coordinate $(\tau,x)$ 
such that $w=x+i\tau$. If the CFT has a gravity dual, it is dual to gravity in the  Poincare AdS$_3$ metric (\ref{po}). However, below we will analyze both holographic and non-holographic CFTs to compare their results.

\subsection{Reduced Density Matrix for Single Interval and CFT Wedges}

We choose the subsystem $A$ to be an interval $0\leq x\leq L$ at $\tau=0$. 
The extremal surface $\Gamma_A$ in the bulk AdS is given by the semi circle $(x-L/2)^2+\eta^2=L^2/4$. 
Therefore the entanglement wedge $M_A$ is given by 
\ba
(x-L/2)^2+\eta^2\leq L^2/4.  \label{EWsingle}
\ea
Note that this is also identical to the causal wedge \cite{HR}.

From the viewpoint of CFTs, we consider an excited state by inserting a local operator $O_\ap$ at 
$(w,\bar{w})$ and define the reduced density matrix (\ref{redb}).
We regard the location $(\tau,x)$ of the insertion point as the parameters of $\rho_A$.
Having in mind the AdS/CFT duality, 
the geodesic which connects $(\tau,x)$ and $(-\tau,x)$ intersects with 
the time slice $\tau=0$ at the point $P$ given by $\eta=\tau$.  
Therefore if the entanglement reconstruction is correct, the CFT wedge, based on a proper distance measure, 
should coincide with $|w-L/2|\leq L/2$ or equally
\ba
C_A:\ \ \left(x-\frac{L}{2}\right)^2+\tau^2\leq \frac{L^2}{4}.    \label{inent}
\ea
Accordingly, the information metric should vanish if the intersection $P$ is outside of the CFT wedge i.e.
\ba
\ov{C_A}:\ \left(x-\frac{L}{2}\right)^2+\tau^2> \frac{L^2}{4},  \label{outent}  
\ea
while it is non-vanishing in the inside the wedge (\ref{inent}).

Below, in this section, we focus on calculating the CFT wedge $C^{(I)}$ for the measure $I(\rho,\rho')$  (\ref{Irho}).

\subsection{Calculation of $I(\rho,\rho')$}

\begin{figure}
  \centering
  \includegraphics[width=10cm]{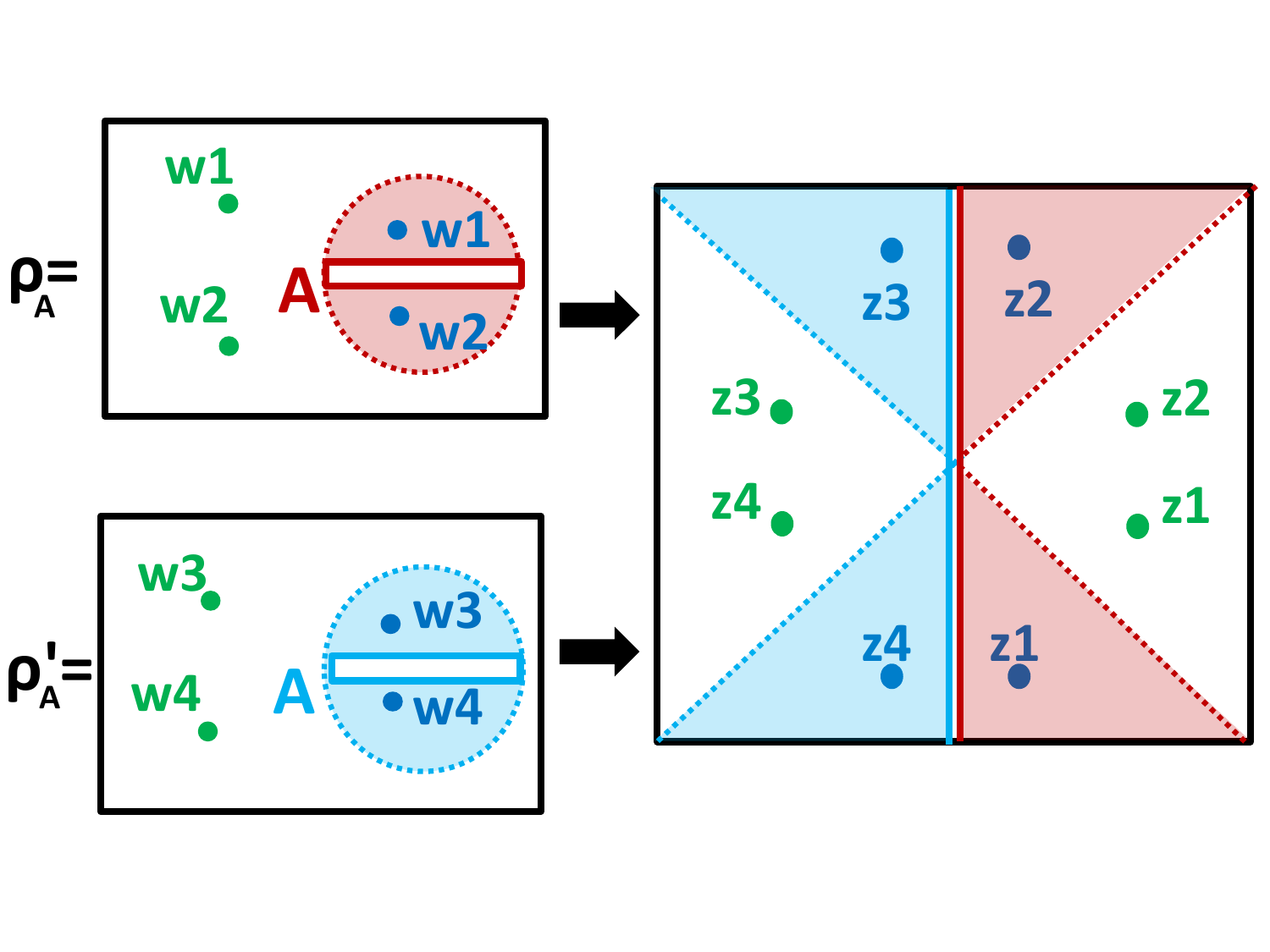}
  \caption{We sketched the conformal mapping for the calculation of Tr$[\rho\rho']$. Green Points 
(or bule points) describe the local excitations in the CFT which are dual to bulk local excitations outside (or inside) of  the CFT wedge. }
\label{fig:singlePO}
  \end{figure}

Let us calculate $I(\rho,\rho')$ (\ref{Irho}) for the two density matrices:
\be
\rho=\rho_A(w,\bar{w}),\ \ \ \rho'=\rho_A(w',\bar{w}').
\ee
To calculate Tr$[\rho\rho']$,  consider the conformal transformation (the calculations are similar to \cite{Nozaki,HNTW}):
\ba
z^2=\frac{w}{w-L},  \label{sintr}
\ea
which maps two flat space path-integrals for $\rho(w,\bar{w})$ and $\rho(w',\bar{w}')$, 
into a single plane. The coordinate of the latter (single plane) is written as $(z,\bar{z})$. The insertion 
points of the local operators $O_\ap$ and $O^\dagger_\ap$ are given by
\ba
w_1=x+i\tau(=w),\  \  w_2=x-i\tau(=\bar{w}),
\ea
for  $\rho(w,\bar{w})$, and 
\ba
w'_3=x'+i\tau'(=w'),\  \  w'_4=x'-i\tau'(=\bar{w}'),
\ea
 for  $\rho(w',\bar{w}')$. Refer to the upper two pictures in Fig.\ref{fig:singlePO}.  The transformation (\ref{sintr}) maps these four points into $z_1.z_2,z'_3$ and $z'_4$ given by
\ba
&& z_1=\s{\frac{-x-i\tau}{L-x-i\tau}}, \ \  \ z_2=\s{\frac{-x+i\tau}{L-x+i\tau}},\no
&& z'_3=-\s{\frac{-x'-i\tau'}{L-x'-i\tau'}}, \ \  \ z'_4=-\s{\frac{-x'+i\tau'}{L-x'+i\tau'}}.
\ea
It is important to note that the boundaries of the CFT wedge $|w-L/2|=L/2$ of the original two flat planes are mapped into the diagonal lines $z=\pm i \bar{z}$ as depicted in Fig.\ref{fig:singlePO}. As we will see soon, this leads to the CFT wedge structure in the distinguishability.

The trace Tr$[\rho\rho']$ is now expressed as a correlation function on the $z$-plane:
\ba
&& \mbox{Tr}[\rho\rho']=\left|\frac{dz_1}{dw_1}\right|^{2h_\ap}\left|\frac{dz_2}{dw_2}\right|^{2h_\ap}
\left|\frac{dz'_3}{dw'_3}\right|^{2h_\ap}\left|\frac{dz'_4}{dw'_4}\right|^{2h_\ap}\cdot
H(z_1,z_2,z'_3,z'_4)\cdot\frac{Z^{(2)}}{(Z^{(1)})^2}, \no
&&
H(z_1,z_2,z'_3,z'_4)\equiv\frac{\la O_\ap^\dagger(z_1,\bar{z}_1)O_\ap(z_2,\bar{z}_2)
O_\ap^\dagger(z'_3,\bar{z}'_3)O_\ap(z'_4,\bar{z}'_4)\lb}{\la O_\ap^\dagger(w_1,\bar{w}_1)O_\ap(w_2,\bar{w}_2)\lb
\la O_\ap^\dagger(w'_3,\bar{w}'_3)O_\ap(w'_4,\bar{w}'_4)\lb},
\no
\ea
where $\la \ddd \lb$ denotes the normalized correlation function such that $\la 1\lb=1$ and 
we also write the vacuum partition function on a $n$-sheeted complex plane by $Z^{(n)}$.

Thus we obtain
\ba
&& I(\rho,\rho')  \no
&& =\left|\frac{dz_1/dw_1}{dz'_1/dw'_1}\right|^{2h_\ap}
\left|\frac{dz_2/dw_2}{dz'_2/dw'_2}\right|^{2h_\ap}\left|\frac{dz'_3/dw'_3}{dz_3/dw_3}\right|^{2h_\ap}
\left|\frac{dz'_4/dw'_4}{dz_4/dw_4}\right|^{2h_\ap}\cdot\frac{F(z_1,z_2,z'_3,z'_4)}
{\s{F(z_1,z_2,z_3,z_4)F(z'_1,z'_2,z'_3,z'_4)}},\no \label{dcfp}
\ea
where $F$ is the (normalized) four point function 
\ba
F(z_1,z_2,z'_3,z'_4)=\la O_\ap^\dagger(z_1,\bar{z}_1)O_\ap(z_2,\bar{z}_2)
O_\ap^\dagger(z'_3,\bar{z}'_3)O_\ap(z'_4,\bar{z}'_4)\lb. \label{Fzc}
\ea

Because we have the relations
\ba
&& z_1=-z_3=z,  \ \ z_2=-z_4=\bar{z},  \no
&& z'_1=-z'_3=z',  \ \ z'_2=-z'_4=\bar{z}', 
\ea
we can simplify (\ref{dcfp}) as follows
\ba
 I(\rho,\rho')= \frac{F(z,\bar{z},-z',-\bar{z}')}{\s{F(z,\bar{z},-z,-\bar{z})F(z',\bar{z}',-z',-\bar{z}')}}.
\label{dcsi}
\ea

Below we will study this quantity for both a holographic CFT and a free scalar CFT.

\subsection{Holographic CFTs}

First let us evaluate (\ref{dcsi}) in holographic CFTs. We assume the range (\ref{rangeh}) of conformal dimension 
$h_\ap$. In this case, the large $N$ (or large $c$) factorization property justifies the generalized free field approximation \cite{ElShowk:2011ag}.
Namely, in the large $c$ limit, the leading contribution to the correlation function (\ref{Fzc})
 is given by a simple Wick contraction based on the two point function 
\ba
\la  O_\ap^\dagger(z,\bar{z})O_\ap(z',\bar{z}')\lb=|z-z'|^{-4h_\ap}.
\ea

The generalized free field prescription leads to the simple expression of four point function:
\ba
F(z_1,z_2,z'_3,z'_4)&\simeq& |z_1-z_2|^{-4h}\cdot |z'_3-z'_4|^{-4h}+ |z_1-z'_4|^{-4h}|z_2-z'_3|^{-4h} \no
&\simeq& |z-\bar{z}|^{-4h}\cdot |z'-\bar{z}'|^{-4h}+ |z+\bar{z}'|^{-8h},  \label{wick}
\ea
in the final line we remember $z_1=z$ and $z'=z_3$. In the right-hand side of (\ref{wick}),
the first term comes from the Wick contraction  $\la O^\dagger(1)O(2)\lb
\la O^\dagger(3)O(4)\lb$, which we call the trivial Wick contraction. The second term arises from 
the other Wick contraction  $\la O^\dagger(1)O(4)\lb
\la O^\dagger(3)O(2)\lb$, which we call the non-trivial Wick contraction. 

First, consider the case where the local operator is inserted outside of the CFT wedge 
(\ref{outent}). This is mapped into the uncolored region in the Fig.\ref{fig:singlePO}
given by the wedge region $|\mbox{Im}[z]|<|\mbox{Re}[z]|$. When both $w$ and $w'$ are outside of 
the wedge, the lengths $|z_1-z_2|=|z-\bar{z}|$ and 
$|z'_1-z'_2|=|z'-\bar{z}'|$ are shorter than $|z_1-z'_4|=|z_2-z'_3|=|z+\bar{z}'|$. Therefore 
the four point function  (\ref{Fzc}) is approximated by the first term, which 
comes from the trivial Wick contraction. Therefore, we finally obtain
\ba
&& \mbox{If $w$ and $w'$ are outside, \ \ then}\  I(\rho,\rho')\simeq 1.  \label{siewira}
\ea
This tells us that we cannot distinguish between $\rho$ and $\rho'$ when the local excitations are outside 
the CFT wedge.

Next we turn to the case where both $w$ and $w'$ are inside the CFT wedge (\ref{inent}).
In this case, the lengths $|z_1-z_2|=|z-\bar{z}|$ and 
$|z'_1-z'_2|=|z'-\bar{z}'|$ are larger than $|z_1-z'_4|=|z_2-z'_3|=|z+\bar{z}'|$.
Therefore the four point function  (\ref{Fzc}) is approximated by the second term, which 
comes from the non-trivial Wick contraction. Therefore, we finally obtain
\ba
I(\rho,\rho')\simeq |z+\bar{z}'|^{-8h}\cdot |z+\bar{z}|^{4h}\cdot |z'+\bar{z}'|^{4h}.
\ea
Since we always have $|z+\bar{z}||z'+\bar{z'}|\leq |z+z'|^2$ and take the limit
$h_\ap\gg 1$, this quantity $I(\rho,\rho')$ is vanishing except $z=z'$:
\ba
&& \mbox{If $w$ and $w'$ are inside and $w=w'$, then}\ I(\rho,\rho')\simeq 1,  \label{siewirb} \\
&& \mbox{If  $w$ and $w'$ are inside and $w\neq w'$, then}\ I(\rho,\rho')\simeq 0,  \label{siewirc}
\ea

Finally when either of $w$ or $w'$ is inside the CFT wedge, we find that 
$I(\rho,\rho')$ is vanishing:
\ba
\mbox{If $w$ is inside and $w'$ is outside (or vice versa), then}\ I(\rho,\rho')\simeq 0, \label{siewird}
\ea

These behaviors (\ref{siewira}), (\ref{siewirb}), (\ref{siewirc}) and (\ref{siewird}) confirm our expectations (\ref{igmews}) and this shows the CFT wedge $C^{(I)}_A$ agrees with the entanglement 
wedge in AdS/CFT in the present example.  Refer to Appendix \ref{app:single}  for more detailed calculations of $I(\rho,\rho')$ in this example.

\begin{figure}
  \centering
  \includegraphics[width=6cm]{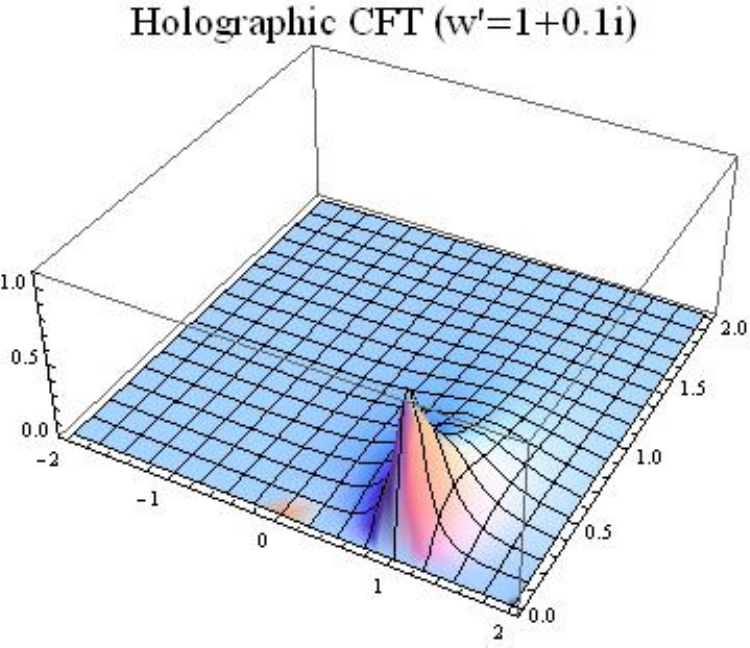}
 \includegraphics[width=6cm]{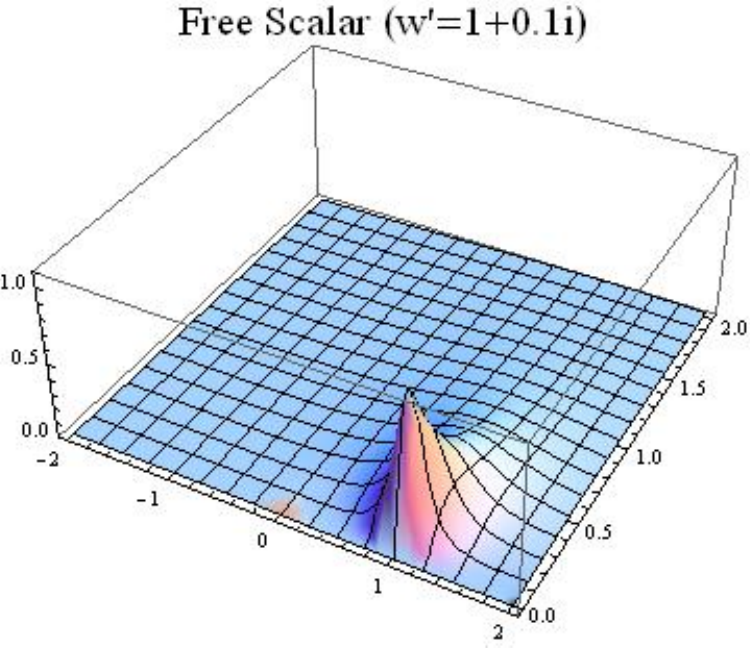}
  \caption{The value of $I(\rho,\rho')$ as a function of $\text{Re}[w]$ (horizontal axis) 
and $\text{Im}[w]$ (depth axis) when $w'$ is fixed inside the CFT wedge. In particular, we chose $h_\ap=1/2$ and $w'=1+0.1i$ and 
$A=[0,2]$ (i.e.$L=2$).  The left and right graph describe the result for the holographic CFT and the $c=1$ free scalar CFT, respectively.}
\label{fig:Insidew}
  \end{figure}

\begin{figure}
  \centering
  \includegraphics[width=6cm]{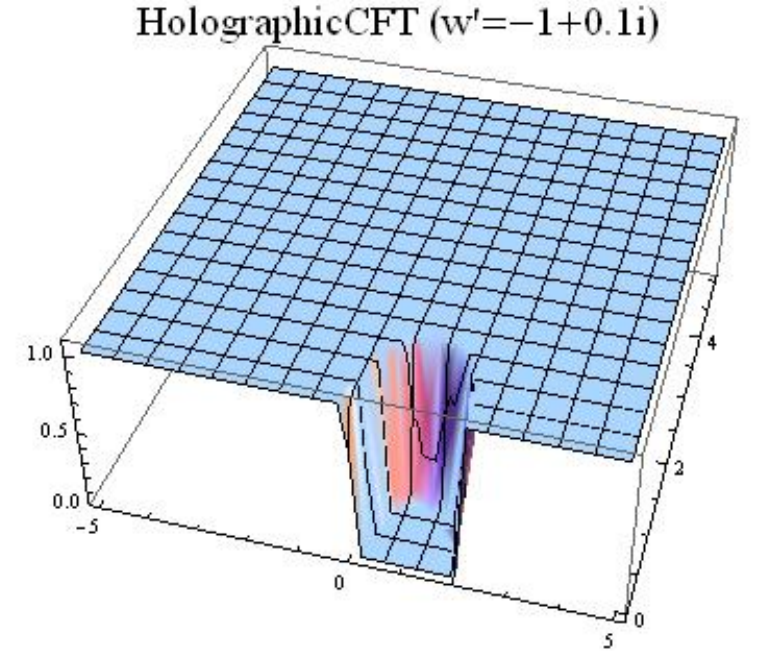}
 \includegraphics[width=6cm]{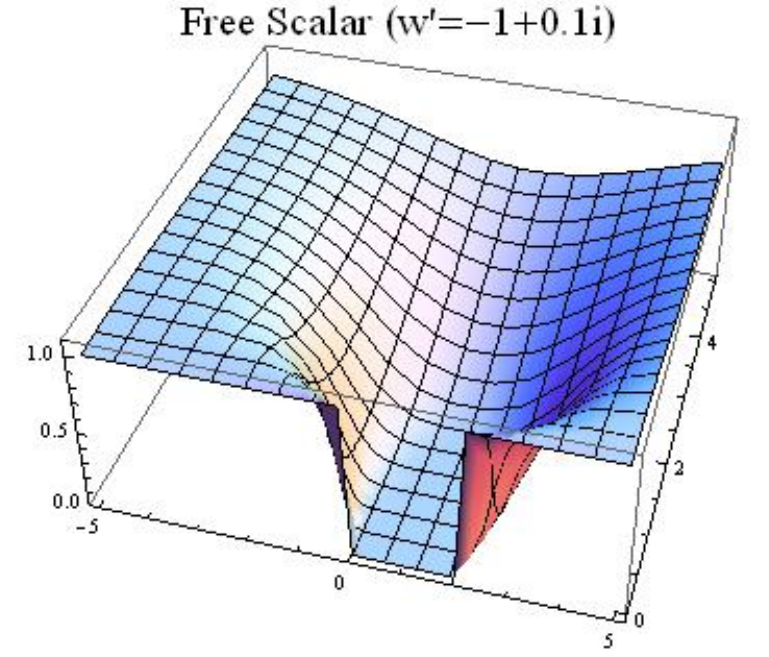}
 \includegraphics[width=6cm]{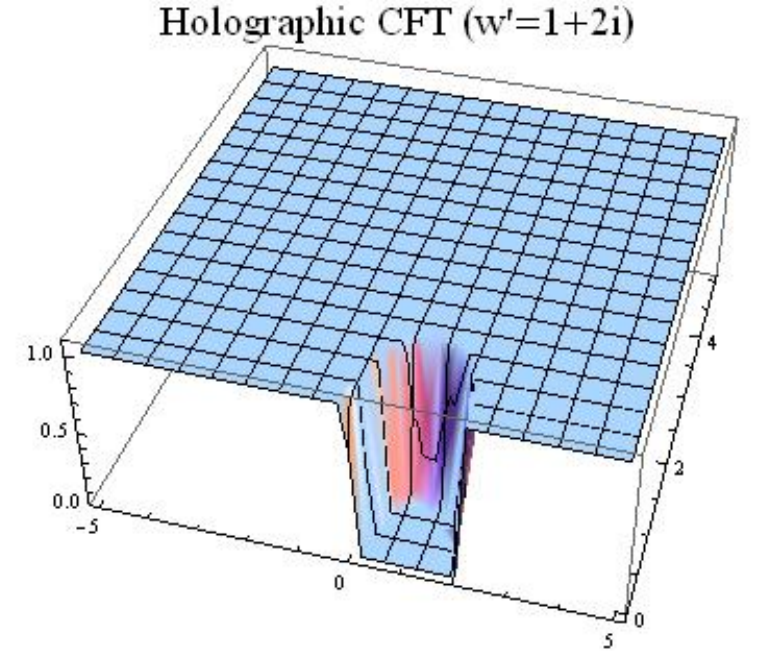}
 \includegraphics[width=6cm]{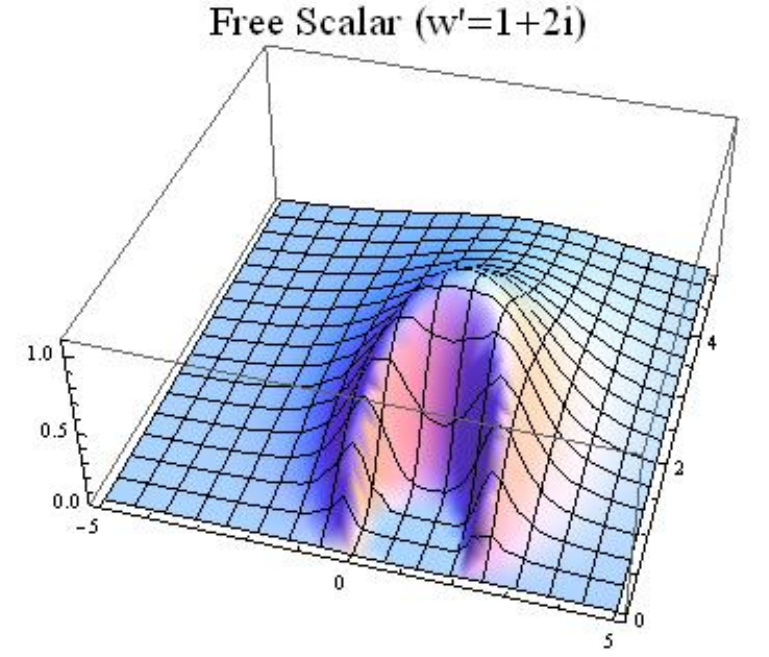}
  \caption{The value of $I(\rho,\rho')$ as a function of $\text{Re}[w]$ (horizontal axis) 
and $\text{Im}[w]$ (depth axis) when $w'$ is fixed outside the CFT wedge. 
In particular, we chose $h_\ap=10$ 
and $A=[0,2]$ (i.e.$L=2$).  The upper two graphs are for  $w'=-1+0.1i$ and 
the lower ones are for $w'=1+2i$, both of which are outside of the wedge. 
The left and right graphs describe the result for the holographic CFT and the $c=1$ free scalar CFT, respectively.
We find that the wedge structure is sharp only in the holographic CFT. For free scalar CFT, we can detect an excitation even outside of the wedge.}
\label{fig:Outsidew}
  \end{figure}

We also plotted the profiles of $I(\rho,\rho')$ in left graphs of Fig.\ref{fig:Insidew} and Fig.\ref{fig:Outsidew}.
The left one in Fig.\ref{fig:Insidew} shows $I(\rho,\rho')$ as a function of $w$, where $w'$ is fixed inside the CFT wedge. We observe the clear peak at $w=w'$, which will be highly localized in the limit $h_\ap\gg 1$. 
In the left ones of Fig.\ref{fig:Outsidew}, we fixed $w'$ outside of the CFT wedge. We can observe a clear 
entanglement wedge structure, where we have $I\simeq 0$ inside and $I\simeq 1$ outside.

\subsection{Free Scalar $c=1$ CFT}

To understand how the properties of holographic CFTs are relevant to the emergence of entanglement wedges in the gravity duals, consider the free massless scalar CFT ($c=1$ CFT) in two dimension. We choose the operator $O_\ap$ to be 
\ba
O_\ap(w,\bar{w})=e^{ip(\phi(w)+\phi(\bar{w}))}, \label{qopfree}
\ea
where $\phi(w)$ and $\phi(\bar{w})$ are chiral and anti-chiral massless scalar field.  
Note that the conformal dimension of 
the above operator is $h_\ap=\bar{h}_\ap=\frac{p^2}{2}$. In this case we obtain
\be
F(z,\bar{z},-z',-\bar{z}')=\frac{|z+z'|^{8h}}{|z-\bar{z}|^{4h}|z'-\bar{z}'|^{4h}|z+\bar{z}'|^{8h}}.
\ee
We can easily estimate (\ref{dcsi}) analytically and obtain 
\ba
I(\rho,\rho')= \left(\frac{|z+z'|^2|z+\bar{z}||z'+\bar{z}'|}{4|z||z'||z+\bar{z}'|^2}\right)^{4h},
\ea
for any values of $w$ and $w'$. Note that in these excited states, we always have $\mbox{Tr}[\rho^2]=\mbox{Tr}[\rho'^2]=1$
as they do not generate entanglement between left and right moving modes \cite{Nozaki,HNTW}.

Thus in this free scalar CFT, there is no sharp CFT wedge structure as 
expected for non-holographic CFTs.
The numerical plots are in the right graphs in  Fig.\ref{fig:Insidew} and Fig.\ref{fig:Outsidew}.
Even though we can observe a peak when $w$ is inside the CFT wedge (see Fig.\ref{fig:Insidew})
which is similar to the holographic case, 
we do not find any sharp CFT wedge  when $w$ is outside the wedge (see Fig.\ref{fig:Outsidew}).  
In this way we can conclude that there is no emergence of
entanglement wedge in $c=1$ CFT as expected.

\subsection{Two Different Operators}

So far we assumed that both $\rho_A$ and $\rho'_A$ are created by the same local operator $O_\ap$ 
as in  (\ref{redb}).
It is also instructive to consider the case where  $\rho_A$ and $\rho'_A$
are created by two orthogonal operators 
$O_\ap$ and $O_\beta$, respectively (each chiral conformal dimension $h_\ap$ and $h_\beta$) such that the two point function $\la O_\ap O_\beta \lb$ vanishes.  
We would like to calculate $I(\rho_A,\rho'_A)$ in this case.  Again we can use the expression (\ref{dcfp}) as
\ba
I(\rho_A,\rho'_A)=\frac{\la O^\dagger_\ap(z_1) O_\ap(z_2) O^\dagger_\beta(z'_3) O_\beta(z'_4) \lb}{\s{
\la O^\dagger_\ap(z_1) O_\ap(z_2) O^\dagger_\ap(z_3) O_\ap(z_4) \lb.
\la O^\dagger_\beta(z'_1) O_\beta(z'_2) O^\dagger_\beta(z'_3) O_\beta(z'_4) \lb.}},  \label{ratisame}
\ea
where we can write $z_1=z, z_2=\bar{z}, z_3=-z, z_4=-\bar{z}$ etc.

Now we would like to evaluate this in holographic CFTs, by applying the large $c$ factorization (generalized free field prescription).
First of all, we can always estimate 
\ba
\la O^\dagger_\ap(z_1) O_\ap(z_2) O^\dagger_\beta(z'_3) O_\beta(z'_4) \lb\simeq |z-\bar{z}|^{-4h_\ap}\cdot |z'-\bar{z'}|^{-4h_\beta}.
\ea
Depending on whether $z\simeq z'$ is inside or outside of the CFT wedge 
(\ref{inent}) or (\ref{outent}) we find
\ba
&& \mbox{Inside EW:}\  \ \la O^\dagger_\ap(z_1) O_\ap(z_2) O^\dagger_\ap(z_3) O_\ap(z_4) \lb\simeq  |z-\bar{z}|^{-8h_\ap}, \no
&& \mbox{Outside EW:}\ \ \la O^\dagger_\ap(z_1) O_\ap(z_2) O^\dagger_\ap(z_3) O_\ap(z_4) \lb\simeq  |z+\bar{z}|^{-8h_\ap}.
\ea
Thus we can evaluate (\ref{ratisame}) as follows
\ba
&& \mbox{Inside EW:}\ \  I(\rho_A,\rho'_A)\simeq \left|\frac{z+\bar{z}}{z-\bar{z}}\right|^{4h_\ap}\cdot \left|\frac{z'+\bar{z'}}{z'-\bar{z'}}\right|^{4h_\beta}\simeq 0, \no
&& \mbox{Outside EW:}\ \ I(\rho_A,\rho'_A)\simeq 1.
\ea

This nicely fits with the entanglement wedge structure in AdS/CFT: we can distinguish two different operators inside the wedge, while we cannot outside.
In particular, since this analysis can be applied to the case $O_\beta$ is the identity operator, $\rho_A$ cannot be distinguished from the vacuum one (no insertions of operators), if 
the insertion of $O_\ap$ is outside the wedge.

\section{Bures Metric in the Single Interval Case}

So far we studied the measure $I(\rho,\rho')$.  Instead, here we would like to calculate the Bures distance $D_B(\rho,\rho')$ 
define by (\ref{disb}) 
and Bures metric $G_B$ defined by (\ref{infometr}) in the same 
setup. This problem is essentially the computation of  the following trace
\ba
A_{n,m}(\rho,\rho')=\mbox{Tr}[(\rho^m\rho'\rho^m)^n]. \label{amn}
\ea
By analytically continuing $n$ and $m$ and setting $n=1/2$ and $m=1/2$, we obtain the fidelity. 
\ba
A_{1/2,1/2}(\rho,\rho')=\mbox{Tr}[\s{\s{\rho}\rho'\s{\rho}}]=F(\rho,\rho').
\ea
Below we will employ this replica-like method below to calculate the fidelity.

\begin{figure}
  \centering
  \includegraphics[width=10cm]{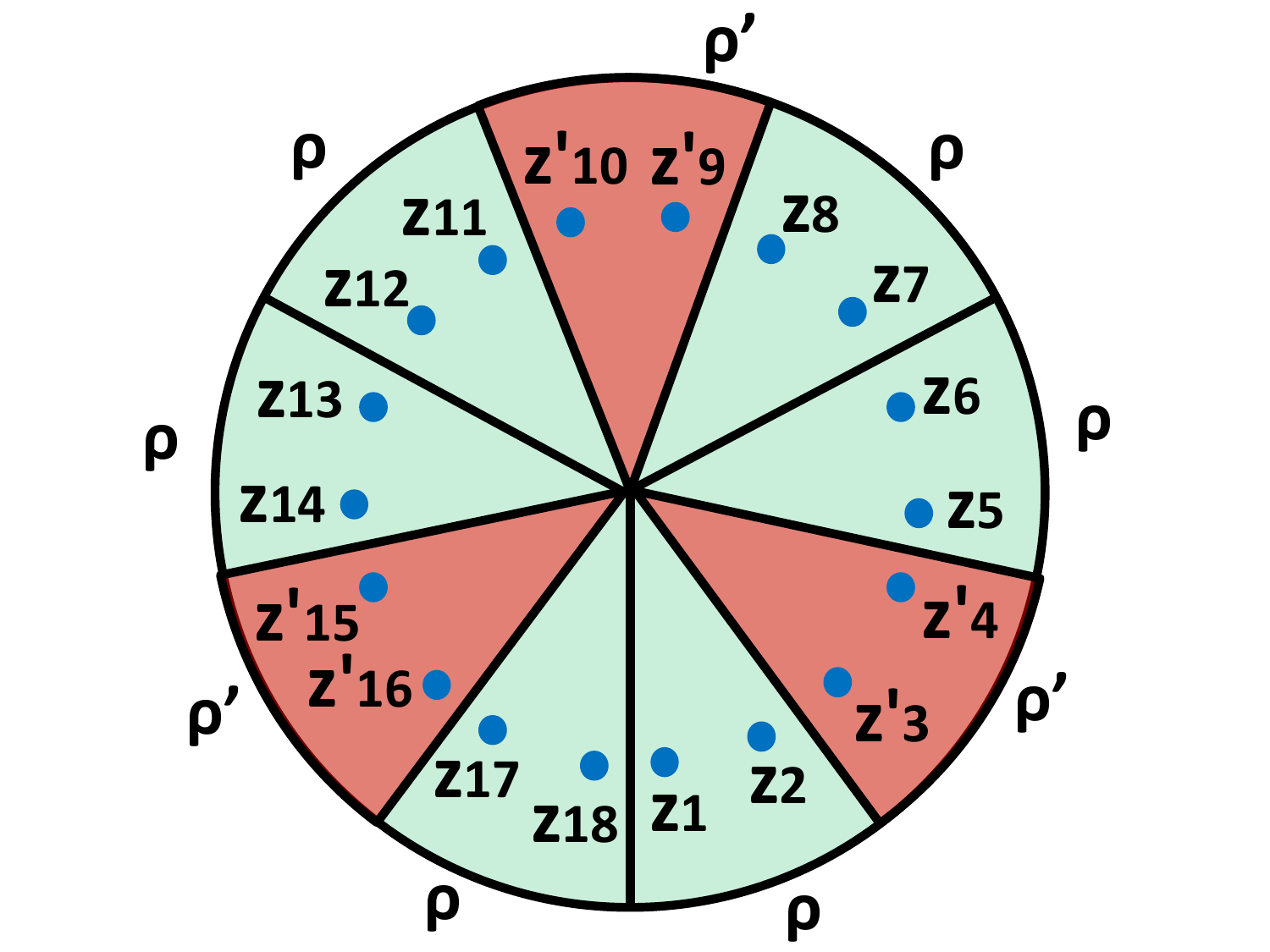}
  \caption{The complex plane which describes the path-integral which calculates the trace $A_{n,m}=\mbox{Tr}[(\rho^m\rho'\rho^m)^n].$ i.e. (\ref{amn}), where we performed the conformal transformation (\ref{confglk}). Here we choose $m=1$ and $n=3$ for convenience.}
\label{fig:BuresS}
  \end{figure}

For this we apply the conformal transformation 
\be
z^{k}=\frac{w}{w-L},  \label{confglk}
\ee
where
\ba
k=(2m+1)n,  \label{kdefp}
\ea
so that the path-integrals for $2mn$ $\rho$s and $n$ $\rho'$s are mapped into that on a single plane,
with the correct order of $\rho$s and $\rho'$s specified by  (\ref{amn}).  Refer to Fig,\ref{fig:BuresS} for 
a sketch of the geometry after the conformal transformation. This map is similar to the ones employed 
for the calculations of relative entropy \cite{Nima,Ug}.

Then $A_{n,m}$ is written as the $2k$-point function divided by the 
normalization of $\mbox{Tr}[\rho]$ and  $\mbox{Tr}[\rho']$ i.e, two point functions:
\ba
A_{n,m}=\frac{\la O_\ap^\dagger(w_1)O_\ap(w_2)\ddd O_\ap^\dagger(w_{2k-1})O_\ap(w_{2k})\lb}
{\prod_{i=1}^k \la O_\ap^\dagger(w_{2i-1})O_\ap(w_{2i})\lb}\cdot
\f{Z^{(k)}}{(Z^{(1)})^k}. \label{ratq}
\ea 
Here $Z^{(k)}$ is the vacuum partition function with $k$-replicated space.
The $2k$-point function in the $w$-plane is mapped into that in the $z$-plane as follows
\ba
&& \la O_\ap^\dagger(w_1)O_\ap(w_2)\ddd O_\ap^\dagger(w_{2k-1})O_\ap(w_{2k})\lb \no
&& =\prod_{i=1}^{2k}\left|\frac{dz_i}{dw_i}\right|^{2h}\cdot
\la O_\ap^\dagger(z_1)O_\ap(z_2)\ddd O_\ap^\dagger(z_{2k-1})O_\ap(z_{2k})\lb. 
\ea
Since we have 
\ba
\frac{dz}{dw}=-\frac{z^{1-k}(z^k-1)^2}{kL},
\ea
and 
\ba
\la O_\ap^\dagger(w)O_\ap(w')\lb=\left|\frac{(z^k-1)(z'^k-1)}{L(z'^k-z^k)}\right|^{4h_\ap},
\ea
the ratio (\ref{ratq}) can be rewritten as 
\ba
A_{n,m}=\prod_{i=1}^{2k}\left|\frac{(z_i)^{1-k}}{k}\right|^{2h_\ap}\cdot \prod_{j=1}^k |(z_{2j-1})^k-(z_{2j})^k|^{4h_\ap}
\cdot \la O_\ap^\dagger(z_1)O_\ap(z_2)\ddd O_\ap^\dagger(z_{2k-1})O(z_{2k})\lb\cdot
\f{Z^{(k)}}{(Z^{(1)})^k}. \no  \label{cosing}
\ea
Note that we have 
\ba
&& z_1=\left(\frac{-x-i\tau}{L-x-i\tau}\right)^{1/k},\ \ \ z_2(=\bar{z}_1)=\left(\frac{-x+i\tau}{L-x+i\tau}\right)^{1/k}, \no 
&& z_{2s+1}=e^{\frac{2\pi i}{k}s}z_1,\ \ \ z_{2s+2}=e^{\frac{2\pi i}{k}s}z_2,\ \ \ (s=1,2,\ddd,k-1).
\ea

As we will see in explicit evaluations, the analytical continuation $m=1/2$ is rather straightforward.
This allows us to define the convenient ratio:
\ba
A_n(\rho,\rho')=\frac{\mbox{Tr}[(\s{\rho}\rho'\s{\rho})^n]}{\s{\mbox{Tr}[\rho^{2n}]\mbox{Tr}[\rho'^{2n}]}}.  \label{antr}
\ea
We immediately find $A_{1}(\rho,\rho')=I(\rho,\rho')$ and $A_{1/2}(\rho,\rho')=F(\rho,\rho')$.

\subsection{Bures Metric in Holographic CFT for Poincare AdS$_3$}

Let us focus on a holographic 2d CFT.  The leading contribution is again given by the generalized free field prescription. When $w$ and $w'$ are outside the CFT wedge (\ref{outent}), we can approximate the $2k$ point function as 
\ba
\la O_\ap^\dagger(z_1)O_\ap(z_2)\ddd O_\ap^\dagger(z_{2k-1})O_\ap(z_{2k})\lb
\simeq  \prod_{j=1}^k \la O_\ap^\dagger(z_{2j-1})O_\ap(z_{2j})\lb
\simeq  \prod_{j=1}^k |z_{2j-1}-z_{2j}|^{-4h_\ap}.\ \  \ 
\ea
In this case we get the trivial Bures distance 
\ba
D_B(\rho,\rho')^2=2(1-A_{1/2,1/2})\simeq 0,    \label{outburesm}
\ea
where note that $k\to 1$ in this limit. Thus the Bures metric $G_{Bij}$ are
all vanishing in the outside wedge case.

On the other hand, when $w$ and $w'$ are inside the CFT wedge (\ref{inent}), we can approximate 
\ba
 \la O^\dagger(z_1)O(z_2)\ddd O^\dagger(z_{2k-1})O(z_{2k})\lb & \simeq&  \prod_{j=1}^k \la O^\dagger(z_{2j-2})O(z_{2j-1})\lb\no
& \simeq & \prod_{j=1}^k |z_{2j-2}-z_{2j-1}|^{-4h_\ap}, \no
& \simeq & |\bar{z}-e^{\frac{2\pi i}{k}}z'|^{-8h_\ap n} |\bar{z}-e^{\frac{2\pi i}{k}}z|^{-4h_\ap(2m-1)n},
\ea
where we regard $z_{0}=z_{2k}$. Thus we have 
\ba
A_{n,m}\simeq \prod_{i=1}^{2k}\left|\frac{(z_i)^{1-k}}{k}\right|^{2h_\ap}\cdot 
|z^k-\bar{z}^k|^{8h_\ap mn}|z'^k-\bar{z}'^k|^{4h_\ap n} 
 |\bar{z}-e^{\frac{2\pi i}{k}}z'|^{-8h_\ap n} |\bar{z}-e^{\frac{2\pi i}{k}}z|^{-4h_\ap(2m-1)n}\cdot
\f{Z^{(k)}}{(Z^{(1)})^k}.\no
\ea
In the limit $m=n\to 1/2$ ($k\to 1$), we find 
\ba
A_{1/2,1/2}=|z-\bar{z}|^{2h_\ap}|z'-\bar{z}'|^{2h}|z'-\bar{z}|^{-4h_\ap}
=|w-\bar{w}|^{2h_\ap}|w'-\bar{w}'|^{2h_\ap}|w'-\bar{w}|^{-4h_\ap},\no  \label{www}
\ea
where $z$ and $w$ are related by $z=\frac{w}{w-L}$ in the $k\to 1$ limit.
By assuming $dz=z'-z$ is infinitesimally small, we obtain the Bures metric
\ba
D_B(\rho,\rho')^2\simeq \frac{h_\ap}{\tau^2}(dx^2+d\tau^2).  \label{wwww}
\ea
Interestingly, this Bures metric coincides with that for the pure state (\ref{pureburesc}). 
Therefore, it is proportional to the metric on a time slice of AdS$_3$.
Remember that the original Euclidean time coordinate $\tau$ can be regarded as the radial coordinate 
$\eta$ via the intersection between the geodesic and the time slice as in Fig.\ref{fig:EW}.
This agreements between the information metric with the bulk metric is 
natural if we think the distinguishability 
in the quantum estimation theory is related to the bulk locality resolution. 
At the same time the agreement between the Bures metric for $\rho_A$ with local excitation inside the CFT wedge 
and that for the pure state, tells us us that we can perfectly reconstruct the information in the entanglement wedge 
from $\rho_A$. This supports the entanglement wedge reconstruction.

\subsection{Bures metric in Holographic CFT for Global AdS$_3$}

Next we turn to a holographic CFT dual to the Euclidean global AdS$_3$
\ba
ds^2=R^2(\cosh^2\rho d\tau^2+d \rho^2+\sinh^2\rho dx^2).  \label{gadsm}
\ea
This is a 2d holographic CFT with the space coordinate compactified on a circle $x\sim  x+2\pi$.
We choose the subsystem $A$ to be the interval $0\leq x\leq l$ at $\tau=0$.

By acting the conformal transformation $w=e^{\xi}$ with $\xi=\tau+ix$, we find  
\ba
A_{1/2,1/2}=\frac{|w-1/\bar{w}|^{2h_\ap}|w'-1/\bar{w}'|^{2h_\ap}}{|w-1/\bar{w'}|^{2h_\ap}|w'-1/\bar{w}|^{2h_\ap}}=
\left[\frac{2\cosh\tau\cosh\tau'}{\cosh(\tau+\tau')-\cos(x-x')}\right]^{2h_\ap}.  \label{geogl}
\ea
This leads to the following Bures metric inside the CFT wedge:
\ba
D_B^2=\frac{h_\ap}{\sinh^2\tau}(d\tau^2+dx^2).  \label{bgsl}
\ea

Since the geodesic in global AdS$_3$ which connects the two points $(\tau_0,x_0)$ and $(-\tau_0,x_0)$ at the boundary 
$\rho\to \infty$ looks like
\ba
e^{2\tau}=\frac{\sinh\rho+\s{\frac{\cosh^2\rho}{\cosh^2\rho_*}-1}}{\sinh\rho-\s{\frac{\cosh^2\rho}{\cosh^2\rho_*}-1}},
\ea
where $\rho_*$ is the intersection point of the time slice $\tau=0$ and this geodesic in the bulk AdS.  By taking the boundary limit 
$\rho\to \infty$
we find the relation
\ba
\sinh\tau_0=\frac{1}{\sinh\rho_*}.
\ea

By relating the boundary point $(\tau,x)$ to the bulk point $(\rho,x)$ on the time slice $\tau=0$ using this relation we can rewrite the metric (\ref{bgsl}) as follows:
\ba
D_B^2=h_\ap(d\rho^2+\sinh^2\rho dx^2),
\ea
which agrees with the time slice metric of the global AdS$_3$ (\ref{gadsm}).

\subsection{Bures metric in Holographic CFT for BTZ}

Consider a holographic CFT dual to the Euclidean BTZ (with a non-compact horizon)
\ba
ds^2=R^2\left(\left(\frac{2\pi}{\beta}\right)^2\sinh^2\rho d\tau^2+d \rho^2+\left(\frac{2\pi}{\beta}\right)^2 \cosh^2\rho dx^2\right). \label{btzmete}
\ea
This is given by  a 2d holographic CFT, with the space coordinate compactified on a circle $\tau\sim  \tau+\beta$.

By acting the conformal transformation $w=e^{\frac{2\pi}{\beta}\xi}$ with $\xi=x+i\tau$, we find  the following result in the case of 
non-trivial Wick contraction:
\ba
A_{1/2,1/2}=|w-\bar{w}|^{2h_\ap}|w'-\bar{w}'|^{2h_\ap}|w'-\bar{w}|^{-4h_\ap}=
\left[\frac{2\sin\left(\frac{2\pi}{\beta}\tau\right)\sin\left(\frac{2\pi}{\beta}\tau'\right)}
{\cos\left(\frac{2\pi(\tau+\tau')}{\beta}\right)-\cosh\left(\frac{2\pi(x-x')}{\beta}\right)}\right]^{2h_\ap}.
\ea
Note that we limit the range of $\tau$ to $-\beta/2\leq \tau\leq \beta/2$.

This leads to the following Bures metric inside the wedge:
\ba
D_B^2=h_\ap\frac{\left(\frac{2\pi}{\beta}\right)^2}{\sin^2\left(\frac{2\pi}{\beta}\tau\right)}(d\tau^2+dx^2).  \label{btzbrm}
\ea

Since the geodesic in BTZ which connects the two points $(\tau_0,x_0)$ and $(-\tau_0,x_0)$ at the boundary $\rho=\infty$  looks like
\ba
e^{i\frac{4\pi}{\beta}\tau}=\frac{\cosh\rho+i\s{\frac{\sinh^2\rho}{\sinh^2\rho_*}-1}}{\cosh\rho-i\s{\frac{\sinh^2\rho}{\sinh^2\rho_*}-1}},
\label{gepobtz}
\ea
where $\rho_*$ is the intersection point of the time slice $\tau=0$ and this geodesic in the bulk.  Note that (\ref{geogl}) in global AdS and 
(\ref{gepobtz}) in BTZ are related by the familar coordinate transformation 
\ba
(\rho,\tau,x)\to (\rho+i\pi/2,i\tau,ix).
\ea

By taking the boundary limit $\rho=\infty$
we find the relation
\ba
\sin\left(\frac{2\pi}{\beta}\tau_0\right)=\frac{1}{\cosh\rho_*}.  \label{wwwq}
\ea

By relating the boundary point $(\tau,x)$ to the bulk point $(\rho,x)$ on the time slice $\tau=0$ using this relation we can rewrite the metric (\ref{btzbrm}) as follows
\ba
D_B^2=h_\ap\left(d\rho^2+\left(\frac{2\pi}{\beta}\right)^2\cosh^2\rho dx^2\right), 
\ea
which agrees with the time slice metric of the BTZ (\ref{btzmete}).

Moreover, we can confirm also the CFT wedge in this case agrees with the entanglement wedge in BTZ as follows.
The condition for the non-trivial Wick contraction is $|z-\bar{z}|>|z+\bar{z}|$, where 
\ba
z^2=\frac{e^{\frac{2\pi}{\beta}(x+i\tau)}-1}{e^{\frac{2\pi}{\beta}(x+i\tau)}-e^{\frac{2\pi}{\beta}l}}.
\ea
This leads to the condition 
\ba
\left[e^{\frac{2\pi}{\beta}l}\sin\left(\frac{2\pi\tau}{\beta}\right)\right]^2
+\left(e^{\frac{2\pi}{\beta}l}\cos\left(\frac{2\pi\tau}{\beta}\right)-1\right)\left(e^{\frac{2\pi}{\beta}l}\cos\left(\frac{2\pi\tau}{\beta}\right)-e^{\frac{2\pi}{\beta}l}\right)
\leq 0. \label{ewbtzz}
\ea

On the other hand, the geodesic which connects $x=0$ and $x=l$ (on the slice $\tau=0$) in the BTZ geometry is found as 
\ba
\frac{\cosh\left[\frac{2\pi}{\beta}\left(x-\f{l}{2}\right)\right]}{\sinh\left[\frac{2\pi}{\beta}\left(x-\f{l}{2}\right)\right]}
=\frac{\cosh\rho_*\sinh\rho}{\s{\cosh^2\rho-\cosh^2\rho_*}},  \label{geoffr}
\ea
where 
\be
\cosh\rho_*=\frac{\cosh\left(\frac{\pi l}{\beta}\right)}{\sinh\left(\frac{\pi l}{\beta}\right)}.
\ee
This coincides with the border of (\ref{ewbtzz}) via the relation between $\tau$ and $\rho$ given by (\ref{wwwq}).

\subsection{Bures Distance for Different Operators}

Next we consider the Bures distance $D_B(\rho_A,\rho'_A)$, where $\rho_A$ and 
$\rho'_A$ are defined by locally excited operators $O_\ap(w,\bar{w})$ and 
$O_\beta(w',\bar{w}')$, which are orthogonal to each other.
Let us work out the behavior of Bures distance by computing $A_n$ introduced in
(\ref{antr}) and taking the limit $n=1/2$. Using the expression (\ref{cosing}), we eventually 
find
\ba
&&\mbox{If $w$ and $w'$ are both outside the CFT wedge, then} \ A_n\simeq 1,\no
&&\mbox{If $w$ and $w'$ are both inside the CFT wedge,}\no
&& \mbox{then}\ A_n=\left|\frac{z-e^{\frac{\pi i}{n}}\bar{z}}{z-\bar{z}}\right|^{4h_\ap n}\cdot 
\left|\frac{z'-e^{\frac{\pi i}{n}}\bar{z}'}{z'-\bar{z}'}\right|^{4h_\beta n}\simeq 0,\no
&& \mbox{If $w$ is inside and  $w'$ are outside the CFT wedge,}\no
&&  \mbox{then}\ A_n=\left|\frac{z-e^{\frac{\pi i}{n}}\bar{z}}{z-\bar{z}}\right|^{4h_\ap n}\simeq 0. \label{ewdifb}
\ea
Here we used the assumption  $h_\ap,h_\beta\gg 1$ and noted that the inside  CFT wedge region is given by 
$|z-e^{\frac{\pi i}{n}}\bar{z}|<|z-\bar{z}|$. By taking the $n=1/2$ limit, the 
fidelity behaves as follows:
\ba
&&\mbox{If $w$ and $w'$ are both outside the CFT wedge, then} \ F(\rho,\rho')\simeq 1,\no
&&\mbox{If otherwise, then }\ F(\rho,\rho')\simeq 0.
\ea
 The above behaviors precisely agree with what we expect from the 
entanglement wedge reconstruction.

\subsection{Bures Distance in Free Scalar $c=1$ CFT}

\begin{figure}[h]
  \centering
  \includegraphics[width=6cm]{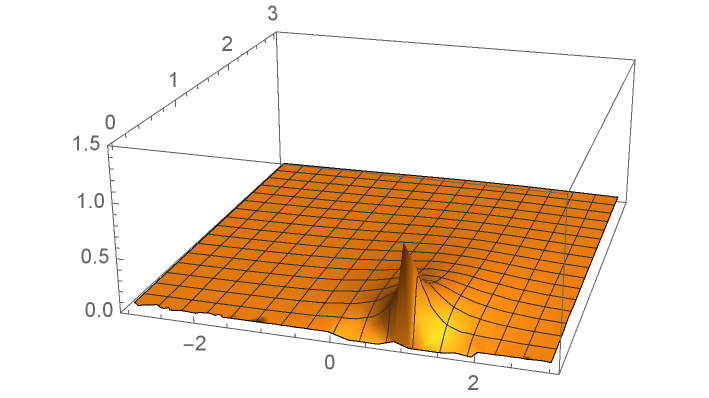}
 \includegraphics[width=6cm]{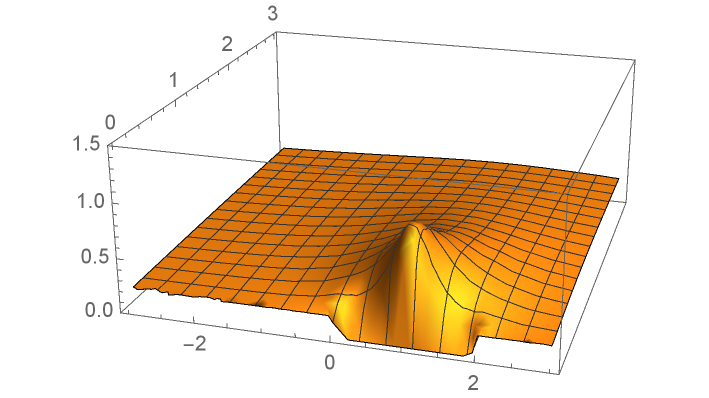}
 \includegraphics[width=6cm]{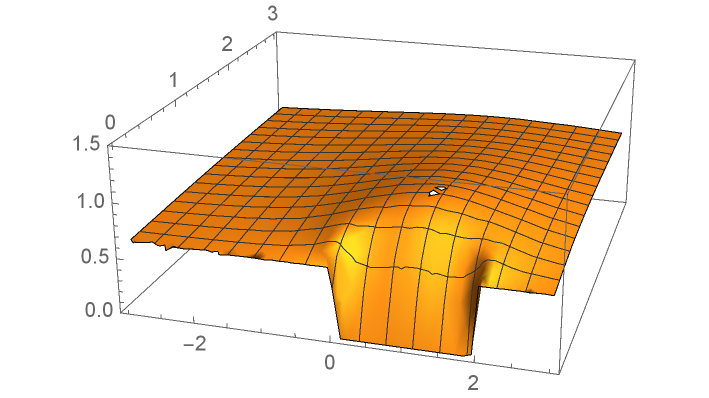}
 \includegraphics[width=6cm]{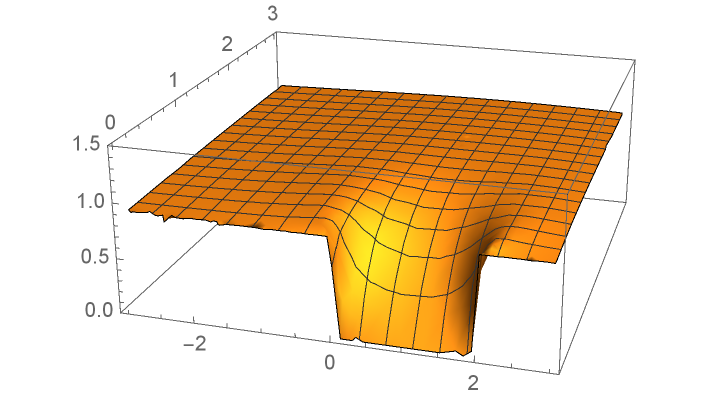}
  \caption{The profile of the Fidelity  $A_{n=1/2,m=1/2}=\mbox{Tr}[\s{\s{\rho}\rho' \s{\rho}}]$ in $c=1$ free scalar CFT for 
the operator $O=e^{i\phi}$ which has the dimension $h=1/2$ when we changes of value of $w$.
The upper left, upper right, lower left and lower right graphs describe  $A_{n=1/2,m=1/2}$  
for $w=1+0.05i$, $w=1+0.2i$, $w=1+0.8i$ and $w=1+2i$, respectively. We plotted 
$A_{n=1/2,m=1/2}$ as a function of $(p,q)$ for $\rho'(w'=p+iq)$. We chose $L=2$.}
\label{fig:c=1Bures2}
  \end{figure}

It is useful to compare the previous Bures metric in holographic CFTs 
with that in free scalar CFT. Consider a $c=1$ free scalar CFT and choose 
the primary operator $O_\ap$ to be (\ref{qopfree}) with $p=1/2$ for the simplification of calculations.
As we explain Appendix \ref{conebr}, in this case we can analytically evaluate $A_{n,m}$ and eventually we find the fidelity:
\ba
A_{1/2,1/2}=\frac{(\s{z}+\s{z'})(\s{\bar{z}}+\s{\bar{z'}})}{(\s{z}+\s{\bar{z}'})(\s{\bar{z}}+\s{z'})}
\cdot\frac{(\s{z}+\s{\bar{z}})(\s{z'}+\s{\bar{z'}})}{4\s{|z||z'|}},  \label{xbx}
\ea
where $z=w/(w-L)$. Several profiles of the fidelity are plotted  in Fig.\ref{fig:c=1Bures2}.

\begin{figure}[h]
  \centering
  \includegraphics[width=5cm]{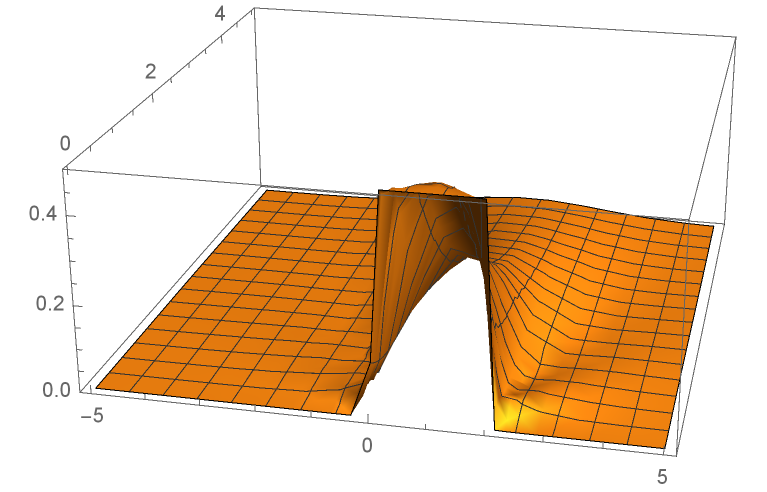}
  \includegraphics[width=5cm]{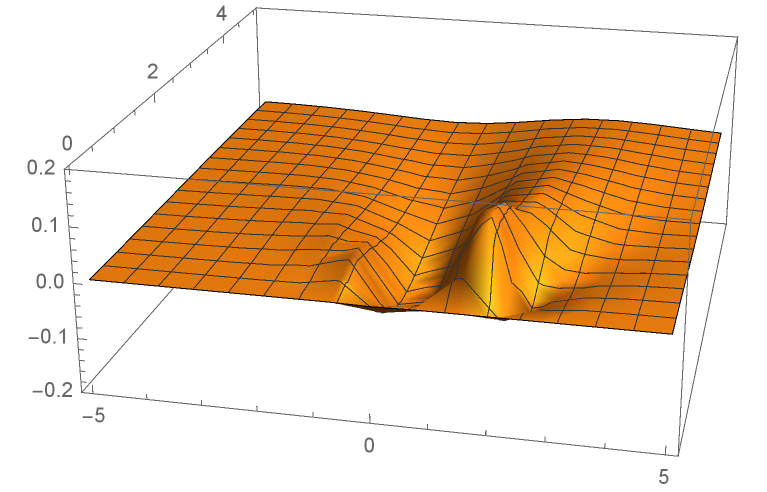}
  \includegraphics[width=5cm]{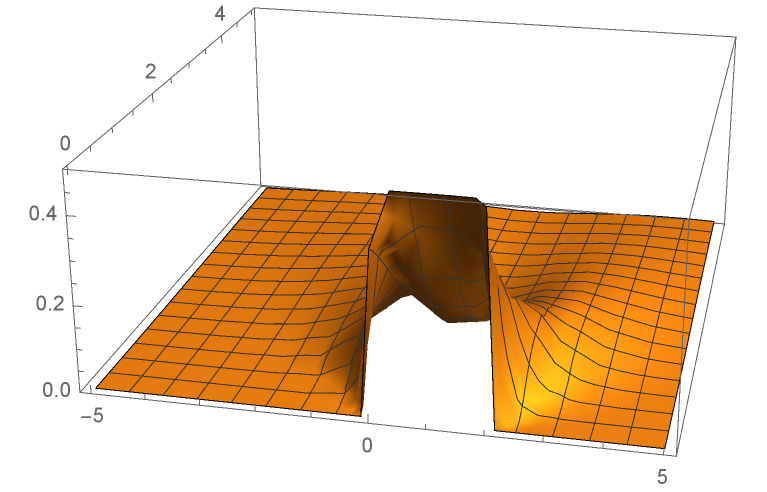}
  \caption{The profile of the Bures metric for $c=1$ free scalar CFT as a function of $w=x+i\tau$.  We plotted $\tau^2 G_{\tau\tau}$ (left), 
 $\tau^2 G_{\tau x}$ (middle), and  $\tau^2 G_{xx}$ (right) as a function of $x$ and $\tau$.  We chose $L=2$.
At the boundary $\tau\to 0$, we find $\tau^2 G_{\tau\tau,xx}\to\frac{1}{2}$ and  $\tau^2 G_{\tau x}\to 0$. }
\label{fig:BM}
  \end{figure}

The Bures metric for the free scalar can be found as 
\ba
D_B^2=-\frac{L^2(dw)^2}{16w^2(L-w)^2}-\frac{L^2(d\bar{w})^2}{16\bar{w}^2(L-\bar{w})^2}
+\frac{L^2}{\left(\s{\f{w}{w-L}}+\s{\f{\bar{w}}{\bar{w}-L}}\right)^2}\cdot\frac{(dw)(d\bar{w})}{2|w||w-L|^3}.
\ea
This metric is plotted in Fig.\ref{fig:BM}.  Note that we cannot find any sharp structure of CFT wedge as opposed to 
the holographic CFT. However, in the limit $\tau\to 0$, we find the metric $D_B^2\simeq \frac{h}{\tau^2}(d\tau^2+dx^2)$ 
for $0\leq x\leq L$.

\section{Time-Dependence}

In this section we would like to analyze how we can understand time evolutions of the 
CFT wedges and how they agree with  the AdS/CFT prediction.

Consider insertions of two operators $O_\ap$ and $O^\dagger_\ap$ 
at $w_1=x+i\tau_1$ and $w_2=x-i\tau_2$.
If we choose
\ba
\tau_1=\tau_0+it,\ \ \ \tau_2=\tau_0-it,  \label{timer}
\ea
then we can describe the Lorentzian time evolution of the state $e^{-\tau_0 H}O_\ap(x)|0\lb$.
 
The gravity dual of the two point function $\la O_\ap^\dagger(w_1,\bar{w}_1)O_\ap(w_2,\bar{w}_2)\lb$ 
is given by the geodesic in the Poincare AdS$_3$ which connects the two boundary points, given by 
\ba
\left(\tau-\frac{\tau_1-\tau_2}{2}\right)^2+\eta^2=\frac{(\tau_1+\tau_2)^2}{4}.
\ea
This intersects with the time slice $\tau=0$ at the point $\eta=\s{\tau_1\tau_2}$. 
Therefore the condition of inside the CFT wedge:
\ba
\left(x-\frac{L}{2}\right)^2+\eta^2\leq \frac{L^2}{4},  
\ea
is rewritten in terms of the CFT as follows
\be
x^2-Lx+\tau_1\tau_2\leq 0. \label{cftwt}
\ee
Below we would like to derive this condition from the information metric analysis. The crucial condition of the CFT wedge is 
\ba
|z_2-z_3|\leq |z_1-z_2|,  \label{condwe}
\ea
where
\ba
&& z_1=\s{\frac{-x-i\tau_1}{L-x-i\tau_1}}=-i\s{\frac{x+i\tau_1}{L-x-i\tau_1}},   \no
 &&  z_2=\s{\frac{-x+i\tau_2}{L-x+i\tau_2}}=i\s{\frac{x-i\tau_2}{L-x+i\tau_2}},  \no
&& z_3=-z_1.
\ea
This condition is rewritten as 
\ba
\mbox{Re}[\s{(x+i\tau_1)(x+i\tau_2)(L-x+i\tau_1)(L-x+i\tau_2)}]\geq 0.
\ea
This is equivalent to 
\ba
\mbox{Im}[(x+i\tau_1)(x+i\tau_2)(L-x+i\tau_1)(L-x+i\tau_2)]\geq 0,
\ea
or equally 
\ba
-(\tau_1+\tau_2)L(x^2-Lx+\tau_1\tau_2)\geq 0.
\ea
which finally reproduces the condition (\ref{cftwt}) derived from the
entanglement wedge structure in AdS/CFT.

\begin{figure}
  \centering
  \includegraphics[width=8cm]{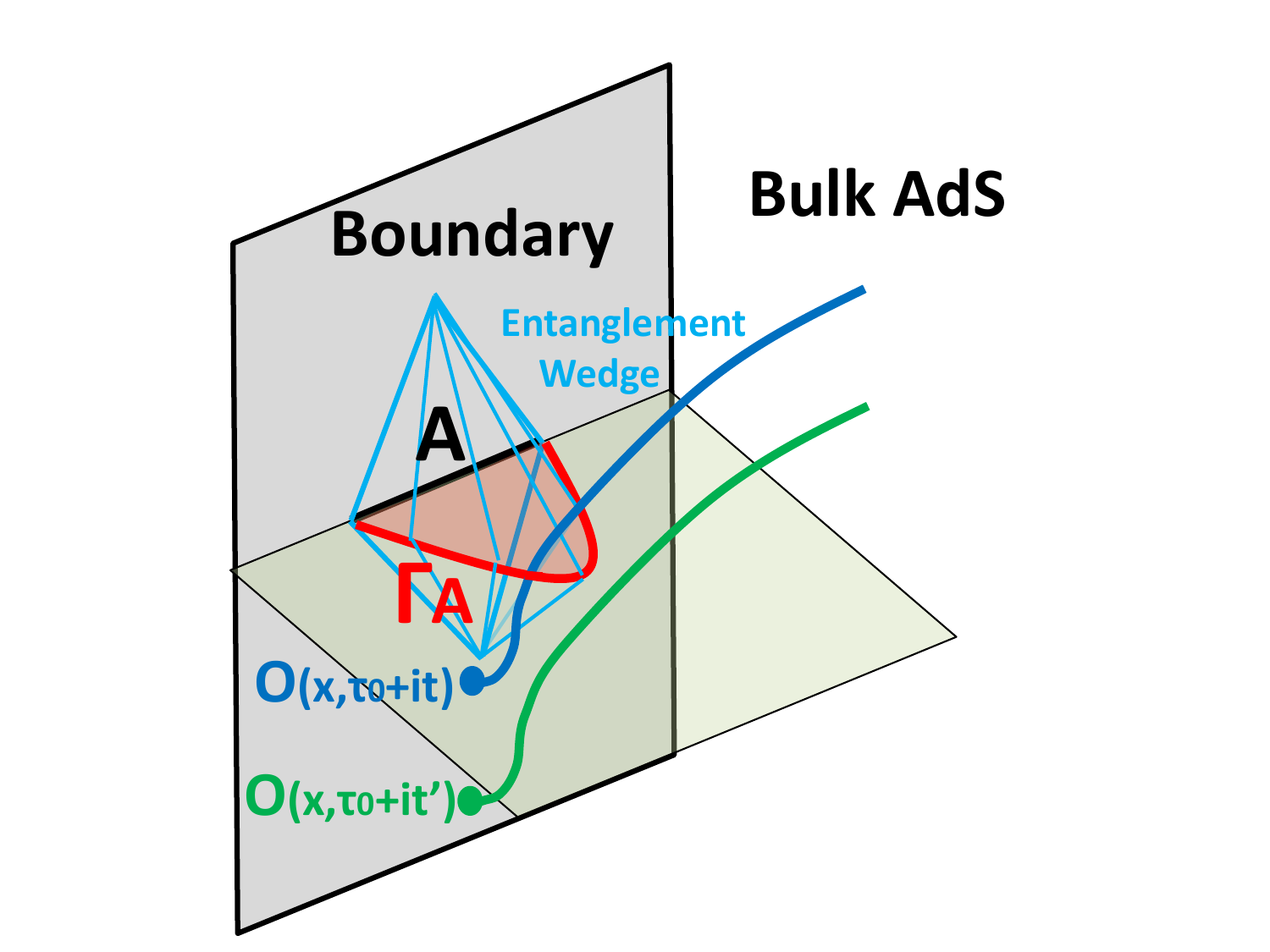}
  \caption{A sketch of time evolution of a local excitation in CFT and entanglement wedge in the gravity dual.}
\label{fig:EWL}
  \end{figure}

After the analytical continuation to the real time evolution (\ref{timer}), the CFT wedge is given by 
\ba
\left(x-\frac{L}{2}\right)^2+\tau_0^2+t^2\leq \frac{L^2}{4},  
\ea
This agrees with the entanglement wedge in AdS/CFT. Refer to Fig.\ref{fig:EWL} for a sketch.

The fidelity $A_{1/2,1/2}=F(\rho,\rho')$ is computed as follows
\ba
A_{1/2,1/2}&=&\left[\frac{|w_2-w_1||w'_2-w'_1|}{|w'_2-w_1||w_2-w'_1|}\right]^{2h_\ap} \no
&=&\left[\frac{|\tau_1+\tau_2|^2|\tau'_1+\tau'_2|^2}{\left((x'-x)^2+(\tau_1+\tau'_2)^2\right)\left((x'-x)^2+(\tau'_1+\tau_2)^2\right)}\right]^{h_\ap}.
\ea
This leads to the Bures metric in Euclidean space
\ba
D_B^2=2(1-A_{1/2,1/2})\simeq \frac{4h}{(\tau_1+\tau_2)^2}(dx^2+d\tau_1d\tau_2).  \label{met12}
\ea
We can actually see that this length coincides with the square of the minimal length between 
the geodesic which connects $w_1$ and $w_2$ and the one which connects $w'_1$ and $w'_2$.

If we substitute (\ref{timer}), then we have the Bures metric under the real time evoution: 
\ba
D_B^2=\frac{h}{\tau^2_0}(dx^2+dt^2).
\ea
Notice that even though we consider the Lorentzian time $t$, the metric is positive definite as 
follows from the definition of Bures metric.  Refer to  the Appendix \ref{generalt} for an analysis of 
Bures metric in more general time-dependent case.

\section{Double Interval Case}

Consider the reduced density matrix $\rho_A$ in a 2d CFT when $A$ consists of two disconnected intervals $A_1$ and $A_2$, which are  parameterized as
\ba
A_1=[0,s],\ \ \  A_2=[l+s,l+2s].
\ea
Owing to the conformal invariance, this parameterization is enough to cover all possible configurations of the 
double intervals. Then as in the single interval case, we insert a local operator $O_\ap$ 
at a point $w=x+i\tau$.
This defines a reduced density matrix $\rho_A$ (\ref{redb}) for the locally excited state.

\subsection{Conformal Map}

We employ the following conformal transformation (analogous to the one in  \cite{Raj}) which maps a complex plane ($w$-plane) with two slits along $A_1$ and $A_2$ into a cylinder (coordinate $z$):
\ba
z=f(w)=-J(\kappa^2)\left(\frac{1}{2K(\kappa^2)}\int^{\ti{w}}_0 \frac{dx}{\s{(1-x^2)(1-\kappa^2 x^2)}}
-\frac{1}{2}\right), \label{zfw}
\ea
where we introduced 
\ba
&& \ti{w}=\frac{2}{l}\left(w-s-\frac{l}{2}\right),\no
&& J(\kappa^2)=2\pi\frac{K(\kappa^2)}{K(1-\kappa^2)},\no
&& K(\kappa^2)=\int^1_0\frac{dx}{\s{(1-x^2)(1-\kappa^2 x^2)}},\no
&& \kappa=\frac{l}{l+2s}.
\ea
Note that we have 
\ba
\frac{dz}{dw}=-\frac{2\pi}{lK(1-\kappa^2)\s{(1-\ti{w}^2)(1-\kappa^2 \ti{w}^2)}}.
\ea
Also notice that we are considering the analytical continuation of the integral given by the Jacobi elliptic function:
\be
\int^{\ti{w}}_0 \frac{dx}{\s{(1-x^2)(1-\kappa^2 x^2)}}=\mbox{sn}^{-1}(\ti{w},\kappa^2).
\ee
It is useful to note the relation
\ba
\mbox{sn}^{-1} (\ti{w},0)=\arcsin(\ti{w}).  \label{sinr}
\ea

\begin{figure}
  \centering
  \includegraphics[width=10cm]{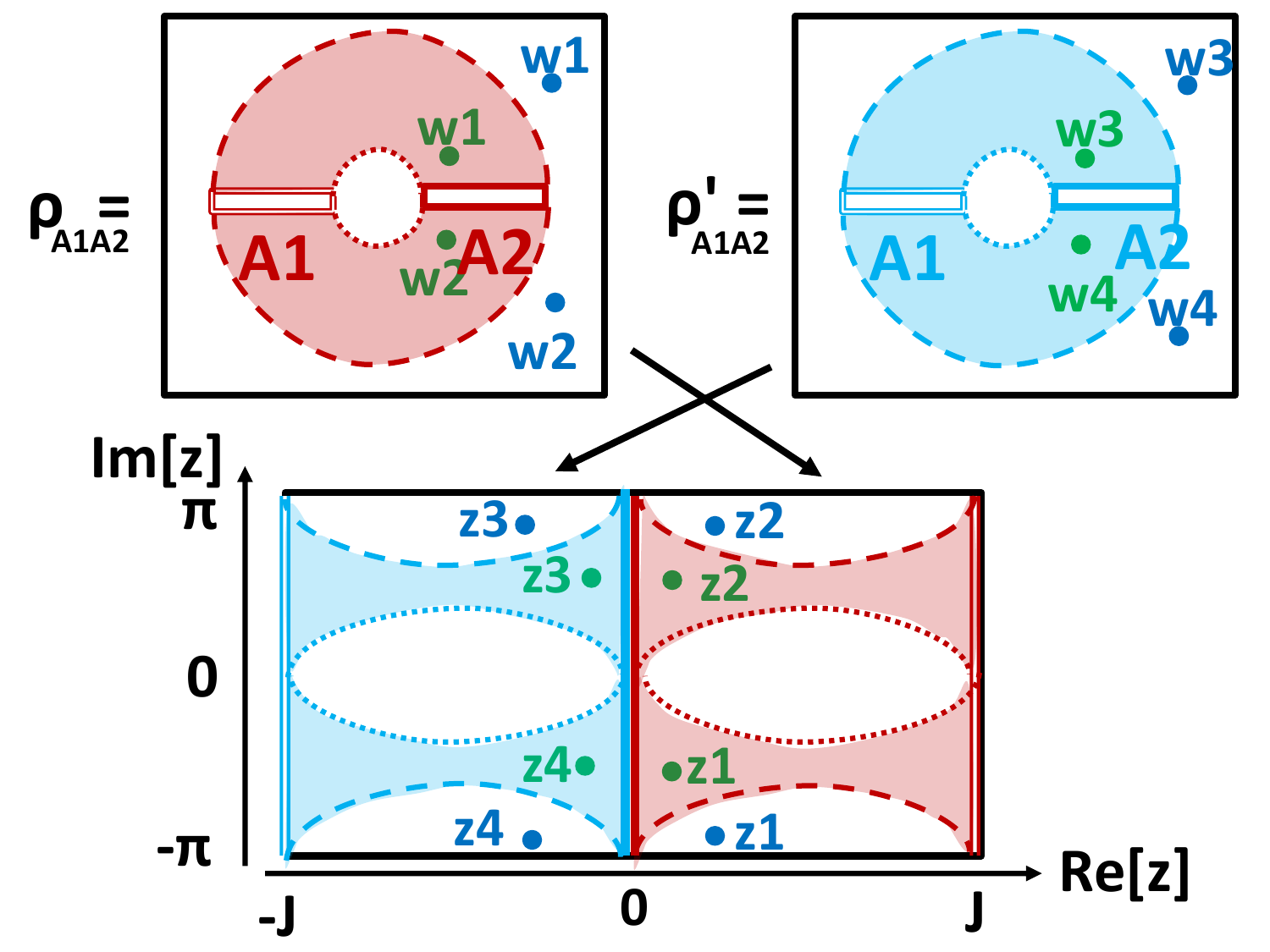}
  \caption{We sketched the conformal mapping for the calculation of Tr$[\rho_A\rho'_A]$ 
in the double interval case. Here we chose the phase (i), where  
the entanglement is connected, as shown by the colored region.
The lower picture described the geometry after the mapping and represents a torus by identifying Im$[z]\sim$ Im$[z]+2\pi$ and  Re$[z]\sim$ Re$[z]+2J$.
Green points 
(or bule points) describe the local excitations in the CFT which are dual to bulk excitations inside (or outside) of entanglement wedge $M_A$. }
\label{fig:double}
  \end{figure}

Consider the calculation of Tr$[\rho\rho']$, where $\rho=\rho_A(w,\bar{w})$ and $\rho'=\rho_A(w',\bar{w}')$.
Each of  $\rho$ and $\rho'$ is described by the path-integral on the complex plane with the two slits.
We can compute Tr$[\rho\rho']$ as the partition function on the space obtained by gluing the two complex planes along the slits. This is conformally mapped into a torus. This torus is constructed by gluing two cylinders: one of them describes $\rho$ and is obtained by performing the transformation $z=f(w)$ 
in (\ref{zfw}). Another one corresponds to $\rho'$ and is obtained from another transformation $z=-f(w)$.
These conformal maps the original two sheeted geometry into a torus is depicted in Fig.\ref{fig:double}. The horizontal and vertical length of the torus are given by $2J$ and $2\pi$, respectively.

Finally we find that $I(\rho,\rho')$ is given by the same formula as in the single interval case
(\ref{dcsi}), where $F$ is the torus four point function. Below in coming subsection, we will study the CFT wedge geometry by focusing on holographic CFTs.

\subsection{CFT Wedges from $I(\rho,\rho')$ in Holographic CFTs}

In holographic CFTs, we need to distinguish two phases depending on the moduli of the torus  \cite{He}:
\ba
&& (i)\  \mbox{Connected phase}: J<\pi \ \ \mbox{or equally}\ \kappa < 3-2\s{2} ,\no
&& (ii)\ \mbox{Disconnected phase}: J>\pi \ \ \mbox{or equally}\ \kappa > 3-2\s{2} .\no
\ea
In the first phase $(i)$, the entanglement wedge gets connected because $s^2>(2s+l)l$ 
i.e. $S_{A_1}+S_{A_2}>S_{A_1A_2}$. In this case, the AdS$_3/$CFT$_2$ duality
tells us  the entanglement wedge $M_A$ 
in the Poincare AdS (\ref{po}) looks like
\ba
M^{Con}_A:\ \  \frac{l^2}{4}\leq \left(x-s-\frac{l}{2}\right)^2+\eta^2 \leq \left(\frac{l}{2}+s\right)^2,  \label{ewdb}
\ea
on the time slice $\tau=0$.
In terms of the location of the local operator $O_\ap$ insertion, the corresponding CFT wedge
is expected to be  
\ba
C^{Con}_{A}:\ \ 
\frac{l^2}{4}\leq \left(x-s-\frac{l}{2}\right)^2+\tau^2 \leq \left(\frac{l}{2}+s\right)^2,  \label{ewdbb}
\ea

On the other hand, in the latter phase $(ii)$, the entanglement wedge gets disconnected as $s^2<(2s+l)l$,
i.e. $S_{A_1}+S_{A_2}<S_{A_1A_2}$. In this case, the entanglement wedge $M_A$ 
in the Poincare AdS (\ref{po}) is found to be $M^{Dis}_A=M^{Dis(1)}_A\cup M^{Dis(2)}_A$, where
\ba
&& M^{Dis(1)}_A:\ \  \left(x-\frac{s}{2}\right)^2+\eta^2\leq \frac{s^2}{4}, \no
&& M^{Dis(2)}_A:\ \  \left(x-\frac{3s}{2}-l\right)^2+\eta^2\leq \frac{s^2}{4}. \label{ewdisb}
\ea
The corresponding CFT wedge reads 
\ba
&& C^{Dis(1)}_A:\ \  \left(x-\frac{s}{2}\right)^2+\tau^2\leq \frac{s^2}{4}, \no
&& C^{Dis(2)}_A:\ \  \left(x-\frac{3s}{2}-l\right)^2+\tau^2\leq \frac{s^2}{4}. \label{ewdcftw}
\ea

Now let us work out  the CFT wedge from the calculation of $I(\rho,\rho')$ in holographic CFTs.
The two point functions on the torus in the phase $(i)$ and $(ii)$ behave like 
\ba
&& \la O^\dagger_\ap(z,\bar{z})O_\ap(z',\bar{z}')\lb_{(i)}\simeq 
\left|\sin\left(\frac{\pi(z+2\pi in_1-z')}{2J}\right)\right|^{-4h_\ap},\no
&& \la O^\dagger_\ap(z,\bar{z})O_\ap(z',\bar{z}')\lb_{(ii)}\simeq \left|\sinh\left(\frac{(z+2J n_2-z')}{2}
\right)\right|^{-4h_\ap},\label{corholdb}
\ea
where we assumed that $\left|\sin\left(\frac{\pi(z+2\pi in-z')}{2J}\right)\right|$ takes the smallest value among all integer $n$ at $n=n_1$ for the phase $(i)$ and that $\left|\sinh\left(\frac{(z+2J n_2-z')}{2}
\right)\right|$ takes the smallest value among all integer $n$ at $n=n_2$ for the phase $(ii)$.

This expression of two point functions (\ref{corholdb}) follows from the standard fact in AdS$_3/$CFT$_2$
that the gravity dual of the torus is given by a solid torus. We can construct the dual solid torus by filling 
the inside of the torus such that the circle Re$[z]$ (or Im$[z]$) shrinks to zero size in the bulk 
when we consider the phase $(i)$ (or $(ii)$), respectively. This is due to the well-known Hawking-Page phase transition \cite{HaPa} and matches perfectly with the large $c$ CFT analysis \cite{He}. 
 .

In holographic CFTs, we can rewrite the value of $I(\rho,\rho')$ in holographic CFTs using the generalized free field approximation:
\ba
I(\rho,\rho')\simeq \frac{F(z_1,z_2,z'_3,z'_4)}{\s{F(z_1,z_2,z_3,z_4)F(z'_1,z'_2,z'_3,z'_4)}},
\ea
where
\ba
&& F(z_1,z_2,z'_3,z'_4)  \no
&&=\mbox{Min}\Biggl[\la O^\dagger_\ap (z_1,\bar{z}_1)O_\ap(z_2,\bar{z}_2)\lb
\la O^\dagger_\ap(z'_3,\bar{z}'_4)O_\ap(z'_4,\bar{z}'_4)\lb,\no
&& \ \ \  \la O^\dagger_\ap(z_1,\bar{z}_1)O_\ap(z'_4,\bar{z}'_4)\lb 
\la O^\dagger_\ap (z_2,\bar{z}_2)O_\ap(z'_3,\bar{z}'_3)\lb\Biggr].\label{mintwo}
\ea
The locations $z_1,z_2$ and $z'_3,z'_4$ of the operator insertions are depicted in Fig.\ref{fig:double}, 
explicitly obtained via the map (\ref{zfw})  from the original insertion locations $w_1,w_2$ and $w'_3,w'_4$ in the double sheeted geometry which describes the path-integral of  $\mbox{Tr}[\rho\rho']$.

When the true minimum is the first one in (\ref{mintwo}), i.e. the trivial contraction, we simply find 
$I(\rho,\rho')=1$ and we cannot detect the local operator insertions. 
On the other hand, if the other one is favored as the minimum (i.e.  the non-trivial contraction), then 
$I(\rho,\rho')$ becomes a non-trivial function of the locations of operator insertions.

The condition that the non-trivial contraction is favored is given by 
\ba
\mbox{Min}\left[\left|\sin\left(\frac{\pi}{2J}(z_2-z_1)\right)\right|, \ 
 \left|\sin\left(\frac{\pi}{2J}(z_2-z_1-2\pi i)\right)\right|\right]
\geq \left|\sin\left(\frac{\pi}{2J}(z_3-z_2)\right)\right|,\no
\ea
in the connected case $(i)$, and by 
 \ba
\left|\sinh\left(\frac{1}{2}(z_2-z_1)\right)\right| \geq 
\mbox{Min}\left[\left|\sin\left(\frac{1}{2}(z_2-z_3)\right)\right|, \  \left|\sin\left(\frac{1}{2}(z_2-z_3-2J)\right)\right|\right], \no
\ea
in the disconnected case $(ii)$.

\begin{figure}
  \centering
  \includegraphics[width=6cm]{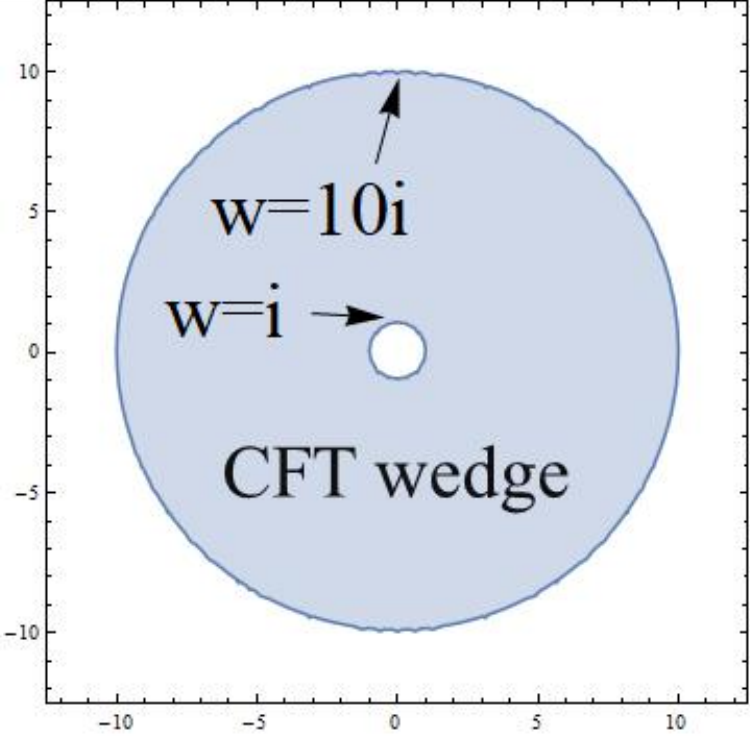}
  \includegraphics[width=4cm]{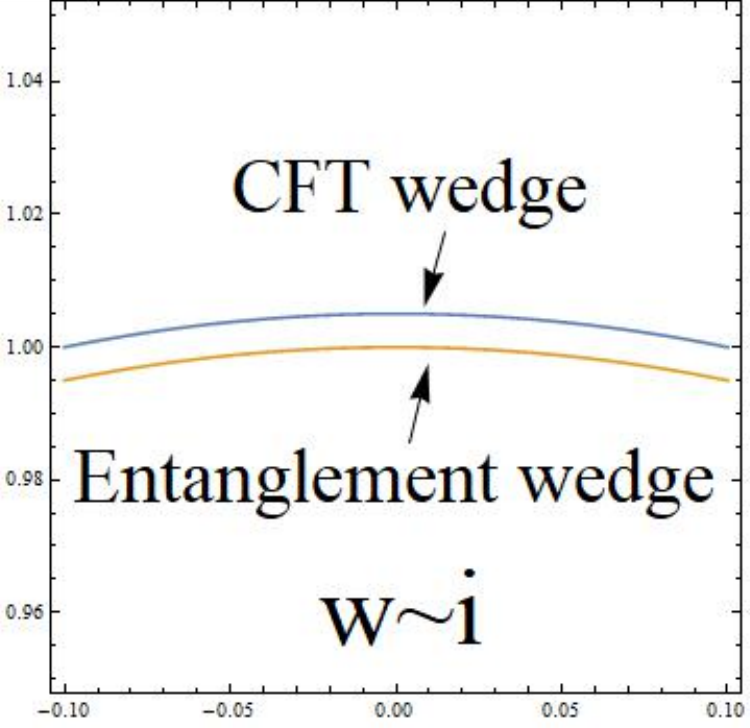}
\includegraphics[width=4cm]{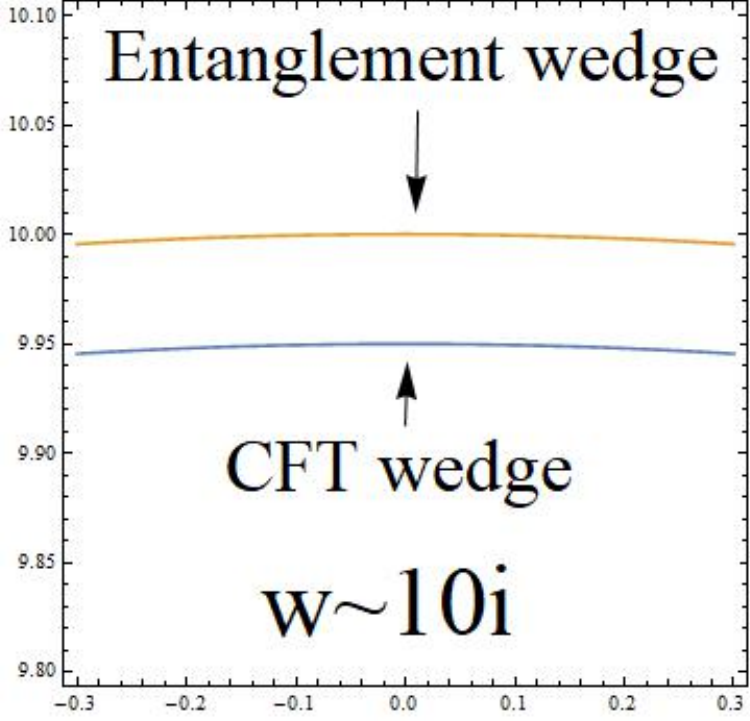}
  \caption{The plot of the location of local operator on $\ti{w}-$plane where the non-trivial contraction is favored (left) and its deviation from the entanglement wedge (middle and right). We set $\kappa=0.1$ where the entanglement wedge is connected i.e. phase (i). Blue curves are the borders between the non-trivial and trivial contraction. The orange line in the right picture describes the entanglement wedge.}
\label{fig:condb}
  \end{figure}

\begin{figure}
  \centering
  \includegraphics[width=6cm]{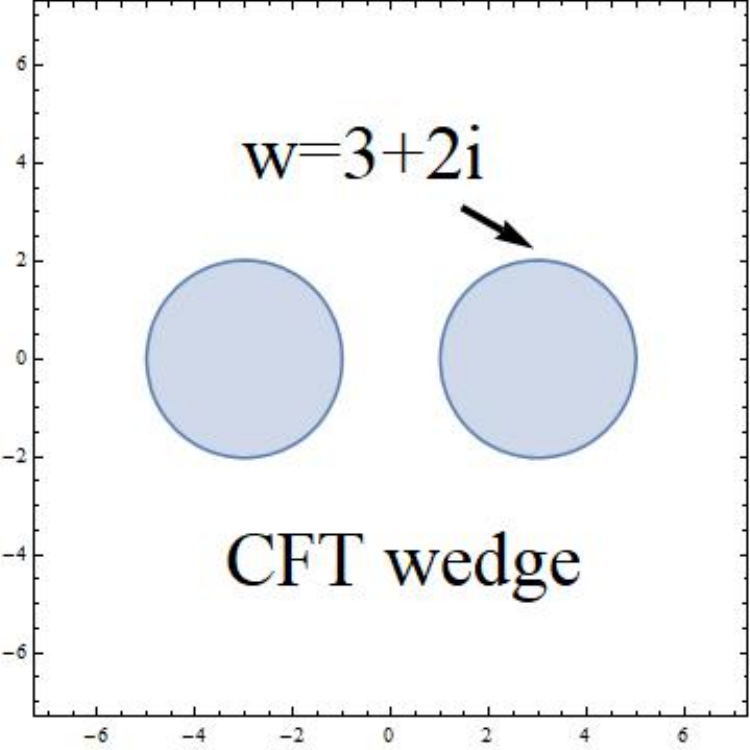}
 \includegraphics[width=6cm]{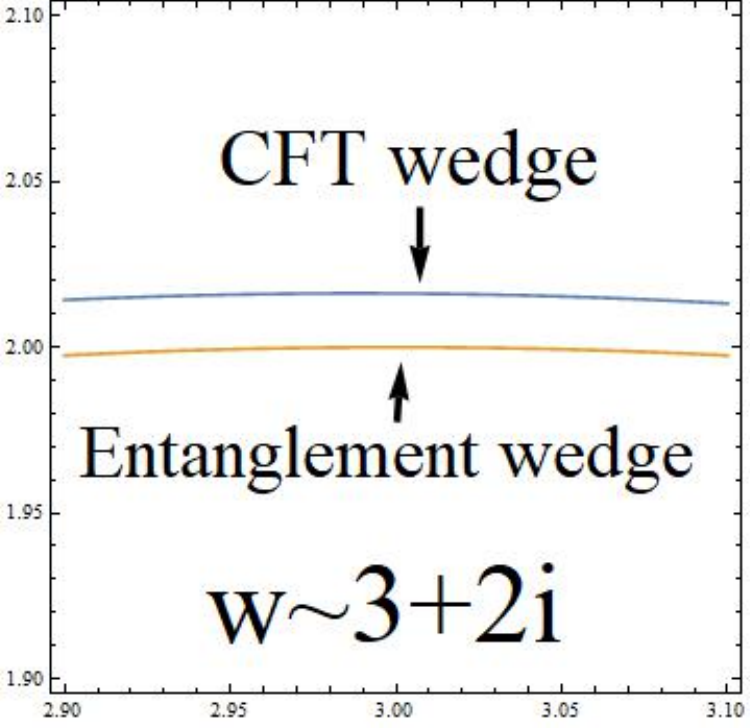}
  \caption{The plot of the location of local operator on $\ti{w}-$plane where the non-trivial contraction is favored (left) and its deviation from the entanglement wedge (right). We set $\kappa=0.2$, where the entanglement wedge is disconnected i.e.  phase (ii). Blue curves are the borders between the non-trivial and trivial contraction. The orange line in the right picture describes the entanglement wedge.}
\label{fig:discondb}
  \end{figure}

We plotted the parameter region of $(x,\tau)$, where 
the non-trivial contraction is favored,  in Fig. \ref{fig:condb} for the connected phase $(i)$ 
 and Fig. \ref{fig:discondb} for the disconnected phase $(ii)$.

\begin{figure}
  \centering
  \includegraphics[width=6cm]{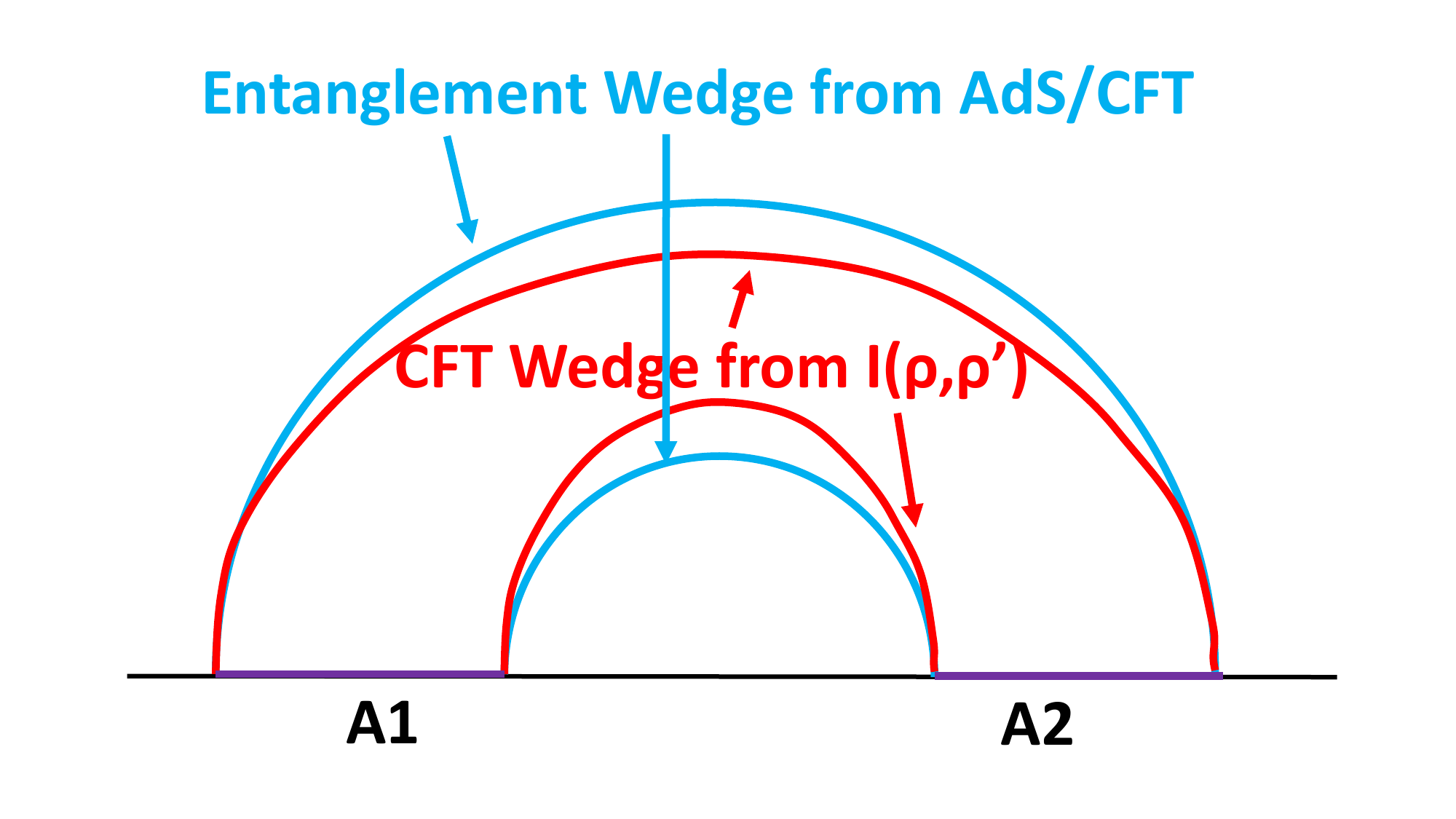}
 \includegraphics[width=6cm]{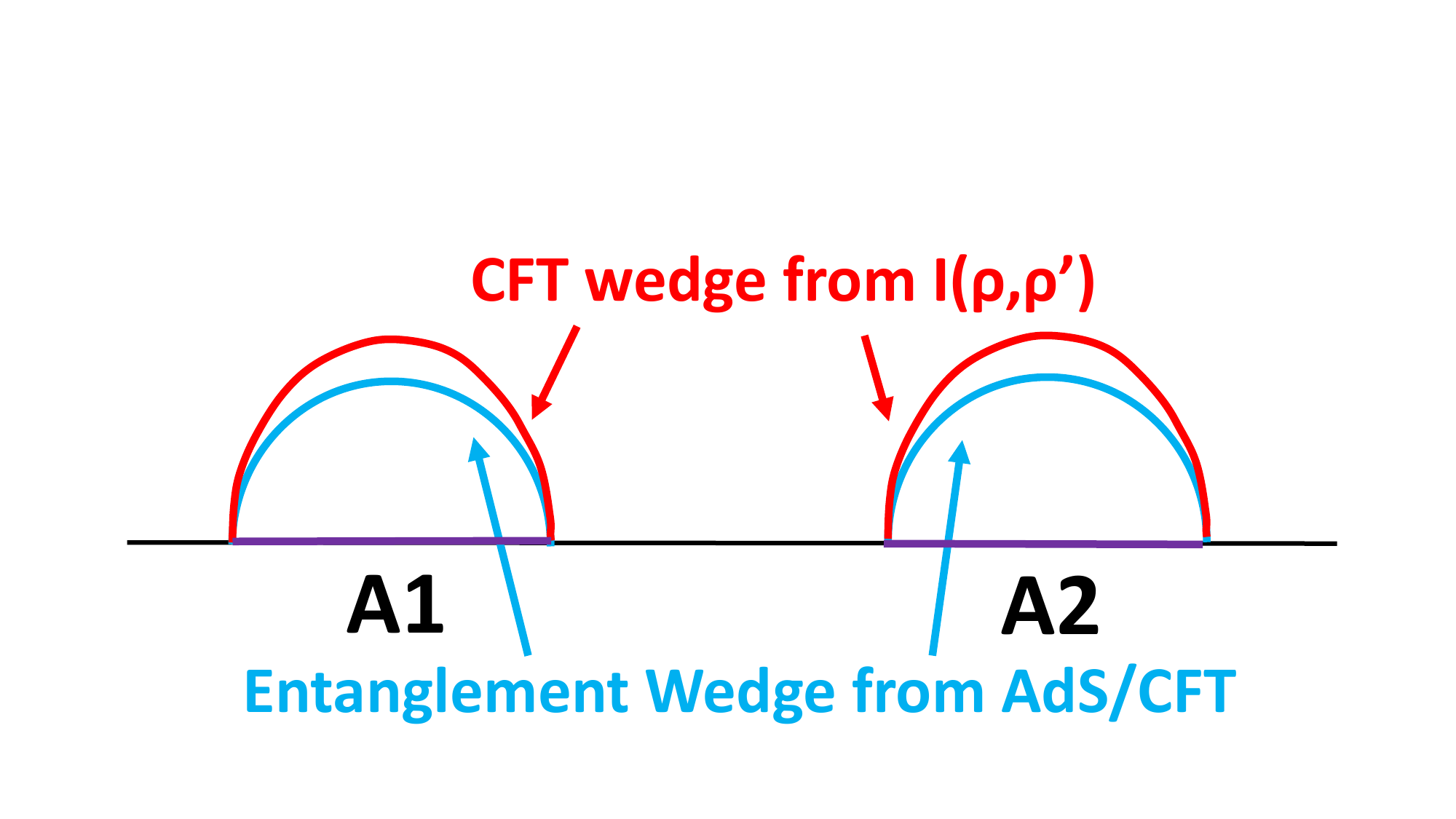}
  \caption{A sketch which emphasizes the small deviation between the CFT wedge (red) based on 
$I(\rho,\rho')$ and the correct entanglement wedge in AdS/CFT.
The left and right picture correspond to the connected and disconnected phase.}
\label{fig:deviation}
  \end{figure}

In both cases, the region is very close to the true entanglement wedge (\ref{ewdbb}). The deviation is interestingly very small (within a few percent) and as sketched in Fig.\ref{fig:deviation}. 
The wedge derived from $I(\rho,\rho')$ in the holographic CFT
can be both larger and smaller than the true entanglement wedge in AdS/CFT depending on the situations.
Notice that these deviations are leading order in our computational scheme i.e. $1/c$ expansions and
thus we cannot regard them as quantum corrections in gravity. Rather it is essential feature of the Renyi-like measure $I(\rho,\rho')$.  We will comment possible interpretations of this phenomena later subsections.

\subsection{Plots of $I(\rho,\rho')$ in Holographic CFTs}

\begin{figure}
  \centering
  \includegraphics[width=6cm]{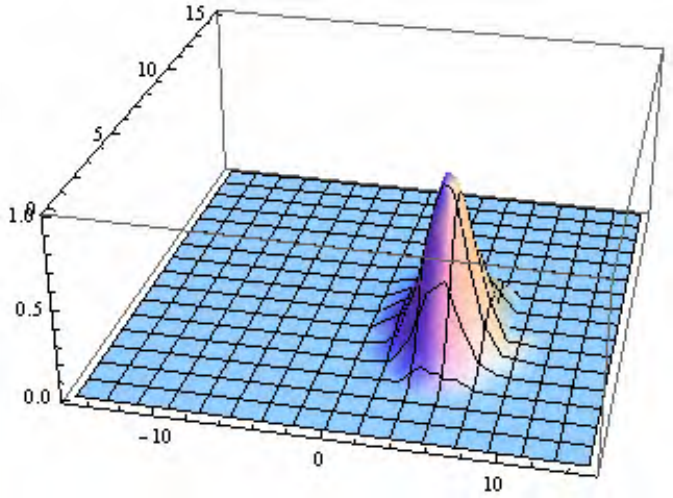}
 \includegraphics[width=6cm]{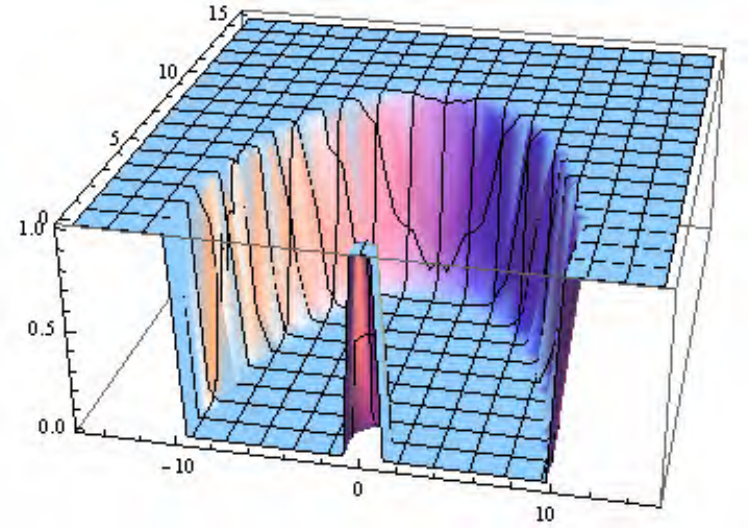}
   \includegraphics[width=6cm]{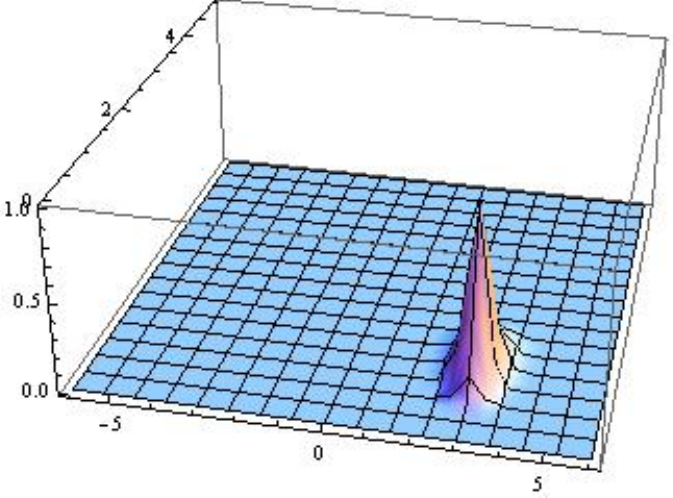}
 \includegraphics[width=6cm]{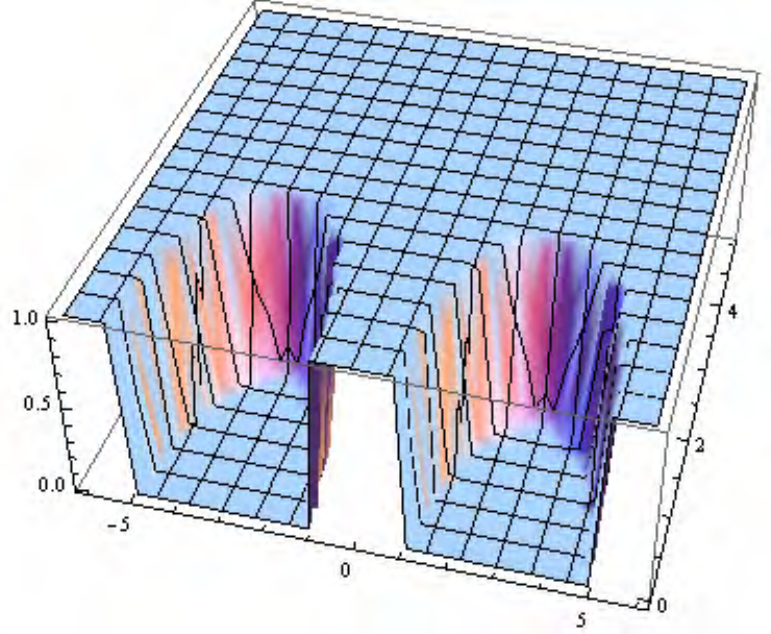}
  \caption{The values of $I(\rho,\rho')$ as a function of Re$[w']$ (horizontal) and Im$[w']$ (depth) for 
  fixed values of $w$ when the subsystem $A$ consists of the double intervals.
  The upper two pictures we set $\kappa=0.1$ (connected phase) and the lower ones we set $\kappa=0.2$ (disconnected phase). In the upper left and right picture, we chose $w=5+5i$ (inside the wedge) 
  and $w=5+20i$ (outside the wedge), respectively. In the lower left and right picture, we set
  $w=3+i$ (inside the wedge) and $w=i$ (outside the wedge), respectively.}
\label{fig:DBLplot}
  \end{figure}

We also explicitly plot the values of $I(\rho,\rho')$ as a function of $w'$ (the location of operator insertion of $\rho'_A$) when we fix $w$ (the location of operator insertion of $\rho_A$) for both the connected (upper two pictures) and the disconnected (lower two pictures) case in Fig.\ref{fig:DBLplot}. In both plots, the left graphs show the plots when we fix $w$ to be inside the wedge. In this case we find a sharp peak of 
$I(\rho,\rho')$, which reaches the maximum $I(\rho,\rho')=1$ only when $w'=w$. In the right graphs we chose 
$w$ to be outside of the wedge. We see that $I(\rho,\rho')=1$ when $w'$ is also outside the wedge, while 
we have $I(\rho,\rho')=0$ when $w'$ is inside the wedge. All of these agree with the expectation from AdS/CFT, neglecting the small deviation we discussed before.

\subsection{CFT Wedge from $I(\rho,\rho')$ for Complement}

\begin{figure}
  \centering
  \includegraphics[width=10cm]{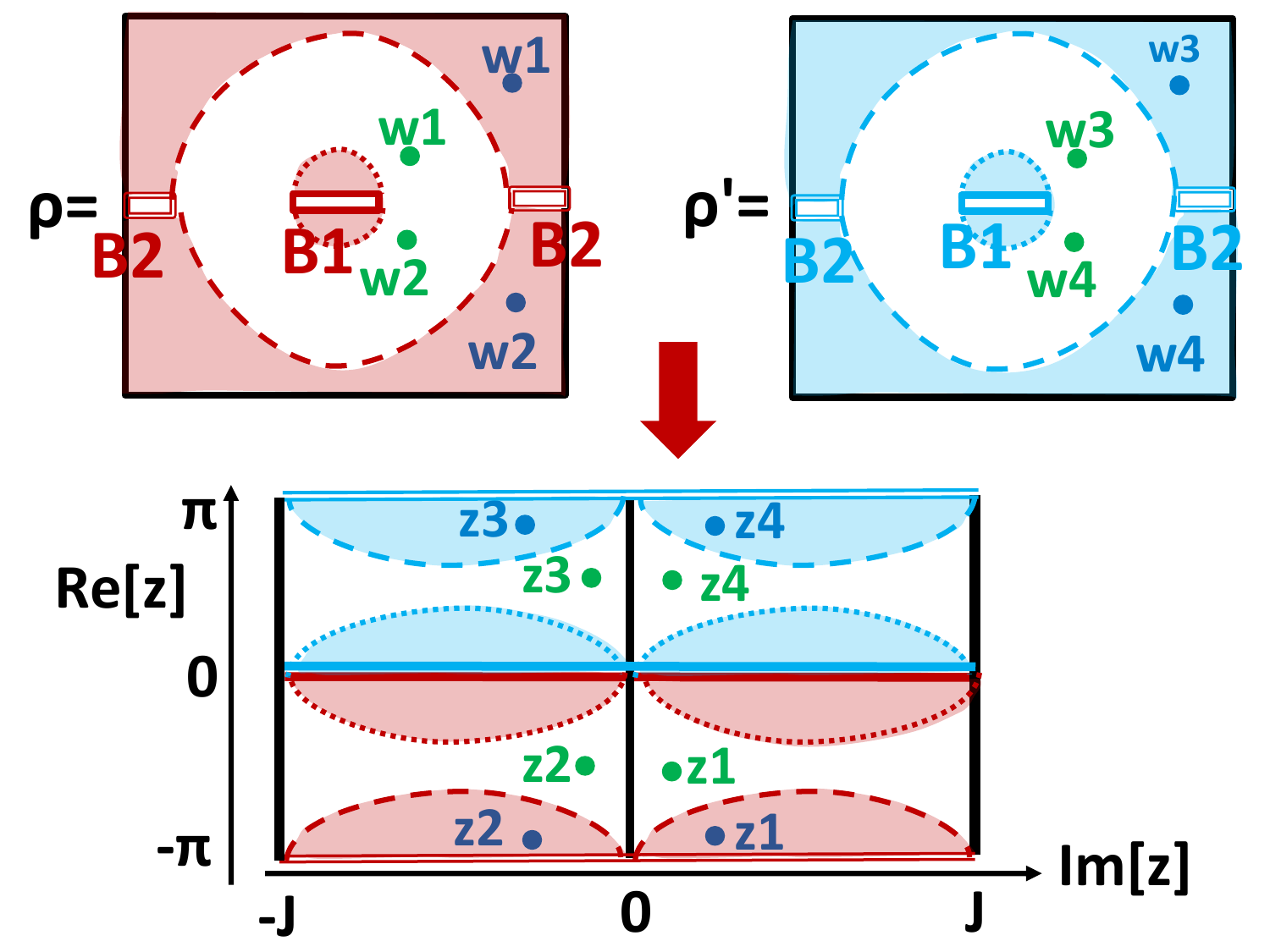}
  \caption{A sketch of  conformal transformation for the calculation of Tr$[\rho_B\rho'_B]$ 
in the double interval case. We assumed the phase (i), where  
the entanglement wedge $B$ is disconnected, as depicted by the colored region.
The lower picture described the geometry after the transformation and is given by a torus by identifying Im$[z]\sim$ Im$[z]+2\pi$ and  Re$[z]\sim$ Re$[z]+2J$.
Green Points 
(or bule points) correspond to the local excitation in the CFT which is dual to bulk excitation outside (or inside) of entanglement wedge $M_B$. }
\label{fig:double2}
  \end{figure}

It is instructive to consider also the behavior of CFT wedges for the reduced density matrix $\rho_B$, where 
$B$ is the complement of the subsystem $A$. We again focus on CFT wedges based on $I(\rho,\rho')$. 
The calculation of Tr$[\rho_B\rho'_B]$ is very similar to the previous one of Tr$[\rho_A\rho'_A]$ as depicted in Fig.\ref{fig:double2}.
The only but very important difference is that the location of $z_2$ and $z_4$ are flipped with each other. 
Therefore the condition of non-trivial Wick contraction is simply opposite to each other: when we need to take the non-trivial 
one for Tr$[\rho_A\rho'_A]$, we need to take the trivial one for Tr$[\rho_B\rho'_B]$ and vise versa. Therefore 
the CFT wedge for $\rho_B$ is just the complement of that for $\rho_A$. 

This relation helps us to understand  the behavior in Fig.\ref{fig:deviation}.
First of all when the CFT wedge for $A=A_1\cup A_2$  is disconnected, it is clear that the CFT wedge $C_A$
should be larger or equal to that for the union of the CFT wedges $C_{A_1}$ and
$C_{A_2}$, as the information included in $\rho_{A}$ is greater than that of the union of $\rho_{A_1}$ and $\rho_{A_2}$. This explains the right picture of Fig.\ref{fig:deviation}. Also this requirement is trivially satisfied in the left picture. 

To better understand the left picture of Fig.\ref{fig:deviation}, let us consider the complement of $A$ i.e. $B=B_1\cup B_2$. Since the wedge of $B$ is disconnected when
that for $A$ is connected, we can apply the same rule i.e. $C_B$ should be larger or equal to that for the union of $C_{B_1}$ and
$C_{B_2}$. As we showed just before, we also know $C_B$ is the complement of $C_A$. These two facts lead to the behavior of the left picture of Fig.\ref{fig:deviation}.

\subsection{Bures Distance in Holographic CFTs}

In the double interval case we found that the CFT wedge defined by the distinguishablity 
measure $I(\rho,\rho')$ does not precisely agree with the expected entanglement wedge from 
AdS/CFT, though the deviations are very small. This motivates us to study CFT wedges for the Bures distance $D_B(\rho,\rho')$ (\ref{disb}) or equally the Fidelity $F(\rho,\rho')$ (\ref{fidelity}), which is expected to be the ideal  distinguishablity measure. As we will see soon below, we will be able to find that the CFT wedge for $D_B$ precisely agrees with the correct entanglement wedge.

The fidelity can be computed from the analytically continuation $A_{1/2,1/2}$ of 
$A_{n,m}$ (\ref{amn}) via the replica-like method. Even though it is very difficult to evaluate 
$A_{n,m}$ for general integers $n$ and $m$, we can heuristically obtain analytical results in the limit 
$n\to 1/2$ and $m\to 1/2$ as follows.  First notice the useful property shown  in \cite{Hartman2013a} that the vacuum replica partition function of a holographic CFT with $k\sim1$ can be approximated by
\footnote{
If $k$ is enough large, then we need to take the contributions from the descendants into account.
We can consider it by making use of Virasoro conformal blocks.
}
\begin{equation}\label{eq:mutual}
\begin{aligned}
Z_{\Sigma_k ([0,s]\cup [l+s,l+2s])}&\ar{c\to \infty}\left\{
    \begin{array}{ll}
    Z_{\Sigma_k ([0,l+2s])}  Z_{\Sigma_k ([s,l+s])}   ,& 
\mbox{(i) connected phase}: s^2>(2s+l)l   ,\\
    Z_{\Sigma_k ([0,s])}  Z_{\Sigma_k ([l+s,l+2s])}    ,& 
\mbox{(ii) disconnected phase}: s^2<(2s+l)l ,\\
    \end{array}
  \right.\\
\end{aligned}
\end{equation}
where $\Sigma_k([a,b])$ means the $k$-sheeted manifold with a cut along the interval $[a,b]$ and 
$Z_{\Sigma_k([a,b])}$ is the vacuum partition function on that manifold.

Indeed, the limit of fidelity $n\to 1/2$ and $m\to 1/2$ corresponds to $k\to 1$ 
as is clear from the relation (\ref{kdefp}). Therefore we can factorize the computation of the fidelity
$F(\rho,\rho')$  into two correlation functions, each of which includes a single interval. In this sense the calculations are reduced to the fidelity in  the single interval  case, which we already worked out before
as in e.g. (\ref{outburesm}) and (\ref{www}). A CFT wedge in the single interval case is bounded by the semicircle, which agrees with the correct entanglement wedge.

\begin{figure}[t]
  \centering
  \includegraphics[width=10cm]{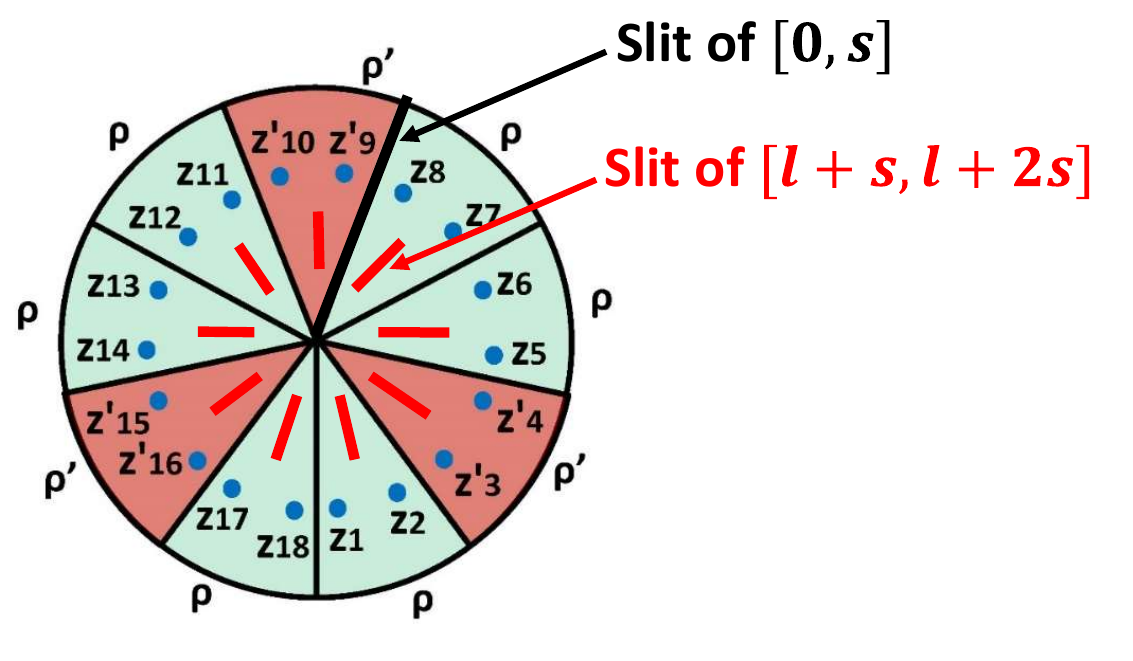}
  \caption{The complex plane which describes the path-integral which calculates the trace $A_{n,m}=\mbox{Tr}[(\rho^m\rho'\rho^m)^n].$ i.e. (\ref{amn}), where we performed the conformal transformation (\ref{confglk}) with $L=s$. Here we choose $m=1$ and $n=3$ for convenience.
Now that we consider the double interval case, we have cuts associated to the slit $[l+s,l+2s]$ (the red solid lines).
}
\label{fig:BuresS2}
  \end{figure}

We can illustrate this factorization from another view point.
If one wants to probe the disconnected entanglement wedge $[0,s]$, one may consider the conformal transformation (\ref{confglk}) with $L=s$. It leads to the geometry shown in Fig.\ref{fig:BuresS2}, which has ``cuts'' associated to the slit $[l+s,l+2s]$ (the red solid lines in the figure).
Although these cuts give nontrivial contributions to the $2k$-point function in general, these contributions can be neglected in the limits $n=m \to 1/2$. Therefore, we can evaluate this $2k$-point function in the same way as the single interval case, which means that the result just reduces to (\ref{www}).

In this way, owing to the factorization (\ref{eq:mutual}), we can conclude that the CFT wedges $C^{(B)}$
calculated from the Bures distance (or equally fidelity), coincide with the expectations from 
the entanglement wedges: (\ref{ewdbb}) in the connected case and (\ref{ewdcftw}) in the disconnected case.
It is also clear that the Bures metric in the double interval case also agrees with the AdS metric as in the single interval case, when the locations of operator insertions are inside the wedge.

\subsection{Interpretation of Two Different CFT Wedges $C^{(I)}_A$ and $C^{(B)}_A$ }\label{sec:difwe}

So far we have seen the calculations of two distinguishability measures $I(\rho,\rho')$ and $F(\rho,\rho')$
in the double interval case. Entanglement wedges in AdS/CFT are precisely reproduced from the latter 
i.e. the fidelity, while the former predicts CFT wedges which are slightly distorted from the actual entanglement wedges. Here we would like to discuss why CFT wedges depend on the choice of these
distinguishability measures.

First remember that  $I(\rho,\rho')$ is essentially the calculation of Tr$[\rho\rho']$ and the fidelity $F(\rho,\rho')$ is  equal to Tr$[\s{\s{\rho}\rho'\s{\rho}}]$. In this sense the total power of the density matrices(for this we identify  $\rho$ and $\rho'$) is two for the former and one for the latter.  

A measurement of a physical quantity is described by $\la O_i\lb=\mbox{Tr}[\rho O_i]$. 
In the classical gravity limit of AdS/CFT, we restrict the operators $O_i$ to low energy ones.
Therefore we expect that the entanglement wedge should be determined by the distinguishability of 
low energy states (or so called code subspaces \cite{ADH}).
  
In this sense, the quantity Tr$[\rho\rho']$ goes beyond the low energy approximation 
as $O_i=\rho'$ is a highly excited operator. A reduced density matrix can be expressed as $\rho_A=e^{-H_A}$ in terms of modular Hamiltonian $H_A$. For a CFT vacuum, for example, $H_A$ is given by an integral of energy stress tensor. Therefore $\rho_A=e^{-H_A}$ includes an infinite number of energy stress tensors, which are clearly outside of low energy states.

On the other hand, the fidelity $F(\rho,\rho')$ distinguishes low energy states when $\rho$ is very close to 
$\rho'$, when we calculate the Bures metric. We would like to argue that the above different 
property of distinguishing states causes the difference of CFT wedges between $I(\rho,\rho')$ 
and  $F(\rho,\rho')$. This also explains why the latter agrees with the expectation from the actual 
entanglement wedge in AdS/CFT. We will explore differences of CFT wedges for various other distance measures later in section \ref{otherme}.

\section{Entanglement Wedges from AdS/BCFT}

\begin{figure}
  \centering
  \includegraphics[width=10cm]{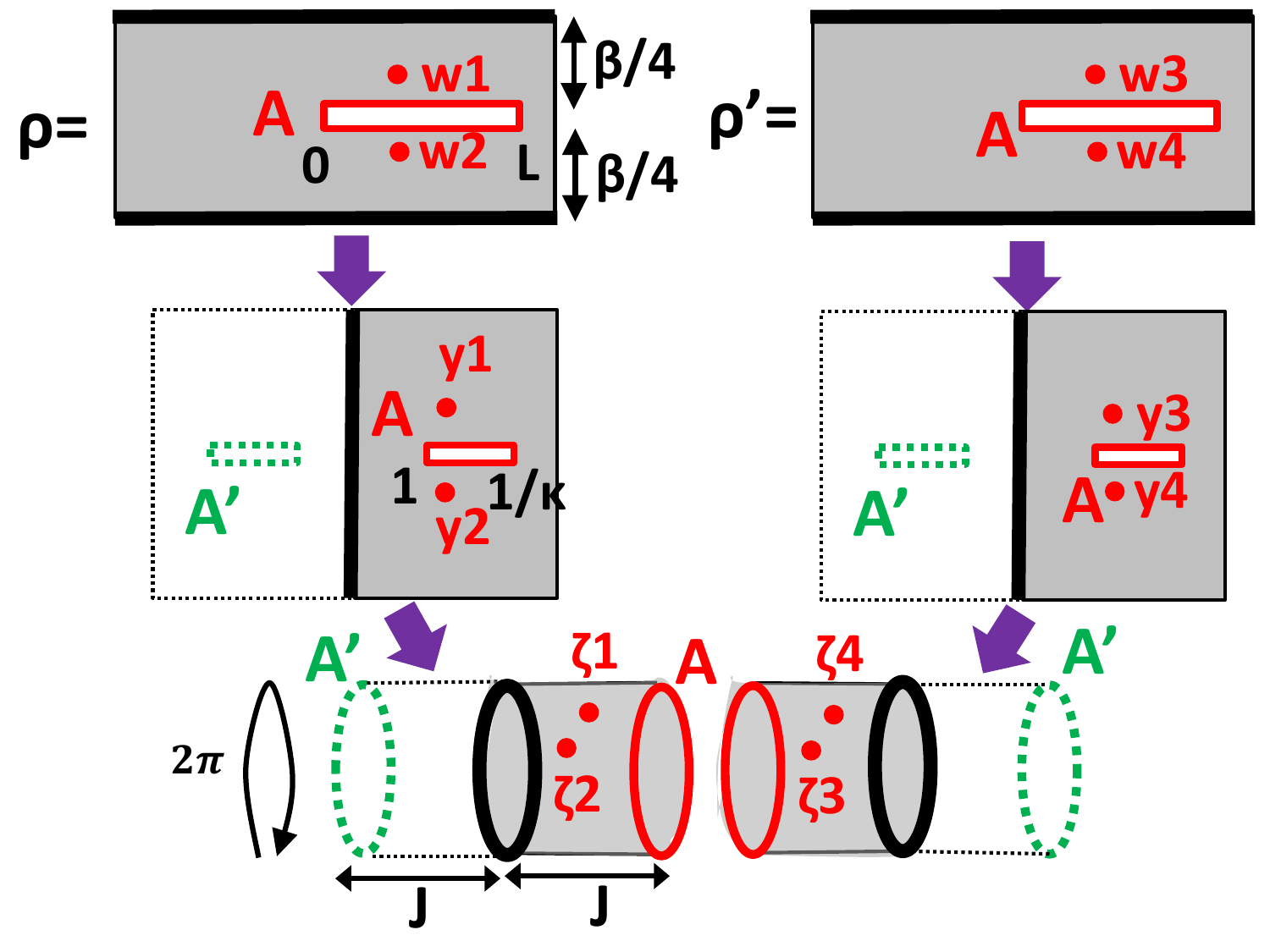}
  \caption{A sketch of  conformal transformation for the calculation of Tr$[\rho_A\rho'_A]$ 
in the BCFT setup. The upper pictures describe the setup in the original $w$ coordinate.
The red slit describes the subsystem $A$. The thick black lines describe the boundaries.
They are  mapped into $y$ coordinate as shown in the middle pictures. Finally they are mapped into 
cylinders as shown in the lower pictures. To calculate  the trace Tr$[\rho_A\rho'_A]$, we
identify two red circles, which describe the subsytem $A$, and the final geometry becomes a cylinder.}
\label{fig:BCFTmap}
  \end{figure}

Here we would like to consider a quantum state $|\Psi\lb$ in a CFT on a 2d space with boundaries, called boundary conformal field theory (BCFT), given by
\ba
|\Psi_{bdy}\lb=e^{-\frac{\beta}{4}H}|B\lb.  \label{GQB}
\ea
Its gravity dual is given by the AdS/BCFT construction \cite{AdSBCFT}  via the holography, 

This is the initial state of the global quantum quench \cite{CCG} using the boundary state $|B\lb$ 
(i.e. Cardy state \cite{Cardy:1989ir}).
We choose the subsystem $A$ to be the interval $[0,L]$ as before. 
The reduced density matrix $\rho_A=\mbox{Tr}_B[|\Psi_{bdy}\lb \la \Psi_{bdy}|]$ is computed as the 
path-integral on a strip $-\frac{\beta}{4}\leq \tau\leq \frac{\beta}{4}$. We describe this space by the coordinate $w=x+i\tau$. Refer to the upper pictures in Fig.\ref{fig:BCFTmap}.

Next we transform by the conformal map:
\ba
y=e^{\frac{2\pi}{\beta}w},
\ea
so that the $w$ plane is mapped into a half plane depicted as the middle pictures in Fig.\ref{fig:BCFTmap}.
In this coordinate, the subsystem $A$ is the interval $[1,e^{\frac{2\pi L}{\beta}}]$.

Finally we introduce a new cylindrical coordinate $\zeta$ via the elliptic map
\ba
\zeta=\frac{\pi}{K(1-\kappa^2)}\int^y_0 \frac{d\ti{y}}{\s{(1-\ti{y}^2)(1-\kappa^2\ti{y}^2)}}
=\frac{\pi}{K(1-\kappa^2)}\cdot \text{sn}^{-1}(y,\kappa^2),
\ea
where we defined
\ba
\kappa=e^{-\frac{2\pi L}{\beta}}(< 1).
\ea
Refer to the lower pictures in Fig.\ref{fig:BCFTmap}.

\subsection{Phase Transitions of Entanglement Wedges in AdS/BCFT}

We expect that the state (\ref{GQB}) is dual to a half of eternal BTZ geometry \cite{Hartman:2013qma}.
In the Euclidean setup, it is identical to the geometry given by the metric (\ref{btzmete}).
In the AdS/BCFT (refer to \cite{AdSBCFT} for details), the gravity dual of a BCFT state is found by adding a boundary surface into a AdS space, which extends to the bulk.

There are two phases in the holographic calculation of the entanglement entropy $S_A$ which follows from the prescription of AdS/BCFT : (a) The connected geodesic $\Gamma_{con}$ is favored and (b) the disconnected geodesics $\Gamma_{dis}$ which end on the horizon are favored. Accordingly the geometry of entanglement wedge changes between (a) and (b). Since the length of connected and disconnected geodesic is computed from the explicit form of the geodesic (\ref{geoffr}) as follows
\ba
&& |\Gamma_{con}|=2\int^{\rho_\infty}_{\rho_*}d\rho\frac{\cosh\rho}{\s{\cosh^2\rho-\cosh^2\rho_*}}
=\left[\mbox{Arctanh}\left(\frac{\sinh\rho}{\s{\cosh^2\rho-\cosh^2\rho_*}}\right)\right]^{\rho_\infty}_{\rho_*} \no
&&=\rho_\infty-\log\sinh\rho_*, \\
&&  |\Gamma_{dis}|=2\int^{\rho_\infty}_{0}d\rho=\rho_{\infty},
\ea
where the constant $\rho_*$ is related to $L$ via
\ba
\cosh\rho_*\tanh\left(\frac{\pi L}{\beta}\right)=1.
\ea

Therefore, the phase (a) and (b) correspond to the regions 
\ba
&&\mbox{Phase (a) $\Gamma_{con}$:}\ \  \sinh\rho_*>1\ \  \lr\ \  \sinh\left(\frac{\pi L}{\beta}\right)<1\ \  \lr\ \  \kappa=e^{-\frac{2\pi L}{\beta}}>3-2\s{2}, \no
&& \mbox{Phase (b) $\Gamma_{dis}$:}\ \  \sinh\rho_*<1\ \  \lr\ \  \sinh\left(\frac{\pi L}{\beta}\right)>1\ \  \lr\ \  \kappa=e^{-\frac{2\pi L}{\beta}}<3-2\s{2}, \no
\ea
This is the same condition which we encounter in the case of double interval. This is not a coincidence and indeed we find the  ratio of the horizontal length and vertical length of the cylinder of $\zeta$ coordinate
in Fig.\ref{fig:BCFTmap} is given by $\frac{\pi}{J}=\frac{K(1-\kappa^2)}{2K(\kappa^2)}$, which is the same ratio as that appeared in Fig.\ref{fig:double}. 
Indeed it is a cylinder with the circumference $2\pi$ and the length $J=2\pi \frac{K(\kappa^2)}{K(1-\kappa^2)}$. Via the doubling trick this can be extended as a torus with the periodicities given by $2\pi$ and $2J$.

\begin{figure}
  \centering
  \includegraphics[width=10cm]{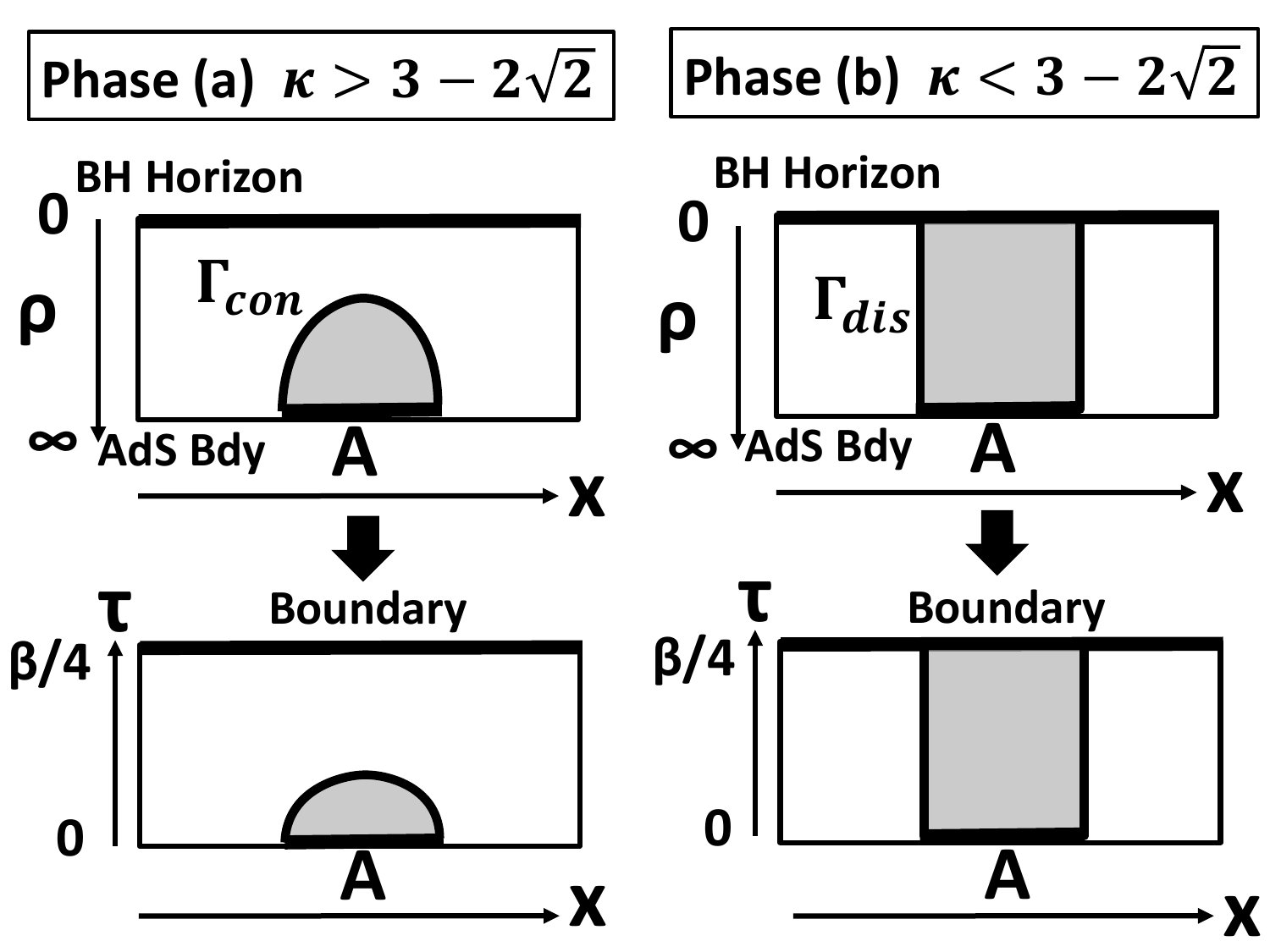}
  \caption{A sketch of  entanglement wedges in AdS/BCFT in phase (a) and (b). The upper pictures describe the geometry of entanglement wedge (gray region) in the time slice of BTZ blackhole. The lower pictures 
show the wedge geometry in the CFT dual (\ref{ewbtzz}) in the $w$-plane by the geodesic projection.}
\label{fig:BCFTEW}
  \end{figure}

In this way, the reduced density matrix analysis provides the phase transition of the entanglement wedge at the correct value of subsystem size. We sketched the expected entanglement wedge geometry from AdS/BCFT in Fig.\ref{fig:BCFTEW}.

\subsection{Wick Contractions and Distinguishability}

Now we come back to the evaluation of $I(\rho_A,\rho'_A)$. This is given by the four point functions as 
\ba
I(\rho_A,\rho'_A)=\frac{F(\zeta_1,\zeta_2,\zeta'_3,\zeta'_4)}{\s{F(\zeta_1,\zeta_2,\zeta_3,\zeta_4)F(\zeta'_1,\zeta'_2,\zeta'_3,\zeta'_4)}},
\ea
where $F$ denotes the four point function on the cylinder in the $\zeta$ coordinate
\be
F(\zeta_1,\zeta_2,\zeta_3,\zeta_4)
=\la O^\dagger_\ap(\zeta_1)O_\ap(\zeta_2)O^\dagger_\ap(\zeta_3)O_\ap(\zeta_4)\lb.
\ee 
Note that this four point function is defined on the cylinder.

\begin{figure}
  \centering
  \includegraphics[width=10cm]{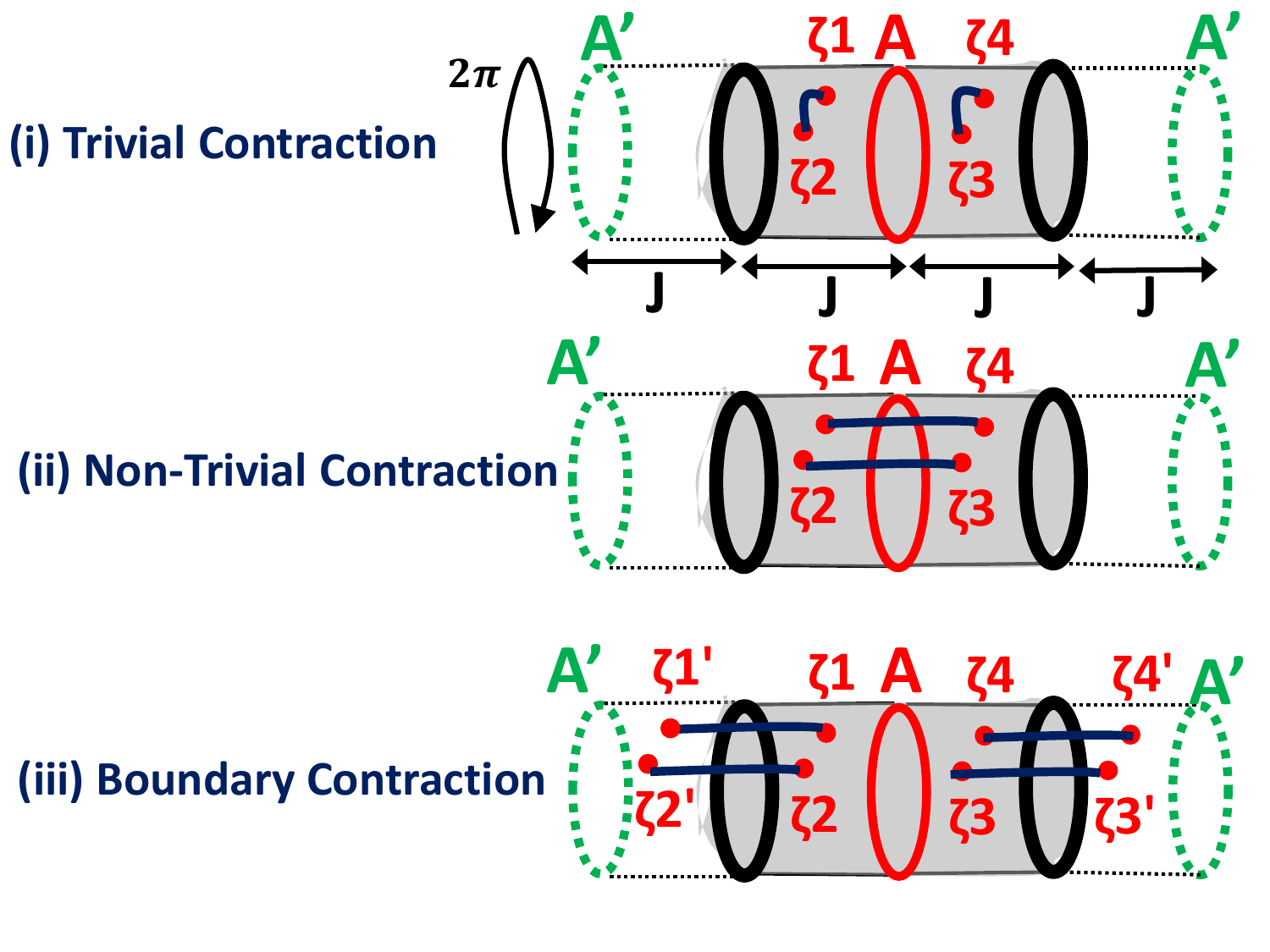}
  \caption{The three possibilities of the Wick contractions in holographic BCFTs.}
\label{fig:BCFTcylinder}
  \end{figure}

In the generalized free field prescription, we can evaluate this four point function via Wick contractions. 
There are three possible Wick contractions: (i) Trivial contraction, (ii) Non-trivial contraction and (iii) Boundary contraction as depicted in 
Fig.\ref{fig:BCFTcylinder}. The third one (iii) is new and is the contraction between each point of $\zeta_i$
($i=1,2,3,4$) and its mirror point $\zeta'_i$ due to the presence of the boundary.

In the phase (a) we have $J>\pi$ and thus the state is dual to BTZ black hole on an interval $-J \leq \mbox{Re}\zeta 
\leq J$, where Im$\zeta$ is the Euclidean time. Therefore the two point function behaves as
\ba
\la O_\ap(\zeta)O_\ap(\zeta')\lb=\sinh\left(\frac{\zeta_1-\zeta_2}{2}\right)^{-4h_\ap}\equiv G_{a}(\zeta-\zeta').  \label{bdybtz}
\ea

In the phase (b), since  $J<\pi$ the state is dual to a global AdS$_3$ 
on an interval $-J\leq \mbox{Re}\zeta \leq J$, where Im$\zeta$ is the Euclidean time. 
Therefore the two point function behaves as
\ba
\la O_\ap(\zeta)O_\ap(\zeta')\lb=\sin\left(\frac{\pi(\zeta_1-\zeta_2)}{2J}\right)^{-4h_\ap}
\equiv G_{b}(\zeta-\zeta'),  \label{bdybtzz}
\ea
where $J=2\pi K(\kappa^2)/K(1-\kappa^2)$.

It is obvious that we obtain $I(\rho_A,\rho'_A)=1$ (i.e. $\rho_A$ and $\rho'_A$ are indistinguishable)
when the contraction (i) or (iii) is favored.  We can distinguish $\rho_A$ and $\rho'_A$ i.e.  $I(\rho_A,\rho'_A)<1$ when the non-trivial contraction (ii) is favored. The condition that the non-trivial contraction (ii)  is favored is the following two inequalities:
\ba
&& \mbox{(ii) is more favored than (i):}\ \  G(\zeta_1-\zeta_4)\gg G(\zeta_1-\zeta_2), \no
&&  \mbox{(ii) is more favored than (iii):}\ \ G(\zeta_1-\zeta_4)\gg G(\zeta_1-\zeta_1'),   
\ea
when $h_\alpha$ is very large. 

\begin{figure}
  \centering
  \includegraphics[width=7cm]{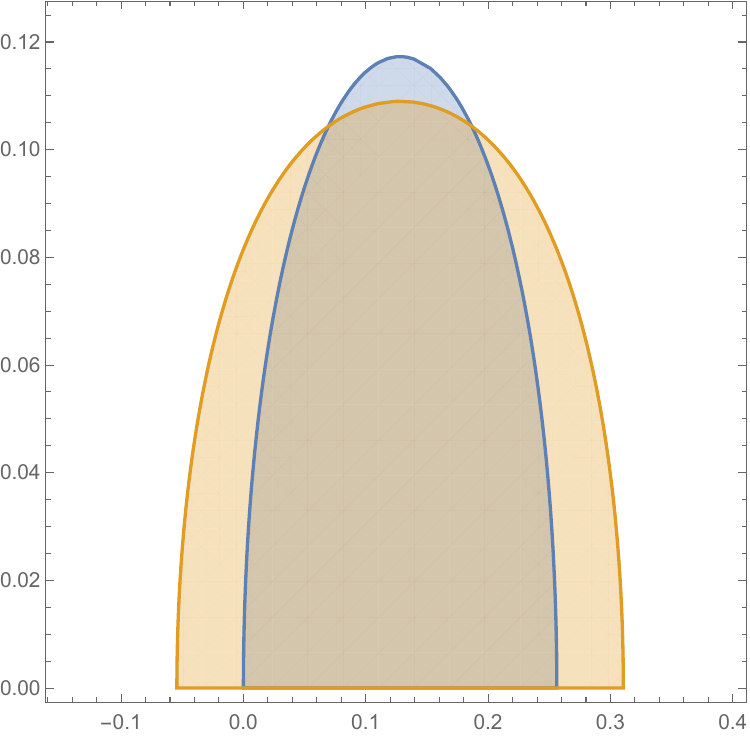}
 \includegraphics[width=7cm]{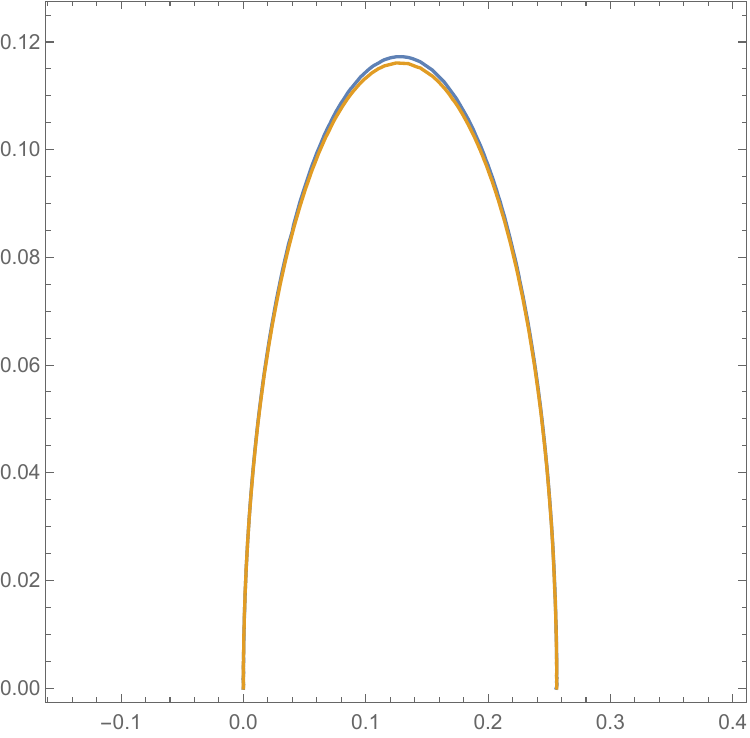}
  \caption{In Phase (a) $\Gamma_{con}$, the region where the non-trivial Wick contraction (ii) is favored, is plotted.
We set $\kappa=1/5$ and $\beta=1$.
In the left picture, the blue region corresponds to (\ref{bcftone}) and the orange region corresponds to 
 (\ref{bcfttwo}). The distinguishable region is the overlap between them.
In the right picture the blue curve is the border of  (\ref{bcftone}), while the orange curve is the expected entanglement wedge profile (\ref{ewbtzz}) from the AdS/CFT. We observe a very small deviation between them.}
\label{fig:BCFTcon}
  \end{figure}

In the phase (a), they are equivalent to the condition 
\ba
&& \mbox{(ii) is more favored than (i):} \left|\sinh\left[\frac{\zeta_1-\zeta_4}{2}\right]\right| < \left|\sinh\left[\frac{\zeta_1-\zeta_2}{2}\right]\right|, \label{bcftone} \\
&&  \mbox{(ii) is more favored than (iii):}\ \ \left|\sinh\left[\frac{\zeta_1-\zeta_4}{2}\right]\right|<\left|\sinh\left[\frac{\zeta_1-\zeta'_1}{2}\right]\right|.  \label{bcfttwo}
\ea
We numerically plotted this region in the left of Fig.\ref{fig:BCFTcon}.
If we ignore the boundary contributions, this CFT wedge is very close to the actual entanglement wedge 
from AdS/CFT as depicted in the right of Fig.\ref{fig:BCFTcon}. This small deviation is because we are actually employing the measure $I(\rho,\rho')$ which has the unwanted 
property that it is also sensitive to high energy states. In other words, if we utilize the Bures metric instead, we can reproduce the expected CFT wedges which agree with the entanglement wedges. This situation is the same as that discussed in section \ref{sec:difwe} for the example of double intervals.

In the phase (b),  they are equivalent to the condition 
\ba
 \mbox{(ii) is more favored than (i)} &:& 
\left|\sin\left[\frac{\zeta_1-\zeta_4}{2}\right]\right|<\left|\sin\left[\frac{\zeta_1-\zeta_2}{2}\right]\right|, \label{bcftonea} \\
&& \left|\sin\left[\frac{\zeta_1-\zeta_4}{2}\right]\right|<\left|\sin\left[\frac{2\pi-\zeta_1+\zeta_2}{2}\right]\right|, \label{bcftoneb} \\
 \mbox{(ii) is more favored than (iii)}& : &
\ \ \left|\sin\left[\frac{\zeta_1-\zeta_4}{2}\right]\right|<\left|\sin\left[\frac{\zeta_1-\zeta'_1}{2}\right]\right|.  \label{bcfttwob}
\ea

\begin{figure}
  \centering
  \includegraphics[width=7cm]{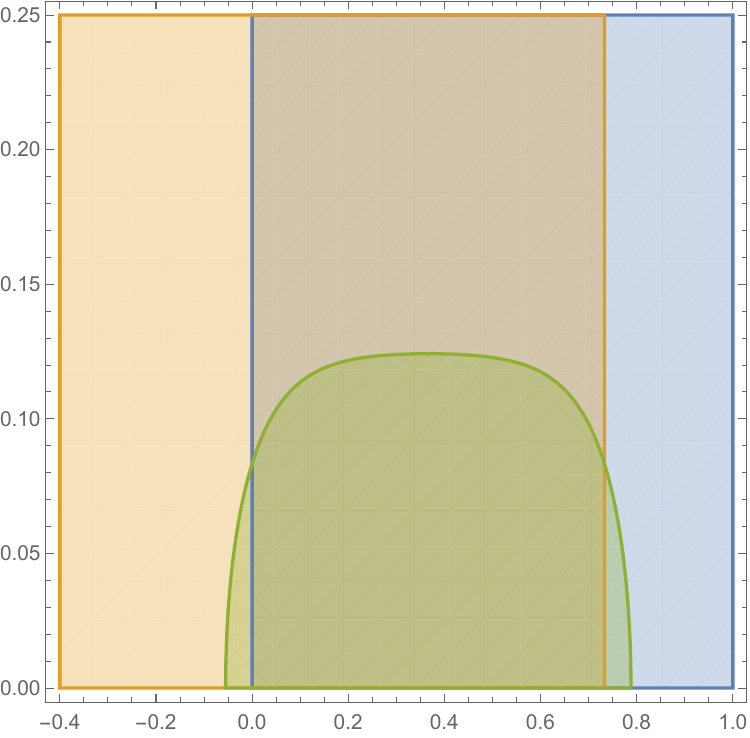}
  \caption{In Phase (b) $\Gamma_{dis}$, the region where the non-trivial Wick contraction (ii) is favored, is plotted.
We set $\kappa=1/10$ and $\beta=1$.
In the left picture, the blue, orange and green region correspond to 
(\ref{bcftonea}), (\ref{bcftoneb}) and (\ref{bcfttwob}), respectively. The distinguishable region is the overlap between these three regions.}
\label{fig:BCFTdis}
  \end{figure}

\begin{figure}
  \centering
  \includegraphics[width=10cm]{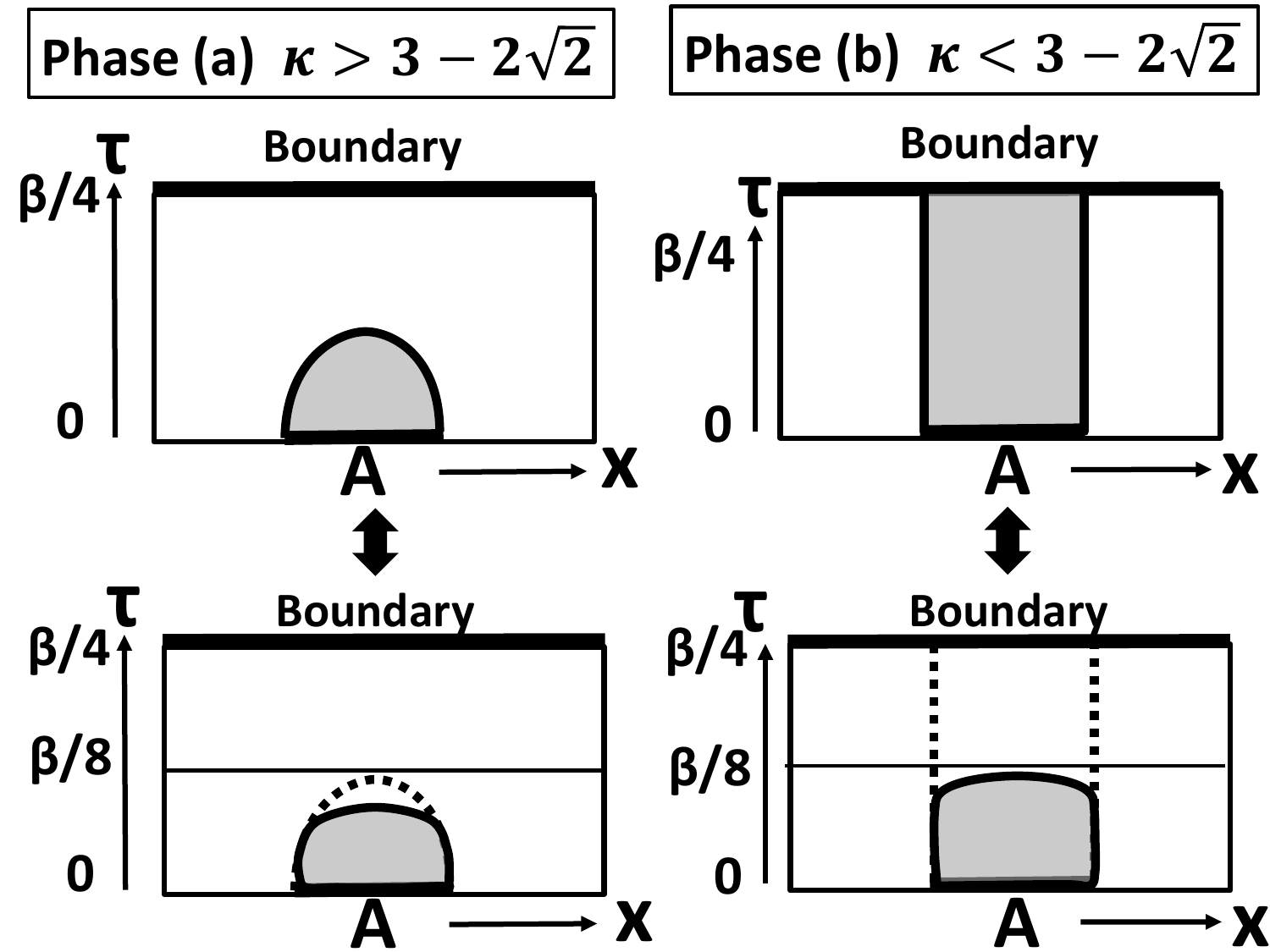}
  \caption{The CFT wedges in the phases (a): left pictures and (b):  right ones.
The upper wedges are obtained from (\ref{bcftone}),  (\ref{bcftonea}) and  (\ref{bcftoneb}). For the lower wedges we impose (\ref{bcfttwo}) and  (\ref{bcfttwob}) in addition. 
 In the phase (b) i.e. the right two pictures,
the upper and lower picture correspond to  $\la O_\ap\lb_{bdy}=0$ and  $\la O_\ap\lb_{bdy}\neq 0$, respectively.}
\label{fig:BCFTEWC}
  \end{figure}

We numerically plotted this region in Fig.\ref{fig:BCFTdis}. The resulting CFT wedge is largely different from that expected from the entanglement wedge.  
However, if we are allowed to ignore the boundary contribution (i.e. the constraint to green region), this CFT wedge is the same as the actual entanglement wedge from AdS/CFT.
In other words, we can reproduce the correct geometry of entanglement wedge only when the one point function $\la O_\ap\lb_{bdy}$ vanishes. This is because in this case the boundary contraction (iii) is not allowed. 
If the boundary one point function does not vanish, then we get the smaller wedge from the holographic CFT
than the correct entanglement wedge. See Fig.\ref{fig:BCFTEWC}. 

Even though when $\la O_\ap\lb_{bdy}\neq 0$ the CFT wedge does not agree with the entanglement wedge in AdS/CFT, this discrepancy is present even when $A$ is the total system 
(i.e. the pure state). In other words, we cannot probe points near the black hole horizon by two point functions dual to the geodesic which connects two boundary points. This is simply because the two point function gets factorized into one point functions when the points are close to the boundaries of BCFT.
Therefore, this means that we cannot employ our original idea that we probe the bulk geometry 
by two point functions when  $\la O_\ap\lb_{bdy}$ does not vanish. In this sense, we should not 
think the above discrepancy shows that the CFT predicts an entanglement wedge  which differs from the
AdS/CFT prediction. Rather we need to find a better CFT quantity which can probe  the bulk geometry.\footnote{If we turn to a setup of pure state black hole created by a heavy operator $O_H$ \cite{Fitzpatrick:2014vua}, we may avoid the mentioned problem because the two point function 
$\la O_H O_{\ap}\lb$ is vanishing.}

The entanglement wedge in AdS/BCFT which ends on the boundary surface as in the upper right 
picture of Fig.\ref{fig:BCFTEW} plays a crucial role in a recent explanation of the black hole information paradox \cite{Penington:2019npb,Almheiri:2019hni,Rozali:2019day,Chen:2019uhq,Penington:2019kki,Almheiri:2019qdq}, where a region of entanglement wedge near by boundary surface is called the Islands.
When $\la O_\ap\lb_{bdy}= 0$, our arguments above supports the entanglement reconstruction relevant to this interesting problem.

\subsection{Thermofield Double State}

The thermofield double (TFD) state also provides a closely related but different setup of AdS/CFT.
It is given by the pure state in the direct product of two identical CFT Hilbert spaces 
${\cal H}_1\otimes {\cal H}_2$:
\ba
|TFD\lb=\frac{1}{Z_{TH}}\sum_n e^{-\beta E_n/2}|n\lb_1|n\lb_2,  \label{TFDS}
\ea 
where $|n\lb$ is the energy eigenstate with energy $E_n$ and $Z_{TH}=\sum_n e^{-\beta E_n}$ is the thermal partition function. When we trace out either one of the Hilbert space, the reduced density matrix coincides with the canonical distribution. As discovered in \cite{Maldacena:2001kr}, this pure state $|TFD\lb$ is dual to the eternal AdS black hole. In the AdS$_3/$CFT$_2$, the dual geometry is given by the eternal BTZ solution which is obtained by continuing the Lorentzian geometry inside the horizon and which has two asymptotically AdS boundaries. The two boundaries correspond to the first and second CFT. In the well-known path-integral formulation, the state (\ref{TFDS}) is described by a strip with the width $\beta/2$ in the Euclidean time direction, while the space direction is an infinite line. The boundary conditions on the two boundaries of the strip are arguments of two CFTs, which totally represent the wave functional of the TFD state.

Let us choose the subsystem $A$ in the first CFT at $\tau=0$ and the subsystem $A'$ in the second CFT at $\tau=-\beta/2$. For simplicity, we choose $A$ and $A'$ to be symmetric with respect to the middle line $\tau=-\beta/4$. In this setup if we artificially take a $Z_2$ quotient $\tau\to \pi/2-\tau$, then we get back to the previous example of the global quantum quench (\ref{GQB}). Thus the mathematical structures are very similar. 

Consider the CFT wedge for the union of these two subsystems $AA'$ in the TFD state.
The entanglement wedge from CFT is simply given by doubling that for the global quench 
(see Fig.\ref{fig:BCFTEW}) across the horizon, utilizing the $Z_2$ symmetry. 

The calculation of the measure $I(\rho,\rho')$ in CFT can be done by doubling the cylinder into a torus 
as depicted in Fig. \ref{fig:BCFTmap}, where the dotted green circle represents the subsystem $A'$.
Therefore we find that the phase transition structure, i.e. the connected phase (a) and the disconnected phase (b), is identical.  Moreover, the CFT wedge is determined by the condition that non-trivial Wick contraction is favored over the trivial one. Notice that boundary contractions are not allowed as we do not have any boundaries in our CFT as opposed to the previous example. Owing to this fact, we find that 
the CFT wedge in the connected phase agrees with the entanglement wedge up to a very small deviation,
which can be confirmed in the right picture of Fig.\ref{fig:BCFTcon}.
In the disconnected phase, the CFT wedge perfectly agrees with the entanglement wedge as confirmed 
from Fig.\ref{fig:BCFTEWC}. This small deviation for the connected case is again due to the measure $I(\rho,\rho')$ and should be absent in the CFT wedges for Bures metric
 as in section \ref{sec:difwe} for the example of double intervals.

\section{Higher Dimensional Case}

Here we would like to derive the entanglement wedge in higher dimensional AdS/CFT.
Consider a $d+1$ dimensional holographic CFT on R$^{d+1}$ dual to AdS$_{d+2}$.
We write the coordinate of R$^{d+1}$ as $(\tau,x_1,\ddd,x_d)$.
Consider the reduced density matrix of locally excited state
$\rho_A=\mbox{Tr}_B\left[O_\ap(\tau,x)|0\lb\la 0|[O_\ap(\tau,x)]^\dagger\right]$ as before.
First we analyze the case where the subsystem $A$ is a half plane and later extend the results to 
the case where $A$ is a round sphere.

\subsection{Half Plane Subsystem}

\begin{figure}
  \centering
  \includegraphics[width=10cm]{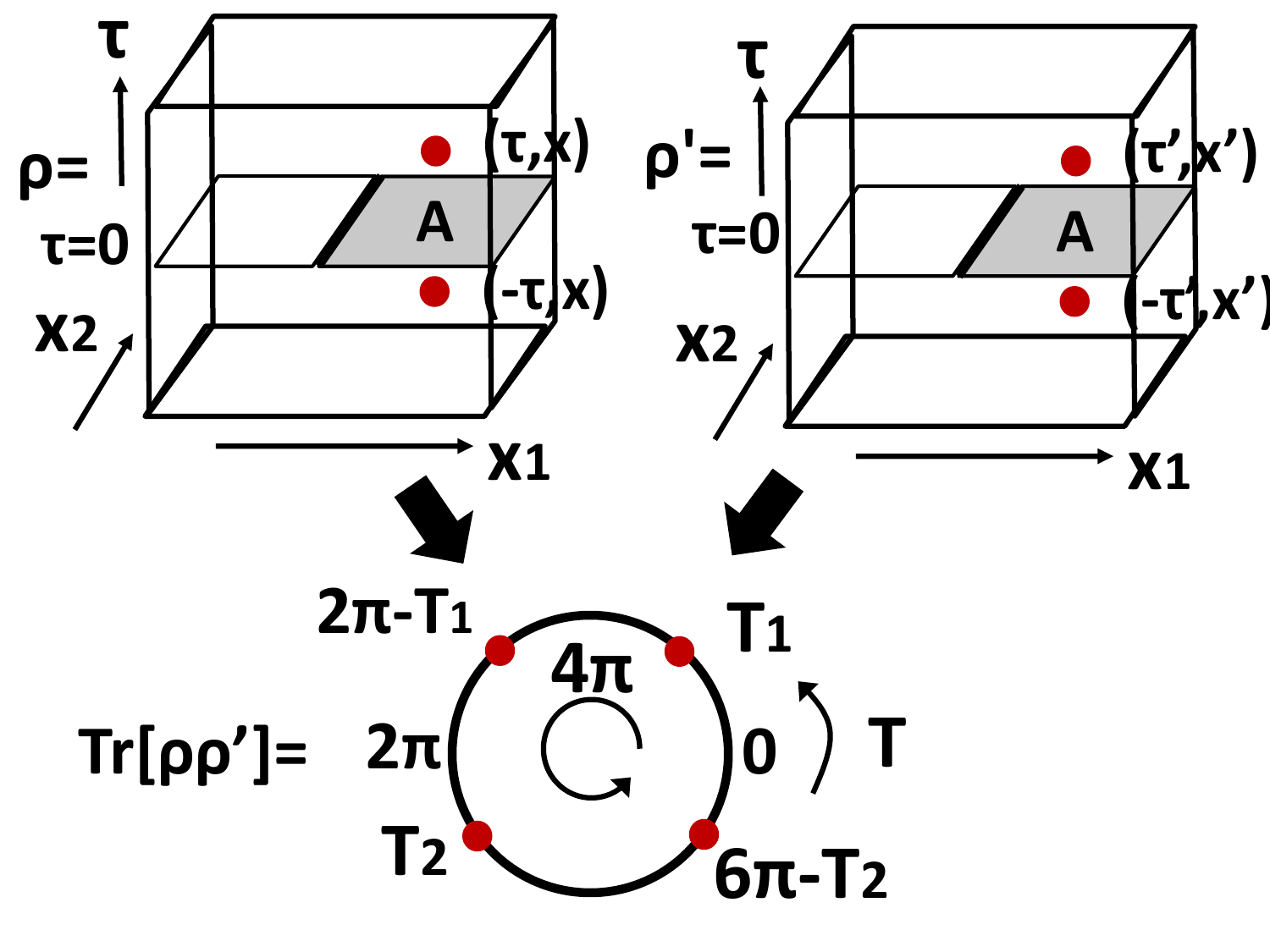}
  \caption{A sketch of computation of Tr$[\rho\rho']$ in a three dimensional CFT.}
\label{fig:higherD}
  \end{figure}

Let us start with the simple example where the subsystem $A$ is given by a half space $x_1>0$ at $\tau=0$. 
A path-integral calculation of the quantity $I(\rho,\rho')$ (\ref{Irho}) can be obtained 
as a natural generalization of our previous analysis in two dimensions and is
depicted in the upper pictures of Fig.\ref{fig:higherD}. To proceed, it is useful to introduce a polar coordinate $(T,\zeta,x_2,\ddd,x_d)$ as follows:
\ba
x_1=\zeta \cos T,\ \ \ \tau=\zeta \sin T,
\ea
where $(x_2,\ddd,x_d)$ are the same as before.
The metric looks like
\ba
ds^2=d\tau^2+\sum_{i=1}^d (dx_i)^2=dT^2+T^2d\eta^2+\sum_{i=2}^d (dx_i)^2.
\ea
By using this polar coordinate, we can express the trace Tr$[\rho\rho']$ as a path-integral on a space illustrated in the lower picture of Fig.\ref{fig:higherD}. Since two spaces R$^d$ are glued with each other along $A$, the periodicity of $T$ is now
$4\pi$.

The gravity dual is given by the topological black hole (refer to \cite{CHM}):
\begin{equation}\label{topbh} 
\begin{aligned}
 ds^2 &=\frac{dz^2+d\tau^2+\sum_{i=1}^d (dx_i)^2}{z^2} \\
&=\frac{dr^2}{f(r)}+f(r)dT^2+r^2\left(\frac{d\eta^2+\sum_{i=2}^d (dx_i)^2}{\zeta^2}\right),  
\end{aligned}
\end{equation}
where $ f(r)\equiv r^2-1-\frac{\mu}{r^{d-2}}$.
The smoothness of the geometry determines the periodicity $\beta_T$ of $T$ as
\be
\beta_T=\frac{4\pi r_+}{(d+1)r_+^2-(d-1)},
\ee
where $r_+$ is the outer horizon $f(r_+)=0$.

We take the periodicity to be $\beta_T=2\pi n$. This leads to
\be
r_+=\frac{1}{n(d+1)}+\s{1-\frac{2}{d+1}+\frac{1}{n^2(d+1)^2}}.
\ee
We can evaluate two point functions in the holographic CFT from this geometry
by applying the standard formula in AdS/CFT: 
\be
\la O_1(a)O_2(b)\lb\sim e^{-\Delta_O L_{ab}},
\ee
where $L_{ab}$ is the geodesic distance between the two points $a$ and $b$ in the gravity dual.
Note that even though the two point functions
on R$^d$ are universal in higher dimensional CFTs, that is not true for two point functions on a curved manifold.
Therefore we need the evaluation of two point functions using the gravity dual.

We consider geodesics described by the form $T=T(r)$, where $\zeta$ and $x_2,\ddd,x_d$ take fixed values.
The geodesic equation in the metric (\ref{topbh}) looks like
\ba
\frac{dT}{dr}=\frac{1}{f(r)\s{\frac{f(r)}{f(r_*)}-1}},
\ea
where $r_*$ is the minimum value of $r$ on the geodesic (or equally the turning point).
By integrating the solution to this equation as
\ba
L_{12}=\int^{r_{\infty}}_{r_*}\s{f(r)\left(\frac{dT}{dr}\right)^2+\frac{1}{f(r)}},
\ea
we can find the geodesic length $L$ between two boundary points $(T,r)=(T_a,r_\infty)$ and  $(T,r)=(T_b,r_\infty)$. $r_{\infty}$ is the cut off at the AdS boundary and
is written as $r_\infty=\zeta/\ep$ in terms of the CFT cut off $\ep$.
The geodesic length $L_{ab}$ is a function of the time difference $T_b-T_a$ and they are parameterized by
$r_*$ as follows:
\ba
&& T_b-T_a=2\int^{r_\infty}_{r_*} \frac{dr}{f(r)\s{\frac{f(r)}{f(r_*)}-1}},   \\
&& L_{ab}=2\int^{r_\infty}_{r_*}\frac{1}{\s{f(r)-f(r_*)}}.
\ea

Now let us consider the evaluation of $I(\rho,\rho')$. As in the two dimensional CFT case, we apply the large $N$ factorization, namely generalized free field calculation. Then the non-trivial Wick contraction is favored when
$L_{ab}>L_{bc}$, where the points $p_1,p_2$ and $p_3$ are the AdS boundary points
$a=(T_1,r_{\infty})$, $b=(2\pi-T_1,r_\infty)$ and $c=(T_2,r_{\infty})$. Since $L_{ab}$ is a monotonically increasing function of $T_b-T_a$, we find that the non-trivial Wick contraction is favored when $L_{ab}>L_{bc}$ holds i.e.
\be
(2\pi-T_1)-T_1>T_2-(2\pi-T_1).  \label{condgh}
\ee
When we calculate the information metric we assume $p_1$ and $p_3$ are almost the same position in each R$^d$. 
This means $T_2\simeq 2\pi+T_1$ (look at the bottom picture of Fig.\ref{fig:higherD}). In this way, the condition of non-trivial Wick contraction (\ref{condgh}) leads to
\be
0\leq T_1< \frac{\pi}{2}.
\ee
In the original coordinate of $(\tau,x_1,\ddd,x_d)$, this is equivalent to
\be
x_1>0.
\ee
This reproduces the correct entanglement wedge of the half plane $A$.

In the Bures distance limit, the replica number $n$ is finally taken to be $n=1$.
Therefore we do not need to worry about the curved space complications
and the two point function takes the standard universal form:
\ba
\la O^\dagger(\tau,x)O(\tau',x')\lb=|(\tau-\tau')^2+\sum_{i=1}^d (x_i-x'_i)^2|^{-2\Delta_O}.
\label{twophi}
\ea
In the same way as that in the two dimensional CFT case, we find in the limit $n=m=1/2$:
\ba
A_{1/2,1/2}=\frac{\la O^\dagger(-\tau,x)O(\tau',x')\lb}{\s{\la O^\dagger(\tau,x)O(-\tau,x)\lb\cdot \la O^\dagger(\tau',x')O(-\tau',x')\lb}},
\ea
where the two point functions are given by (\ref{twophi}).

Thus the final Bures information metric is computed as
\ba
ds^2=\frac{\Delta_O}{2}\cdot \frac{d\tau^2+\sum_{i=1}^{d}(dx_i)^2}{\tau^2}.
\ea
Indeed this agrees with the time slice metric of a $d+2$ dimensional Poincare AdS.

\subsection{Spherical Subsystem}

Next we turn to spherical subsystems. 
Consider a holographic CFT on R$^{d+1}$. 
In polar coordinates, the metric takes
\[
\begin{split}ds^{2}=d\tau^2+dr^{2}+r^{2}d\varOmega_{d-1}^{2}.\end{split}
\]
 We take the subregion A to be inside of the spherical region defined
by $\left\{ \tau=0,r\leq R\right\} $. To apply the replica method, we
use the map  \cite{CHM}
\begin{eqnarray*}
r=R\frac{\sinh\left(u\right)}{\cosh\left(u\right)+\cos\left(\frac{\tau_H}{R}\right)}, & \tau=R\frac{\sin\left(\frac{\tau_{H}}{R}\right)}{\cosh\left(u\right)+\cos\left(\frac{\tau_{H}}{R}\right)}.
\end{eqnarray*}
After this coordinate transformation, the metric looks like
\[
ds^{2}=\frac{1}{\left(\cosh\left(u\right)+\cos\left(\frac{\tau_{H}}{R}\right)\right)^{2}}\left(d\tau_{H}^{2}+R^{2}\left(du^{2}+\sinh^{2}\left(u\right)d\Omega_{d-1}^{2}\right)\right)
\]
 which is conformally equivalent to S$^1\times$H$^d$. S$^{1}$ direction represents Euclidean time
coordinate and its period is $\beta=2\pi R$ and in this map the original
surface $\tau=0^{-}$ and $\tau=0^{+}$ will transform to $\tau_{H}=0^{+}$
and $\tau_{H}=R\beta^{-}$ respectively. 

The gravity dual of the above space is
a topological black hole with the metric (refer to \cite{CHM})
\begin{eqnarray*}
ds^{2}=f\left(\rho\right)d\tau_{H}^{2}+\frac{d\rho^{2}}{f\left(\rho\right)}+\rho^{2}\left(du^{2}+\sinh^{2}ud\Omega_{d-1}^{2}\right), & \ & f\left(\rho\right)=\frac{\rho^{2}}{R^{2}}-1-\frac{M}{R^{2}\rho^{d-1}}. \label{topsp}
\end{eqnarray*}

Around the event horizon we can approximate 
$f\left(\rho\right)\simeq\epsilon f^{\prime}\left(\rho^{+}\right)$
where $\rho^{+}$ is the larger solution of $f\left(\rho\right)=0$.
After substituting this form if we require that this space-time is the regular solution
to the Einstein equation , i.e. we do not admit any conical singularity,
the inverse temperature is fixed as $\beta_{T}=\frac{4\pi\rho_{+}R^{2}}{\left(d+1\right)\rho_{+}^{2}-\left(d-1\right)R^{2}}$.

Now let us consider calculating $I\left(\rho,\rho^{\prime}\right)$. $\rho$
is a state in which operators $O\left(\tau,r\right)$ and
$O^{\dagger}\left(-\tau,r\right)$ are inserted and $\rho^{\prime}$
is also a state in which operators $O\left(\tau^{\prime},r^{\prime}\right)$
and $O^\dagger\left(-\tau^{\prime},r^{\prime}\right)$ are inserted
similarly. If we apply the replica method to evaluate the correlation
function, we have to consider geodesics in the topological
black hole which connect two boundary points and to choose the mass parameter $M$ 
in (\ref{topsp}) such that the periodicity of $\tau_H$ is $4\pi R$. However, as in the previous calculation,
the geodesic length is monotonic with the difference of boundary time coordinates, hence we only have to specify the difference of 
$\tau$ instead of calculating the length of the geodesic directly.

As in the previous argument, we find  that non-trivial contraction is favored when
\be
0\leq\tau_{H}\leq\frac{\pi R}{2}.
\ee
This condition is equivalent to
\be
0\leq r\leq\sqrt{R^{2}-\tau^{2}},
\ee
which indeed reproduces the expected entanglement wedge in AdS$_{d+2}$, perfectly.

Correlation functions on S$^{1}$$\times$H$^{d}$ are
related with those on R$^{d+1}$ by the following:
\begin{eqnarray*}
\left\langle \mathcal{O}\left(\tau_{H},u\right)\mathcal{O^{\dagger}}\left(\tau_{H}^{\prime},u^{\prime}\right)\right\rangle  & = & \left|\frac{\partial\left(\tau,r\right)}{\partial\left(\tau_{H},u\right)}\right|^{\triangle_{\mathcal{O}}}\left|\frac{\partial\left(\tau^{\prime},r^{\prime}\right)}{\partial\left(\tau_{H},u^{\prime}\right)}\right|^{\triangle_{\mathcal{O}}}\left(\varOmega\left(\tau_{E,}u\right)\varOmega\left(\tau_{H}^{\prime},u^{\prime}\right)\right)^{\triangle_{\mathcal{O}}}\left\langle \mathcal{O}\left(\tau,r\right)\mathcal{O^{\dagger}}\left(\tau^{\prime},r^{\prime}\right)\right\rangle ,
\end{eqnarray*}
where $\Omega=\frac{1}{\cosh\left(u\right)+\cos\left(\frac{\tau_{H}}{R}\right)}$
is a conformal factor.

In the above form we just care about Jacobian and conformal transformations
of the correlation functions, whose explicit forms are given by
\begin{eqnarray*}
\left\langle \mathcal{O}\left(\tau,r\right)\mathcal{O^{\dagger}}\left(\tau^{\prime},r^{\prime}\right)\right\rangle =\left|\left(\tau-\tau^{\prime}\right)^{2}+\left(r-r^{\prime}\right)^{2}\right|^{-\triangle_{\mathcal{O}}}, & \\
\left|\frac{\partial\left(\tau,r\right)}{\partial\left(\tau_{H},u\right)}\right|=\left|R\frac{\sinh^{2}u-\sin^{2}\frac{\tau_{H}}{R}}{\left(\cosh\left(u\right)+\cos\left(\frac{\tau_{H}}{R}\right)\right)^{2}}\right|.
\end{eqnarray*}
Then the Bures distance becomes
\begin{eqnarray*}
A_{\frac{1}{2},\frac{1}{2}}=\frac{\left\langle \mathcal{O}\left(-\tau_{H},u\right)\mathcal{O^{\dagger}}\left(\tau_{H}^{\prime},u^{\prime}\right)\right\rangle }{\sqrt{\left\langle \mathcal{O}\left(-\tau_{H},u\right)\mathcal{O^{\dagger}}\left(\tau_{H},u\right)\right\rangle \left\langle \mathcal{O}\left(-\tau_{H}^{\prime},u^{\prime}\right)\mathcal{O}^{\dagger}\left(\tau_{H}^{\prime},u^{\prime}\right)\right\rangle }}\\
=\frac{\left|\left(\tau_{-}-\tau^{\prime}\right)^{2}+\left(r-r^{\prime}\right)^{2}\right|^{-\triangle_{\mathcal{O}}}}{\sqrt{\left|\left(\tau_{-}-\tau\right)^{2}+\left(r_{-}-r\right)^{2}\right|^{-\triangle_{\mathcal{O}}}\left|\left(\tau_{-}^{\prime}-\tau^{\prime}\right)^{2}+\left(r_{-}^{\prime}-r^{\prime}\right)^{2}\right|^{-\triangle_{\mathcal{O}}}}}, & 
\end{eqnarray*}
where
\ba
\tau_{-}=R\frac{\sin\left(\frac{-\tau_{H}}{R}\right)}{\cosh\left(u\right)+\cos\left(\frac{-\tau_{H}}{R}\right)} ,\ \ \ 
 r_{-}=R\frac{\sinh\left(u\right)}{\cosh\left(u\right)+\cos\left(\frac{-\tau_{H}}{R}\right)}.
\ea
Above we neglected the spherical part for simplicity, however we can treat it in a similar way and thus can derive the full Bures metric:
\ba
ds^{2}=\frac{1}{2}\frac{\triangle_{\mathcal{O}}}{\sin^{2}\frac{\tau_{H}}{R}}\left(\frac{1}{R^{2}}d\tau^{2}_H
+du^{2}+\sinh^2 u d\Omega^2_{d-1} \right).  \label{buressxt}
\ea
By considering a geodesic which connects $\tau_H$ at the AdS boundary $\rho=\infty$ and 
the middle point $\tau_H=0$ and $\rho=\rho_*$, the relation between $\tau_H$ and $\rho_*$ is found as 
\ba
\sin\left(\frac{\tau}{R}\right)=\frac{R}{\rho_*}.
\ea
This maps the Bures metric (\ref{buressxt}) into the time slice metric of AdS:
\ba
ds^2=\frac{d\rho^2}{\rho^2/R^2-1}+\rho^2 (du^2+\sinh^2 u d\Omega^2_{d-1}),
\ea
up to a constant factor.

\section{Other Distinguishability Measures}\label{otherme}

In this section, we would like to analyze behaviors of some more distinguishability measures other than 
$I(\rho,\rho')$ and $F(\rho,\rho')$ in our CFT setup. Finally we will summarize which distinguishability measures
can reproduce correct entanglement wedges and discuss possible reasons.

\subsection{Affinity (Hellinger Distance)}

The affinity  $A(\rho,\rho')$ is defined by (\ref{aff}) and the Hellinger distance $D_H(\rho,\rho')$ is 
introduced as in (\ref{heldis}), accordingly. 
The affinity for our density matrix  (\ref{redb}) in 2d CFTs with a single interval $A$ can also be 
evaluated by the analytic continuation of the replica correlation function as
\begin{equation}
A(\rho,\rho')\equiv \lim_{m,n\to \fr{1}{2}} \tr\rho^m \rho'^n  = \lim_{m,n\to \fr{1}{2}} \fr{ Z_{m,n}}{\ca{N}_{m,n}},
\end{equation}
where the correlation function is the same as (\ref{ratq}) with $ k=m+n$ and
\begin{equation}
\begin{aligned}
w_j &=\left\{
    \begin{array}{ll}
      w ,& \text{if }  j=1, \cdots, m ,\\
      w' ,& \text{otherwise }   .\\
    \end{array}
  \right.\\
\end{aligned}
\end{equation}
The normalization is given by
\begin{equation}
\ca{N}_{m,n}=\abs{w-\bar{w}}^{-4mh} \abs{w'-\bar{w'}}^{-4nh}.
\end{equation}
The partition function can be evaluated in a similar manner to the fidelity.
For example, the partition function for the single interval case is
\begin{equation}
\begin{aligned}
Z_{1/2, 1/2} &=\left\{
    \begin{array}{ll}
     \abs{w-\bar{w}}^{-2h}  \abs{w'-\bar{w'}}^{-2h}  ,& \text{outside the CFT wedge }   ,\\
     \abs{w-\bar{w}}^{2h}  \abs{w'-\bar{w'}}^{2h}  \abs{w-\bar{w'}}^{-8h}  ,& \text{inside the CFT wedge}   ,\\
    \end{array}
  \right.\\
\end{aligned}
\end{equation}
where the CFT wedge for the affinity is the same as that for the fidelity. 
In this example we find
\begin{equation}\label{eq:A=F^2}
A(\rho,\rho')=F^2(\rho,\rho').
\end{equation}
Actually, the same relation also holds for the double interval case. 
CFT wedges of affinity in both single and double interval case coincide with those 
of the fidelity and therefore agree with the actual entanglement wedge in AdS.

\subsection{Trace Distance}
From the property (\ref{ineqtra}), we have
\begin{equation}
F(\rho,\rho')\ar{} 1  \Longleftrightarrow  D_{tr}(\rho,\rho')\ar{} 0.
\end{equation}
Therefore, the trace distance has the same transition point as the fidelity, which perfectly matches the entanglement wedge.
It would be interesting to check this conclusion from a direct calculation in holographic CFTs.

\subsection{Chernoff Bound}
The {\it quantum Chernoff bound} is largely discussed as another distinguishability measure, which was first introduced in \cite{Audenaert2006} as
\begin{equation}
Q(\rho,\rho')\equiv \min_{0 \leq m \leq 1}Q_m(\rho,\rho'),
\end{equation}
where $Q_m$ is the {\it quantum Renyi overlaps} \cite{Boca2008},
\begin{equation}
Q_m(\rho,\rho')\equiv \tr\rho^m \rho'^{1-m} \lim_{n\to 1-m}= \fr{  Z_{m,n}}{\ca{N}_{m,n}}.
\end{equation}
The partition function is the same as (\ref{ratq}) with $ k=m+n$ and
\begin{equation}
\begin{aligned}
w_j &=\left\{
    \begin{array}{ll}
      w ,& \text{if }  j=1, \cdots, m ,\\
      w' ,& \text{otherwise }   .\\
    \end{array}
  \right.\\
\end{aligned}
\end{equation}
Note that this quantity is bounded from above by $Q(\rho,\rho')\leq1$, which is saturated if $\rho=\rho'$, and from below by $0 \leq Q(\rho,\rho')$, which saturates if $\rho \rho'=0$.
One important property is that the Chernoff bound gives bounds on the affinity and the fidelity as,
\begin{equation}
F^2(\rho,\rho') \leq Q(\rho,\rho')\leq A(\rho,\rho') (=Q_{1/2}(\rho,\rho'))  .
\end{equation}
Combining with (\ref{eq:A=F^2}), one can easily find for the single and double interval cases,
\begin{equation}
A(\rho,\rho')= Q(\rho,\rho')=F^2(\rho,\rho').
\end{equation}
We can directly check this equality by evaluating the replica partition function.
Note that this equality holds if both two density states $\rho$ and $\rho'$ are pure states, that is, 
\begin{equation}
A(\rho,\rho')= Q(\rho,\rho')=F^2(\rho,\rho')=\tr{\pa{\rho \rho'}}  .
\end{equation}

Note that we have also the following bounds on the trace distance,
\begin{equation}
1-Q(\rho,\rho') \leq D_{tr}(\rho,\rho') \leq \s{1-Q^2(\rho,\rho')},
\end{equation}
which is consistent with our conclusion that the quantum Chernoff bound also plays a role as a probe of the correct entanglement wedge.

\subsection{Super-Fidelity}
In general cases, it is hard to get fidelity and affinity due to the complication involved in evaluating the square root of a density matrix.
Instead, we can rely on {\it super-fidelity}, which is defined by
\begin{equation}
F_N(\rho,\rho')\equiv \tr \rho \rho' +\s{1-\tr \rho^2} \s{1-\tr \rho'^2}.
\end{equation}
This quantity involves only products of density matrices, which greatly simplifies its evaluation in sharp contrast with the fidelity. The super-fidelity does not satisfy the property: $F_N(\rho,\rho')=0  \Leftrightarrow \rho \rho'=0$.

The point is that the super-fidelity gives the upper bound on the fidelity as \cite{Miszczak2008, Mendonca2008}
\begin{equation}
F(\rho,\rho')  \leq  F_N(\rho,\rho')  \leq 1.
\end{equation}
The equality is satisfied when $\rho=\rho'$.
From this inequality, one can find that $F_N(\rho,\rho') <1$ directly implies $F(\rho,\rho') <1$, which means that the super-fidelity is another similarity measure.

Let us focus on holographic CFTs. In fact, one can immediately find that $\tr \rho \rho' \sim \tr \rho^2 \sim \tr \rho'^2 \sim \ex{-\#c} $, which means that the super-fidelity reduces to the trivial upper bound $F_N(\rho,\rho')  = 1$ in the large $c$ limit.
Therefore, we cannot distinguish our two states by making use of the super-fidelity in holographic CFTs.
Note that in CFTs with finite $c$, this also gives a non-trivial bound.

\subsection{$p$-Fidelity }

$p$-fidelity \cite{Liang2018} is a generalization of the fidelity and is defined by
\begin{equation}
F_p(\rho, \rho') \equiv \fr{\abs{\abs{  \s{\rho} \s{\rho'} }}^2_p}{\max \{   \abs{\abs{ \rho }}^2_p ,  \abs{\abs{ \rho' }}^2_p \}},
\end{equation}
where we introduce
\begin{equation}
\abs{\abs{A}}_p=\pa{\tr\BR{ \pa{A\dg{A}}^\fr{p}{2} }  }^{\fr{1}{p}}.
\end{equation}
The fidelity $F(\rho,\rho')$ coincides with $F_1(\rho, \rho')$.
By using the $p$-fidelity, the lower bound on $F_2(\rho,\rho')$ is given by the measure $I(\rho,\rho')$ (\ref{Irho}):
\begin{equation}
F_2(\rho, \rho') \leq I(\rho,\rho').
\end{equation}
Therefore, we cannot utilize the 2-fidelity as a probe of the entanglement wedge in general.

\subsection{Quantum Jensen Shannon Divergence }

The {\it quantum Jensen Shannon divergence} (QJS divergence) is defined in  \cite{Majtey2005}\footnote{
The QJS divergence has also studied in the context of holography in\cite{He2017a}.
} as 
\begin{equation}
JS(\rho, \rho') \equiv H \pa{\fr{\rho+\rho'}{2}}-\fr{H(\rho)+H(\rho')}{2},
\end{equation}
where $H$ is the von-Neumann entropy. This quantity can also be seen in quantum information theory, where it is called as the Holevo information.
As shown in \cite{Majtey2005}, it shares many physical relevant properties with the relative entropy.
Since the relative entropy is well-defined only in some restricted situations, the QJS divergence is more useful as distinguishability measure.
The QJS divergence also satisfies the inequality (which comes from the bound on the Holevo information \cite{nielsen2002quantum}),
\begin{equation}
0 \leq JS(\rho, \rho') \leq 1,
\end{equation}
where the lower bound is saturated if and only if $\rho=\rho'$.

For two neighboring density states, this quantity can be approximated by the fidelity as
\begin{equation}
JS(\rho, \rho') \simeq  1-F(\rho,\rho'), \ \ \ \ \text{if }  \rho \simeq \rho'.
\end{equation}
Through this relation, we can conclude that the QJS divergence can also probe the entanglement wedge in a similar way as the fidelity.

\subsection{Comparison of Distinguishability Measures and Entanglement Wedge Reconstruction} 

Finally we would like to compare the results of above distinguishability measures in addition to $I(\rho,\rho')$ and the fidelity $F(\rho,\rho')$.
CFT wedges defined by the measures  $\{ F,A,Q,D_{tr}, JS \}$ reproduce the correct entanglement wedges 
for 2d holographic CFTs. On the other hand, CFT wedges deviate from the correct entanglement wedges when we employ the measures 
$\{ I,F_N,F_2 \}$. These are summarized in Table \ref{table:EW}.

The fundamental properties of these measures are listed in Table \ref{table:comparison} in Appendix \ref{ap:list}.
By comparing this table with the previous one, we notice that the property ix) i.e. the monotonicity under the CPTP map
seems to be responsible for reproducing correct entanglement wedges.\footnote{The monotonicity is analogous to the 
strong subadditivity of the entanglement entropy  \cite{LR,HeT}.}
At the same time, another common property for the coincidence between CFT wedges and entanglement wedges is 
that the total power of $\rho$ and $\rho'$ is one in the trace as we emphasized in section \ref{sec:difwe}. 
This requirement comes from the probing only low energy states dual to the classical gravity.
On the the other hand, other measures $\{ I,F_N,F_2 \}$, the total power of $\rho$ and $\rho'$ is two.
In this sense the former look analogous to the von-Neumann entropy, while the latter analogous to 2nd Renyi
entropy.   In summary, our results in this paper suggest these two properties are necessary for a distinguishability measure in 
holographic CFTs to reconstruct the correct entanglement wedges.\footnote{
This observation naturally raises a question; can we find a similar deviation of CFT wedge versus entanglement wedge  as $I$ if we employ the Hilbert-Schmidt distance:
in particular, is the wedge from $I$ same as that from the Hilbert-Schmidt distance?
$D_{\text{HS}}(\rho,\rho') \equiv \s{\tr (\rho-\rho')^2}$, which is analogous to the 2nd Renyi entropy.
 It is known that the Hilbert-Schmidt distance is bounded by the trace distance \cite{Coles2012} (see also \cite{Coles2019}),
$0 \leq D_{\text{HS}}(\rho,\rho') \leq \s{2} D_{tr}(\rho,\rho').$
Unfortunately, the Hilbert-Schmidt distance reduces to $0$ in the large $c$ limit for the same reason as the super-fidelity,
therefore, we cannot extract some interesting information from this quantity.
Note that $\{ F_N, D_{\text{HS}} \}$ have the term $\tr\pa{\rho \rho'}$, which means that these two quantities contain the same information as $I$. In fact, if one appropriately normalize them, then we can extract the same wedge as from $I$.}

It would be interesting to note that there are other similarity measures which satisfy the property ix).
For examlpe, the relative entropy satisfies the property ix).
For this reason, we can expect that this quantity can also probe the entanglement wedge.
It would be interesting to investigate whether the relative entropy can actually detect the entanglement wedge and we would like to 
leave this for future works.

\begin{table}
\begin{center}
  \begin{tabular}{|l||c|} \hline
       & EW reproduction \\ 
\hline \hline
   $F$ &  \checkmark  \\ \hline
   $A$ &   \checkmark  \\ \hline
   $Q$ &   \checkmark  \\ \hline
   $D_{tr}$ &   \checkmark  \\ \hline
   $JS$ &   \checkmark  \\ \hline
   $F_N$ &     \\ \hline
   $I$ &    \\ \hline
   $F_2$ &    \\ \hline
  \end{tabular}
\end{center}
\caption{We mark an entry with  $\checkmark$ when a measure enable us to reproduce the entanglement wedge.}
\label{table:EW}
\end{table}

\section{Entanglement Wedges from HKLL Operators}

In this paper we have worked out the shape of entanglement wedge from purely CFT computations by exciting the CFT vacuum by a  local operator 
inserted at various locations. In this sense, a local operator plays the role of a probe for our holographic geometry. However, we need to choose the conformal 
dimension of the operator $O_\ap$ in the range of (\ref{rangeh})  to obtain sensible results. Even though it will be difficult to remove the 
constraint $h_\ap \ll c$ for negligible backreactions, one might think that we can somehow remove the requirement $h_\ap\gg 1$, which was necessary to have a sharp 
resolution of image of CFT wedge by the local operator. The resolution of distinguishability can be estimated by the Bures information metric owing to the Cramer-Rao bound
(\ref{CRaB}), which is given for the local operator result (\ref{wwww}) as follows
\ba
\la (\delta x)^2 \lb \geq \frac{\tau^2}{h_\ap}.
\ea
In this sense, the resolution of our local operator analysis is $O(1/\s{h_\ap})$ in the length scale. Therefore we need the assumption $h_\ap\gg 1$ to probe the geometry.
On the other hand, the classical gravity approximation of AdS/CFT predicts the actual resolution is a scale of $O(1/c)$, which is equivalent to the 
Planck scale. Therefore, the local operator is a slightly coarse-grained probe, especially when $h_\ap$ is not very large. 

A more fined-grained operator for this purpose is known as the HKLL operator \cite{HKLL}. This operator is known as the CFT counterpart of bulk local field operator $\phi_\ap$ and 
thus should be suitable to extract the bulk geometry including the entanglement wedge. Thus in this section we would like to study how we can probe the entanglement wedge 
geometry by the HKLL operator. However, note that the analysis of HKLL operators has a disadvantage that  the computations get highly complicated compared to the local operator ones. Due to this technical issues, our analysis below will rely on heuristic arguments.  

We focus on the simplest setup of AdS$_3/$CFT$_2$, where the global AdS$_3$ is dual to a holographic two dimensional CFT on a cylinder.
The global AdS$_3$ is described by the coordinate $(\rho,x,\tau)$  with the metric (\ref{gadsm}) and the two dimensional cylinder is 
parameterized by the complex coordinate $\xi=\tau+ix$ and $\bar{\xi}=\tau-ix$.
It is useful to employ the state representation of HKLL operators given in \cite{MNSTW,Goto:2017olq}, which is written as 
\ba
|\phi_\ap(\rho,x,\tau)\lb={\cal \ti{N}_\ap} \cdot \sum_{k=0}^\infty (-1)^k e^{-\delta (L^{\xi_0}_{0}+\bar{L}^{\xi_0}_{0})}
\frac{\Gamma(2h_\ap)}{k!\Gamma(k+2h_\ap)}(L^{\xi_0}_{-1})^k(\bar{L}^{\bar{\xi}_0}_{-1})^k  
O_\ap(\xi_0,\ti{\xi}_0)|0\lb,\no  \label{hklls}
\ea 
where ${\cal \ti{N}_\ap}$ is the overall normalization for the unit norm: $L^{\xi_0}_{n}$ and $\bar{L}^{\xi_0}_{n}$ 
are the chiral and anti-chiral Virasoro operators around the point $\xi_0$. 
The term $e^{-(L^{\xi_0}_{0}+\bar{L}^{\xi_0}_{0})\delta}$ represents the regularization of the infinite summation of $k$ over the descendants and the infinitesimally small 
parameter $\delta$ controls this UV regularization of localized excitation. 
More importantly, the location $\xi_0$ on the cylinder is given by the projection along the geodesic which passes through the bulk point $(\rho,x,\tau)$ 
in the global AdS$_3$ (as depicted in Fig.\ref{fig:EW}). This is explicitly given by $\xi_0=\tanh\frac{\rho}{2}\cdot e^{\tau+ix}$.

First note that the state (\ref{hklls}) can be obtained from our original local operator state by replacing the primary operator with a summation over descendants.
In this sense we can effectively estimate  the conformal dimension of the local operator in (\ref{hklls}) as its average $h_\ap\sim 1/\delta$. As argued in \cite{MNSTW},
in large $c$ CFTs, we expect that $\delta$ is $O(1/c)$. This agrees with the resolution expected from the AdS/CFT i.e. the scale larger than 
the Planck scale.  Our previous results for the excited states by local operators imply that the result of Bures information metric for the reduced density matrix $\rho_A$ 
is identical to that for the pure state as long as the excited point is within the CFT wedge. When we consider a pure HKLL state i.e. (\ref{hklls}), the Bures metric is computed as follows
(see \cite{MNSTW}): 
\ba
D^2_B=\frac{1}{8\delta^2}(d\rho^2+\sinh^2\rho dx^2).  \label{BRHK}
\ea
The Cramer-Rao bound from this result indeed agrees with the AdS/CFT prediction $\la (\delta x)^2 \lb \geq O(1/\delta^2)=O(1/c^2)$. 
In other words, the metric (\ref{BRHK}) agrees with the correct time slice metric of the global AdS if we set $\delta=O(c)$ up to an $O(1)$ constant.
 
Moreover, from above heuristic arguments, we expect that the CFT wedge for the Bures metric 
for HKLL states agrees with the correct entanglement wedge as in the local operator case.  In this way, we can reproduce the shape of entanglement wedge from the analysis of Bures metric 
of HKLL states such that the resolution scale agrees with the AdS/CFT expectation. It will be an interesting future problem to confirm the above arguments by explicit CFT calculations and their replica interpretations.

\section{Conclusions and Discussions}

In this paper,  we presented a new method to determine the shape of entanglement wedge from purely CFT calculations. 
Our strategy is to introduce CFT wedges, which are counterparts of entanglement wedges in AdS/CFT and which are defined for a given CFT.
We can view a CFT wedge as a  shadow of an entanglement wedge because the former is obtained from the latter by projecting along a geodesic 
in AdS backgrounds.

To determine the border of CFT wedge, we employed the locally excited states and asked the question whether 
we can distinguish two reduced density matrices $\rho_A$ and $\rho'_A$ with slightly different points excited. If the points are in the CFT wedge, we can 
distinguish them, while we cannot if they are outside the wedge. To quantify this we mainly examined two different distinguishability measures, namely the 
Bure distance (or equally fidelity) $D_B(\rho,\rho')$ and its Renyi-like version denoted by $I(\rho,\rho')$ (called geometric mean fidelity). In general, we find the CFT wedges are sharp 
only for the holographic CFTs, while for generic CFTs the CFT wedges get blurred. This special feature of sharp CFT wedge for holographic CFTs mainly origins from
the large $N$ factorization property.  In a very brief summary, we observed that the CFT wedges for the Bures distance 
perfectly agree with the expected entanglement wedge in AdS/CFT in all examples we studied. Moreover, it turned out that the Bures metric agrees with the 
metric on the entanglement wedge in AdS up to the overall factor. Thus our results in this paper provide a genuine CFT derivation of 
entanglement wedges in AdS/CFT for the first time.

As the first example, we intensively studied the case where the subsystem $A$ is a single interval in 2d CFTs. We found that in holographic CFTs, 
the border of CFT wedge gets sharp and perfectly agrees with the entanglement wedge both for the two different choices of distinguishability measures.
We also studied a free scalar 2d CFT and showed that the CFT wedge structure is obscure, though a part of qualitative features are similar. 
This clearly shows that the geometry of entanglement wedges emerges only in holographic CFTs, being consistent with our understanding of AdS/CFT.
 We also calculated the Bures information metric and found that it is proportional to the metric on the entanglement wedge. Moreover, we studied the time evolution of 
the reduced density matrix and confirmed that the resulting time-dependent CFT wedges agree with the covariant description of entanglement wedges in AdS/CFT.
As a future problem, we can also consider another non-trivial time-dependent setup, the falling-particle geometry, where we can rely on the CFT techniques developed in \cite{Kusuki2019,Kusuki2018}

As the second example, which is more non-trivial, we chose $A$ to be double intervals in 2d holographic CFTs. 
In this case, the standard holographic analysis tells us the phase transition 
between the connected and disconnected entanglement wedge. Our CFT wedge analysis perfectly reproduced this phase transition. However, we found that the resulting 
CFT wedge for the measure $I(\rho,\rho')$ slightly deviated from the expected entanglement wedge. On the other hand, we showed that the CFT wedge for the Bures 
distance reproduces the entanglement wedge in AdS/CFT perfectly. We argued that this difference of CFT wedges between two measures occurs because they are sensitive to 
different part of quantum states in CFT. The Bures distance $D_B(\rho,\rho')$ or fidelity $F(\rho,\rho')$ 
is sensitive to low enery states as the total power $p_{tot}$ of $\rho$ and $\rho'$ (i.e. $\sim \rho^{p_{tot}}$) is one, while the (2nd) Renyi-like measure 
$I(\rho,\rho')$ is also sensitive to high energy modes as the total power $p_{tot}$ is two. This is analogous to the well-known fact 
that the von-Neumann entropy is simply computed as the area in AdS/CFT, 
while the computation of Renyi entropy requires us to take into account back reactions \cite{Hung:2011nu,Dong:2016fnf}. 

We also analyzed an example of 2d boundary conformal field theory (BCFT), which has a gravity dual via the AdS/BCFT.
This example also experiences a phase transition between a connected and disconnected extremal surface. 
We showed that the CFT wedges agree with the expectation from entanglement wedges in AdS/BCFT under the
assumption that the boundary one point function vanishes. The similar argument holds also for the thermofield double (TFD) state without any assumptions. 
 
Moreover, we presented calculations of CFT wedges in higher dimensional CFTs when the subsystem $A$ is given by a round ball or a half space. 
The resulting CFT wedges perfectly agree with the expectation from the entanglement wedge in the higher dimensional AdS/CFT. 
Since this only covers the special example in higher dimensions, it will be intriguing future problem to explore more on the higher dimensional 
CFT wedges.

Since there are many other known distinguishability measures of quantum states, we examined whether such measures can reproduce the 
expected CFT wedges. We found that the affinity (Hellinger distance) $A(\rho,\rho')$,  the trance distance $D_{tr}(\rho,\rho')$, Chernoff bound 
$Q(\rho,\rho')$ and quantum Jensen Shannon divergence $JS(\rho,\rho')$ pass this test, as is so for the Bures distance or fidelity. 
Interestingly, these measures have the common feature of the monotonicity under CPTP maps. Also they share the aforementioned property that
the total power $p_{tot}$ of $\rho$ is one. It will be interesting to understand systematically
 how the difference of this total power affects the CFT wedges. Also it will be an important future problem to extend our analysis of CFT wedges to 
qunatum Fisher metric based on the relative entropy, which we have not discussed in this paper.

In the final part of the present paper we studied the excited states by HKLL operators for the computation of information metric
instead of those created by the local operators in CFTs. This is because when the conformal dimension is not large, the local operator excitations are not sharp 
probes to detect the bulk geometry. The HKLL operators are expected to be localized in a bulk point well even if the conformal dimension is small.
We gave a heuristic argument how we can extract the expected CFT wedge from HKLL states.  This allows us to detect the entanglement wedge 
up to the Planck scale, matching with the AdS/CFT prediction. Moreover, the Bures information metric for the HKLL states agrees with the actual 
metric of AdS up to an $O(1)$ factor, which we could not fix. It will be very interesting to pursuit this agreement more with the precise coefficient. 

All of calculations in this paper were about the leading contribution in the $1/N$ or $1/c$ expansion dual to the classical gravity approximation.
Therefore it will be an interesting future direction to study $1/N$ or $1/c$ corrections dual to the quantum corrections in gravity.
In this context, we may study the emergence of the quantum extremal surfaces \cite{Engelhardt:2014gca}.

Also the present work of deriving the entanglement wedges from CFTs might be related to other approaches to entanglement wedges. 
This involves an emergence of entanglement wedges in the path-integral optimization \cite{Caputa:2017urj}, where the mathematical structure has a 
significant similarity. Also one basic geometrical characterization of entanglement wedges will be the 
entanglement wedge cross section, whose CFT interpretations have been discussed from various viewpoints
 \cite{UT,Nguyen:2017yqw, Umemoto:2018jpc, Kudler-Flam:2018qjo,CMTU,Tamaoka:2018ned,Dutta:2019gen,Kusuki:2019rbk, Umemoto:2019jiz}. 
We hope we come back to these connections in future works.

\section*{Acknowledgements}

We are grateful to Ibrahim Akal, Jose Barbon, Pawel Caputa, Ignacio Cirac, Ben Freivogel , Esperanza Lopez, Robert Myers, Masahiro Nozaki, German Sierra, Erik Tonni, and Xiao-liang Qi for useful discussions.
YK  and KU are supported by the JSPS fellowship.
YK is supported by Grant-in-Aid for JSPS Fellows No.18J22495.
KU is supported by Grant-in-Aid for JSPS Fellows No.18J22888. 
TT is supported by the Simons Foundation through the ``It from Qubit'' collaboration.
TT is supported by World Premier International Research Center Initiative (WPI Initiative) 
from the Japan Ministry of Education, Culture, Sports, Science and Technology (MEXT). 
TT is also supported by JSPS Grant-in-Aid for Scientific Research (A) No.16H02182 and 
by JSPS Grant-in-Aid for Challenging Research (Exploratory) 18K18766.\\

TT would like to dedicate this paper to the memory of Tohru Eguchi, who was TT's great Ph D supervisor and 
kept TT highly stimulated and encouraged.
Among many other important things, TT learned a lot from Tohru Eguchi how string theory is beautiful and elaborated, 
as manifested e.g. in the seminal text book \cite{EGH}. This has always given TT the strong motive power for researches in this field.

\appendix

\section{Details of Calculations of $I(\rho,\rho')$ in Single Interval Case}\label{app:single}

Here we present detailed analysis of the quantity $I(\rho,\rho')$ 
when $w$ and $w'$ take generic values.
We write $z=p+iq(=z_1)$ and 
$z'=p'+iq'(=-z_3)$ such that 
$p,p'>0$ and $q,q'<0$ as we see from Fig.\ref{fig:singlePO}. 
We denote the region inside and outside of the CFT wedge by $W_{in}$ and $W_{out}$.
Note that $W_{out}$ corresponds to $p>-q$ and $p'>-q$. The non-trivial Wick contraction 
for the calculation of the four point function $F(z,\bar{z},-z',-\bar{z}')$ given by 
(\ref{wick}) is favored when 
$|z-\bar{z}||z'-\bar{z}'|>|z+\bar{z}'|^2$ i.e.
\be
4qq'>(p+p')^2+(q-q')^2.
\ee

When $w\in W_{out}$ and $w'\in W_{out}$, we find 
\be
F(z,\bar{z},-z,-\bar{z})\simeq |2q|^{-8h},\ \ \  F(z',\bar{z}',-z',-\bar{z}')\simeq |2q'|^{-8h},
\ee
where the trivial Wick contractions are favored. 
Also since $(p+p')^2+(q-q')^2>(q+q')^2+(q-q')^2>4qq'$, we find 
\ba
F(z,\bar{z},-z',-\bar{z}')\simeq |4qq'|^{-4h},
\ea
where the trivial Wick contractions are favored.
Thus we have $I(\rho,\rho')\simeq 1$.

When $w\in W_{in}$ and $w'\in W_{out}$, we find 
\be
F(z,\bar{z},-z,-\bar{z})\simeq |2p|^{-8h},\ \ \  F(z',\bar{z}',-z',-\bar{z}')\simeq |2q'|^{-8h}.
\ee
When the trivial Wick contraction is favored for $F(z,\bar{z},-z',-\bar{z}')$, we find
\ba
I(\rho,\rho')\simeq 
\frac{|p|^{4h}}{|q|^{4h}}\ll 1,
\ea
in the $h\gg 1$ limit.
When the non-trivial one is favored we obtain
\ba
I(\rho,\rho')\simeq 
\frac{|4pq'|^{4h}}{|(p+p')^2+(q-q')^2|^{4h}}\ll 1,
\ea
where we noted
\ba
(p+p')^2+(q-q')^2>(p-q')^2+(q-q')^2>-4pq'.
\ea
Thus in this case we have  $I(\rho,\rho')\simeq 0$.

\begin{figure}
  \centering
  \includegraphics[width=6cm]{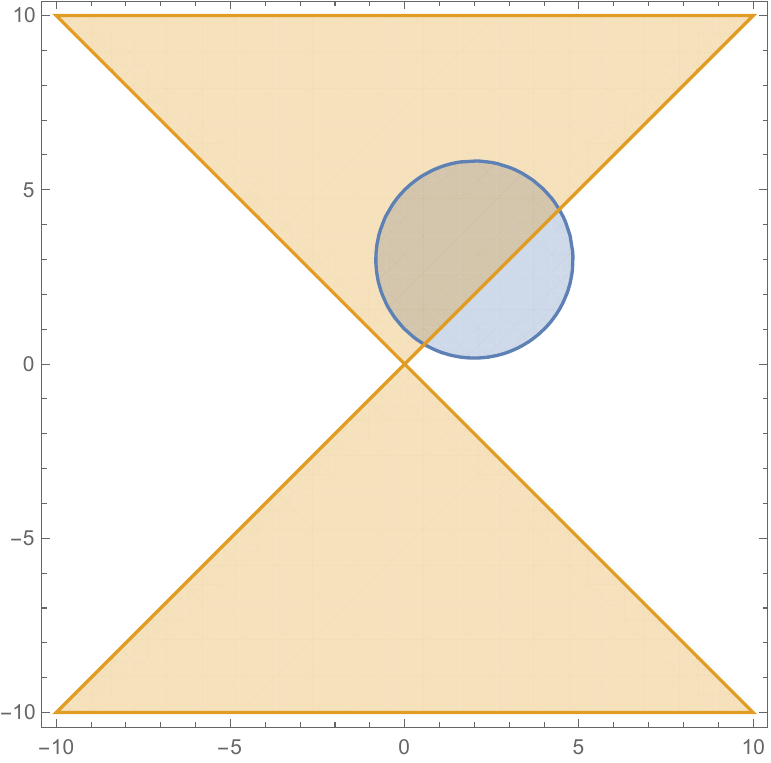}
 \includegraphics[width=6cm]{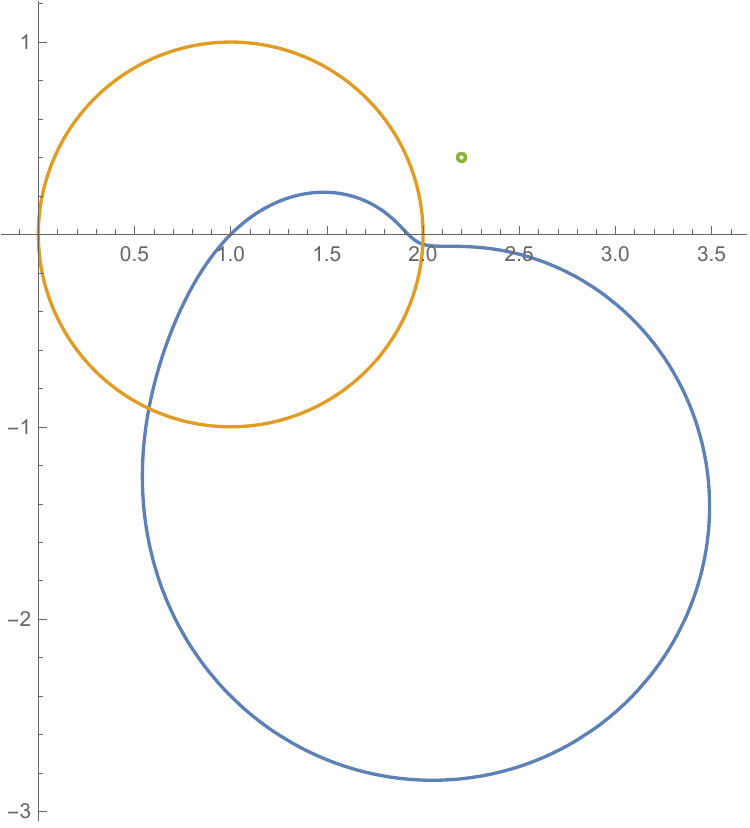}
 \includegraphics[width=6cm]{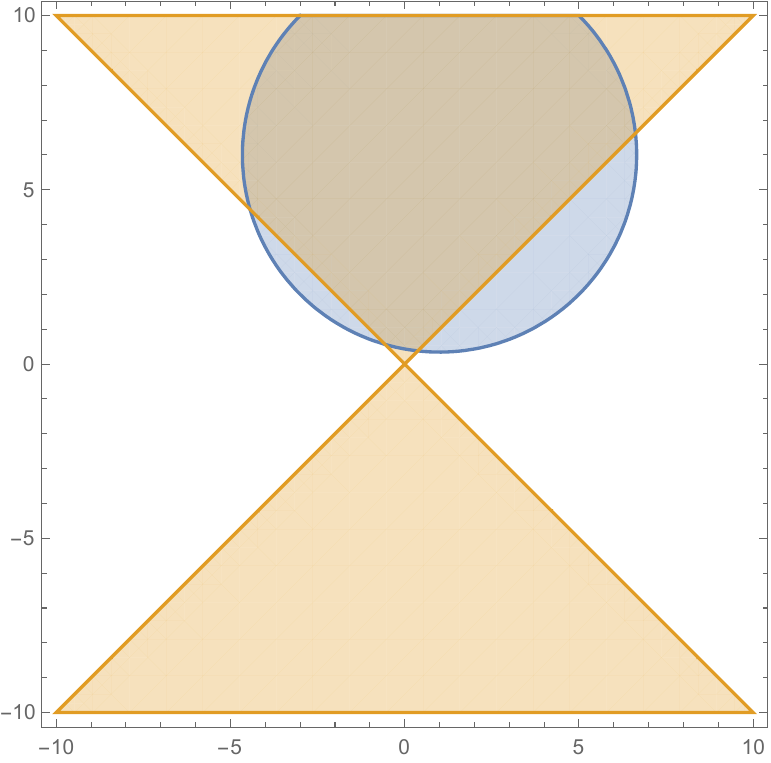}
 \includegraphics[width=6cm]{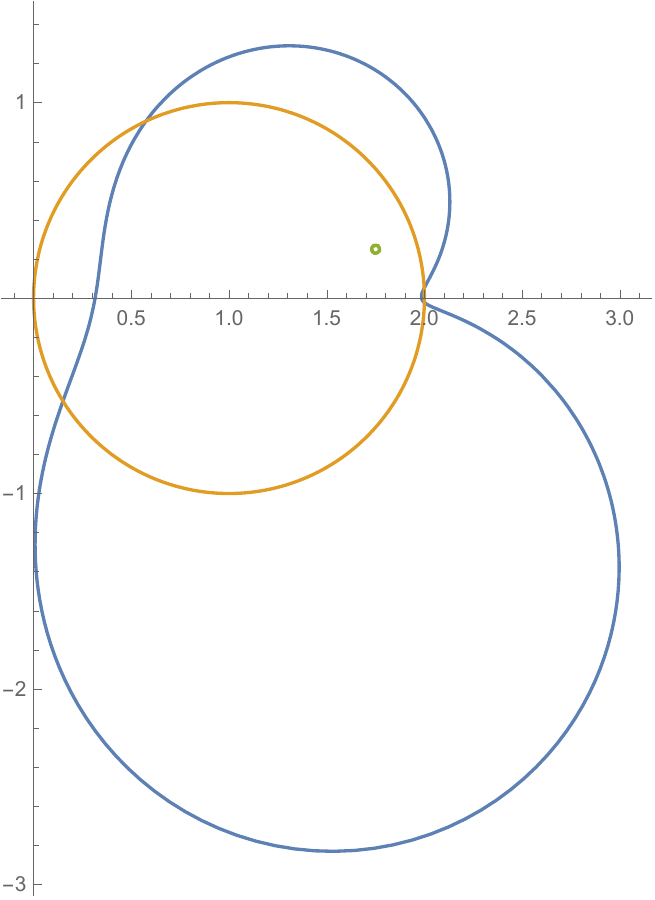}
  \caption{The profile of the regions of $z'$ and $w'$ (surrounded by blue curves) 
where the non-trivial Wick contraction is favored 
i.e. $|z-\bar{z}||z'-\bar{z'}|>|z+\bar{z}'|^2$. In the upper pictures we set $z=2-i$ (outside the wedge) and 
in the lower two pictures we set $z=1-2i$ (inside the wedge). The left ones  and right ones depict the regions in $z'$ and $w'$ plane, respectively. The orange curves describe the borders of the wedges. The green points describe the locations of $w$ and $z$. We took the subsystem $A$ to be $[0,2]$.} 
\label{fig:ZW}
  \end{figure}

Finally when $w\in W_{in}$ and $w'\in W_{in}$, we have 
\be
F(z,\bar{z},-z,-\bar{z})\simeq |2p|^{-8h},\ \ \  F(z',\bar{z}',-z',-\bar{z}')\simeq |2p'|^{-8h}.
\ee
When the trivial Wick contraction is favored for $F(z,\bar{z},-z',-\bar{z}')$, we find
\ba
I(\rho,\rho')\simeq 
\frac{|pp'|^{4h}}{|qq'|^{4h}}\ll 1,
\ea
in the $h\gg 1$ limit, unless $p=p'$ and $q=q'$.
When the non-trivial one is favored we obtain
\ba
I(\rho,\rho')\simeq 
\frac{|4pp'|^{4h}}{|(p+p')^2+(q-q')^2|^{4h}}\ll 1,
\ea
where we noted
\ba
(p+p')^2+(q-q')^2\geq 4pp',
\ea
where the equality holds when  $p=p'$ and $q=q'$.
Thus in this case, we have $I(\rho,\rho')\simeq 0$ except the case $w=w'$.
If $w=w'$ we have $I(\rho,\rho')=1$.
Refer to Fig.\ref{fig:ZW} for plots.

When $\delta z=z'-z$ is infinitesimally small, we can expand 
$D_I(\rho,\rho')\equiv 2-2I(\rho,\rho')$ as follows:
\ba
&& D_I(\rho,\rho')\simeq \frac{4h}{|z+\bar{z}|^2}\cdot |d z|^2\no
&&=\frac{h}{4}\cdot \frac{\left(\s{x(L-x)+i\tau L+\tau^2}
+\s{x(L-x)-i\tau L+\tau^2}\right)^2}{\tau^2\s{x^2+\tau^2}\s{(L-x)^2+\tau^2}}(dx^2+d\tau^2).
\ea
This is the expression of the information metric constructed from the distance measure 
$D_I$.

\section{Detailed Analysis of Bures Metric in $c=1$ CFT}  \label{conebr}

We start with the expression (\ref{cosing}) and consider the free scalar CFT:
\ba
A_{n,m} &=&k^{-4kh}\cdot |z|^{8mnh(1-k)}\cdot |z'|^{4nh(1-k)}\cdot
 |z^k-\bar{z}^k|^{8mnh}
\cdot  |z'^k-\bar{z}'^k|^{4nh}\no
&& \times  \la O_\ap^\dagger(z_1)O_\ap(z_2)\ddd O_\ap^\dagger(z_{2k-1})O(z_{2k})\lb \cdot
\f{Z^{(k)}}{(Z^{(1)})^k}. \no  \label{cosingg}
\ea
Below we set $h=1/2$ by assuming the operator $O=e^{i\phi}$.

We can write the $2k$ point function as follows
\ba
\la O_\ap^\dagger(z_1)O_\ap(z_2)\ddd O_\ap^\dagger(z_{2k-1})O(z_{2k})\lb=f(z)^{k}\cdot g(z,z')^n,
\ea
such that $f(z)^{k}$ corresponds to the computation Tr$\rho^k$ and $g(z,z')^n$ corresponds to the ratio between 
Tr$(\rho^m\rho'\rho^m)^n$ and Tr$\rho^k$. The former one $f(z)$ is computed as  
\ba
f(z)&=&\frac{\prod_{j=1}^{k-1}|z-ze^{\frac{2\pi i}{k}j}|^{4h}}{\prod_{j=0}^{k-1}|z-\bar{z}e^{\frac{2\pi i}{k}j}|^{4h}}, \no
&=&\frac{k^2}{2r^2(1-\cos (k\theta_1))},  \label{fz}
\ea
where we set $h=1/2$. We defined
\ba
&& r=\s{x^2+y^2},\ \ \ r'=\s{x'^2+y'^2},\no
&&  \cos\theta_1=\frac{x^2-y^2}{r^2},\ \ \  \sin\theta_1=\frac{2xy}{r^2},\no
&&  \cos\theta_2=\frac{xx'-yy'}{rr'},\ \ \  \sin\theta_2=\frac{x'y+xy'}{rr'},\no
&& \cos\theta_3=\frac{xx'+yy'}{rr'},\ \ \  \sin\theta_3=\frac{x'y-xy'}{rr'},\no
&&  \cos\theta_4=\frac{x'^2-y'^2}{r^2},\ \ \  \sin\theta_4=\frac{2x'y'}{r'^2}.
\ea

The function $g(z,z')$ is estimated as follows
\ba
&& g(z,z')  \no
&& =\left[\frac{\prod_{j=0}^{k-1}|z-\bar{z}e^{\frac{2\pi i}{k}j}|^{4h}\cdot \prod_{j=1}^{k-1}|z'-ze^{\frac{2\pi i}{k}j}|^{4h}}
{\prod_{j=1}^{k-1}|z-\bar{z'}e^{\frac{2\pi i}{k}j}|^{4h}\cdot \prod_{j=1}^{k-1}|z-ze^{\frac{2\pi i}{k}j}|^{4h}}\right]^2
\cdot \left[\frac{\prod_{l=1}^{n-1}|z-\bar{z'}e^{\frac{2\pi i}{n}l}|^{4h}\cdot \prod_{l=1}^{n-1}|z-ze^{\frac{2\pi i}{n}l}|^{4h}}
{\prod_{l=0}^{n-1}|z-\bar{z}e^{\frac{2\pi i}{n}l}|^{4h}\cdot \prod_{l=1}^{n-1}|z'-ze^{\frac{2\pi i}{n}l}|^{4h}}\right]\no
&& \cdot  \left[\frac{\prod_{l=1}^{n-1}|z-\bar{z'}e^{\frac{2\pi i}{n}l}|^{4h}\cdot \prod_{l=1}^{n-1}|z'-z'e^{\frac{2\pi i}{n}l}|^{4h}}
{\prod_{l=0}^{n-1}|z'-\bar{z'}e^{\frac{2\pi i}{n}l}|^{4h}\cdot \prod_{l=1}^{n-1}|z-z'e^{\frac{2\pi i}{n}l}|^{4h}}\right],\no
&&=\prod_{j=0}^{k-1}\left[\frac{|z-\bar{z}e^{\frac{2\pi i}{k}j}|^{4h}\cdot |z'-ze^{\frac{2\pi i}{k}j}|^{4h}}
{|z-\bar{z'}e^{\frac{2\pi i}{k}j}|^{4h}}\right]^2 \cdot \prod_{l=0}^{n-1}\left[\frac{|z-\bar{z'}e^{\frac{2\pi i}{n}l}|^{4h}}
{|z-\bar{z}e^{\frac{2\pi i}{n}l}|^{4h}|z'-ze^{\frac{2\pi i}{n}l}|^{4h}}\right] \no
&& \cdot  \prod_{l=0}^{n-1}\left[\frac{|z-\bar{z'}e^{\frac{2\pi i}{n}l}|^{4h}}
{|z'-\bar{z'}e^{\frac{2\pi i}{n}l}|^{4h}|z-z'e^{\frac{2\pi i}{n}l}|^{4h}}\right]\cdot\frac
{ \prod_{l=1}^{n-1}\left[|z-ze^{\frac{2\pi i}{n}l}|^{4h}\cdot |z'-z'e^{\frac{2\pi i}{n}l}|^{4h}\right]}
{\prod_{j=1}^{k-1}|z-ze^{\frac{2\pi i}{k}j}|^{8h}}.
\ea

Let us assume $h_\ap=1/2$.
To evaluate $g(z,z')$, the following identities are useful:
\ba
\prod_{j=1}^{n-1}\sin\left(\frac{\pi}{n}j\right)=\frac{n}{2^{n-1}},  \label{ap1}
\ea
and for $w=re^{i\theta}$ and $w'=r'e^{i\theta'}$, 
\ba
\prod_{j=0}^{n-1}|w-w'e^{\frac{2\pi i}{n}j}|^2=r^{2n}+r'^{2n}-2r^n r'^n\cos \left(n(\theta-\theta')\right).
\ea
If we write $w=x+iy$ and $w'=x'+iy'$ we have 
\be
\cos(\theta-\theta')=\frac{xx'+yy'}{rr'},\ \ \ \sin(\theta-\theta')=\frac{x'y-xy'}{rr'}.  \label{ap2}
\ee

By using (\ref{ap1}) and (\ref{ap2}) we can rewrite  $g(z,z')$ as follows:
\ba
&& g(z,z')=\left[\frac{2r^{2k}(1-\cos(k\theta_1))\left(r^{2k}+r'^{2k}-2r^k r'^k\cos(k\theta_3)\right)}
{\left(r^{2k}+r'^{2k}-2r^k r'^k\cos (k\theta_2)\right)(2r)^{2(k-1)}\cdot k^2 \cdot 2^{2(1-k)}}\right]^2 \no
&&\ \ \ \times \frac{\left(r^{2n}+r'^{2n}-2r^n r'^n\cos(n\theta_2)\right)^2 \cdot (2r)^{2(n-1)}(2r')^{2(n-1)}\cdot n^4\cdot 2^{4(1-n)}}
{\left(r^{2n}+r'^{2n}-2r^n r'^n\cos(n\theta_3)\right)^2 \cdot 2(r)^{2n}(1-\cos(n\theta_1)) \cdot 2(r')^{2n}(1-\cos(n\theta_4))}.\no
\label{gz}
\ea

Finally by taking the limit $n=m\to 1/2$ ($k\to 1$) , we find 
\ba
A_{n=1/2,m=1/2}=|z-\bar{z}| \cdot |z'-\bar{z}'|\cdot \frac{1}{4y^2}\cdot g(z,z')^{1/2},
\ea
where $g(z,z')$ in the limit $n=m\to 1/2$ reads
\ba
&& g(z,z')_{n=m=1/2}=\left[\frac{4y^2\cdot \left(r^2+r'^2-2rr'\cos\theta_3\right)}{r^2+r'^2-2rr'\cos\theta_2}\cdot
\frac{r+r'-2\s{rr'}\cos(\theta_2/2)}{r+r'-2\s{rr'}\cos(\theta_3/2)}\right]^2\no
&& \ \ \ \times \frac{(1/16)\cdot (1/rr')}{4rr'(1-\cos(\theta_1/2)(1-\cos(\theta_4/2))}.
\ea
Thus we obtain
\ba
A_{n=1/2,m=1/2}=\frac{r+r'+2\s{rr'}\cos(\theta_3/2)}{r+r'+2\s{rr'}\cos(\theta_2/2)}
\cdot\frac{|y||y'|}{2rr'\s{(1-\cos(\theta_1/2))(1-\cos(\theta_4/2))}}.\no  \label{cone}
\ea

To evaluate (\ref{cone}) we have to be careful with the computations of cosines such as $\cos(\theta_3/2)$.
For this, it is useful to focus on the case $m=1/2$ and $k=2n$ for the integer $n$ in (\ref{fz}) and (\ref{gz}) which corresponds to the calculation of $\mbox{Tr}[(\rho\rho')^n]$. In this case we have 
\ba
&& \cos(n\theta_1)=\frac{1}{2}\left(\zeta+\zeta^{-1}\right), \ \ \ \cos(2n\theta_1)=\frac{1}{2}\left(\zeta^2+\zeta^{-2}\right),\no
&& \cos(n\theta_2)=\frac{1}{2}\left(\zeta^{1/2}\zeta'^{1/2}+\zeta^{-1/2}\zeta'^{-1/2}\right),
 \ \ \ \cos(2n\theta_2)=\frac{1}{2}\left(\zeta\zeta'+\zeta^{-1}\zeta'^{-1}\right), \no
&& \cos(n\theta_3)=\frac{1}{2}\left(\zeta^{1/2}\zeta'^{-1/2}+\zeta^{-1/2}\zeta'^{1/2}\right), 
\ \ \ \cos(2n\theta_3)=\frac{1}{2}\left(\zeta\zeta'^{-1}+\zeta^{-1}\zeta'\right),\no
&& \cos(n\theta_4)=\frac{1}{2}\left(\zeta'+\zeta'^{-1}\right), \ \ \ \cos(2n\theta_4)=\frac{1}{2}\left(\zeta'^2+\zeta'^{-2}\right), 
\ea
where we defined
\ba
\zeta=\frac{z^{2n}}{|z|^{2n}}=\frac{w}{w-L}\cdot \frac{|w-L|}{|w|},\ \ \ \ \ 
\zeta'=\frac{z'^{2n}}{|z'|^{2n}}=\frac{w'}{w'-L}\cdot \frac{|w'-L|}{|w'|}.
\ea
By using this expression we can take the analytical continuation $n\to 1/2$. 
In this way we obtain the final expression (\ref{xbx}).

\begin{figure}
  \centering
  \includegraphics[width=6cm]{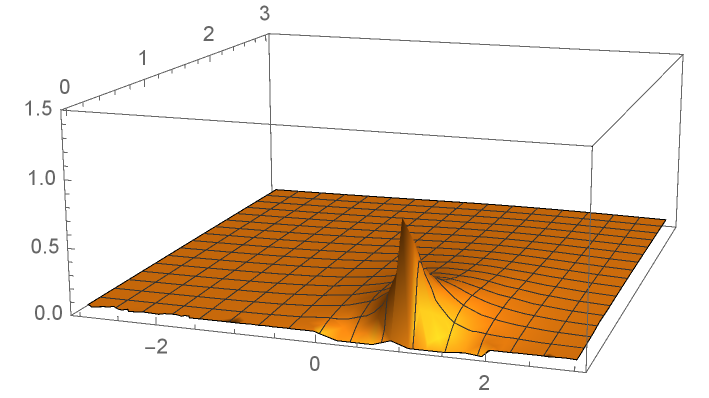}
 \includegraphics[width=6cm]{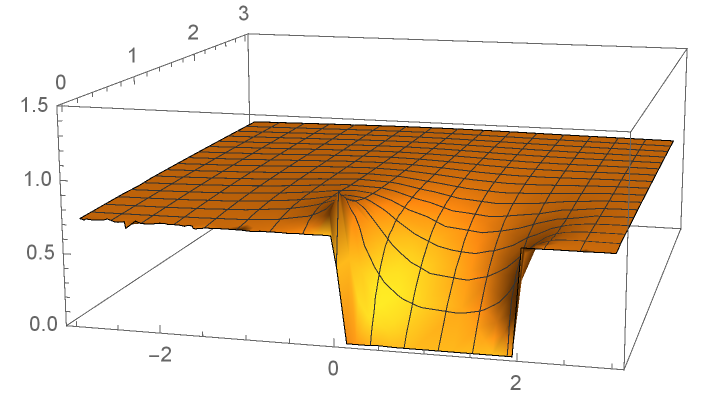}
 \includegraphics[width=6cm]{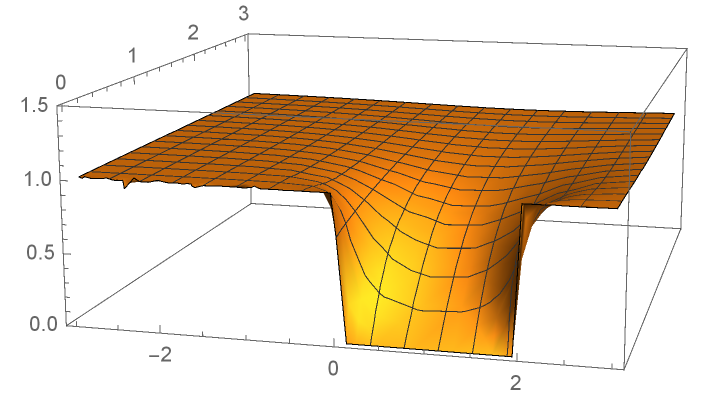}
 \includegraphics[width=6cm]{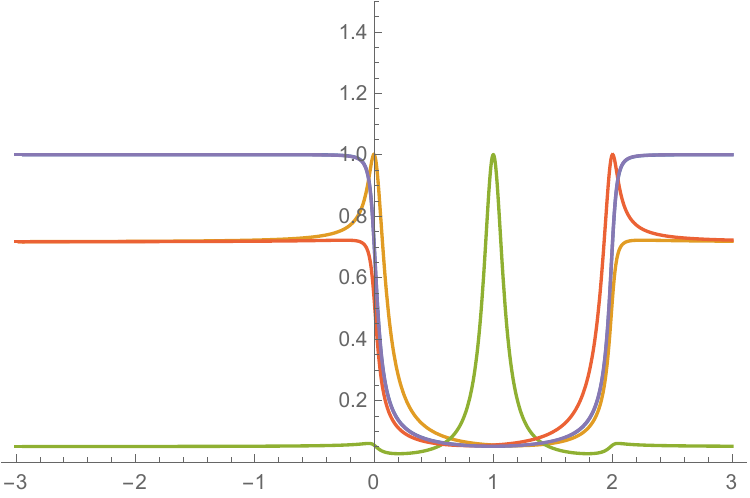}
  \caption{The profile of the $A_{n=1/2,m=1/2}=\mbox{Tr}[\s{\s{\rho}\rho' \s{\rho}}]$ in $c=1$ free scalar CFT for 
the operator $O=e^{i\phi}$ which has the dimension $h=1/2$ for various choices of excited points. 
The upper left, upper right and lower left graphs describe  $A_{n=1/2,m=1/2}$  for $\rho(w=1+0.05i)$, $\rho(w=0.05i)$ and 
$\rho(w=-1+0.05i)$, respectively. We plotted  $A_{n=1/2,m=1/2}$ as a function of $(p,q)$ for $\rho'(w'=p+iq)$. The lower right graphs describe  $A_{n=1/2,m=1/2}$ for $w=s+0.05i$ ($s=-1$(blue), $s=0$ (orange), $s=1$(green) and $s=2$(red)) as a function of $p$ 
such that $w'=p+0.05i$. We chose $L=2$.}
\label{fig:c=1Bures}
  \end{figure}

We plotted $A_{n=1/2,m=1/2}=\mbox{Tr}[\s{\s{\rho}\rho' \s{\rho}}]$ for fixed choices of $w'$ as a function of 
$w'=p+iq$ in Fig.\ref{fig:c=1Bures} and Fig.\ref{fig:c=1Bures2}. 
We find a localized peak $A\simeq 1$ at $w=w'$ when $w$ is close to the center of the subsystem $A$.
However the entanglement wedge is not clear again as opposed to the holographic case.

\section{General Time-dependent  Case}\label{generalt}

For a generic pure state in a holographic CFT with a gravity dual, the Fidelity $F(\rho,\rho')=A_{1/2,1/2}$ is computed from the two point function 
$\la O^\dagger_\ap(w,\bar{w})O_\ap(w',\bar{w}')\lb$ in such a state dual to a geodesic length $L(w,\bar{w}:w',\bar{w}')$ simply written by $L(w:w')$ as follows 
\ba
A_{1/2,1/2}\simeq e^{h\left[L(w_1:w_2)+L(w'_1:w'_2)-L(w'_1:w_2)-L(w_1:w'_2)\right]}.
\ea
By setting $w_1=x_1+i\tau_1$ and $w_2=x_2-i\tau_2$ 
and taking the limit $x'_{1,2}-x_{1,2}=dx^{1,2}\to 0$ and $\tau'_{1,2}-\tau_{1,2}=d\tau^{1,2}\to 0$, 
this leads to the Bures metric given by
\ba
&& D_B^2=2(1-A_{1/2,1/2})\no
&& \simeq (-2h)\cdot
\left[(\de_{x_1}\de_{x_2}L)dx_1dx_2+(\de_{x_1}\de_{\tau_2}L)dx_1d\tau_2
+(\de_{\tau_1}\de_{x_2}L)d\tau_1dx_2+(\de_{\tau_1}\de_{\tau_2}L)d\tau_1d\tau_2\right].\no
\ea

If we set $x_1=x_2=x$ and $\tau_1=\tau_2=\tau$ we get the 2d metric 
\ba
D_B^2\simeq (-2h)\cdot
\left[(\de_{x_1}\de_{x_2}L)(dx)^2+(\de_{x_1}\de_{\tau_2}L+\de_{\tau_1}\de_{x_2}L)d\tau dx
+(\de_{\tau_1}\de_{\tau_2}L)d\tau d\tau \right].\no
\ea

If we plug the geodesic length in Poincare AdS$_3:$ $L=\log[(x_1-x_2)^2+(\tau_1+\tau_2)^2]$, 
we obtain
\ba
&& D_B^2=h\left[G_{xx}dx_1dx_2+G_{tx}(dx_1d\tau_2-dx_2d\tau_2)+G_{tt} d\tau_1d\tau_2\right],  \no
&& G_{xx}=G_{tt}=\frac{4\left[(\tau_1+\tau_2)^2-(x_1-x_2)^2\right]}{\left[(\tau_1+\tau_2)^2+(x_1-x_2)^2\right]^2},  \no
&& G_{tx}=\frac{8\left[(\tau_1+\tau_2)(x_1-x_2)\right]}{\left[(\tau_1+\tau_2)^2+(x_1-x_2)^2\right]^2}.
\ea
If we restrict as $x_1=x_2=x$, then we reproduce the metric (\ref{met12}) as expected.

\section{Distinguishability Measures}\label{ap:list}

Here we would like to list  fundamental properties (including Joza's axioms \cite{Jozsa1994}) of distinguishability measures in Table \ref{table:comparison} (see \cite{Liang2018} in more details).
\begin{table}[h]
\begin{center}
  \begin{tabular}{|l||c|c|c|c|c|c|c|c|c|} \hline
       & $i$  & $ii$  & $iii$  & $iv$  & $v$  & $vi$  & $vii$  & $viii$  & $ix$ \\ 
\hline \hline
   $F$ &  \cm & \cm & \cm & \cm  & \cm  & \cm & $ \cm  $ & \cm & \cm    \\ \hline
   $A$  &  \cm & \cm & \cm & \cm   & \cm  & \cm & $  \cm $ & \cm & \cm    \\ \hline
   $Q$ &  \cm & \cm & \cm & \cm  & \cm  & \cm & \cm & \cm & \cm    \\ \hline
   ${D_{tr}}^{*_1}$ &  \cm & \cm & \cm & \cm  & \cm  & ? &  & \cm & \cm    \\ \hline
   $JS^{*_1, *_2  }$ &  \cm & \cm & \cm & \cm  & \cm  & \cm &  &   & \cm    \\ \hline
   $F_N$&  \cm & \cm &     & \cm & \cm  & \cm & \cm & Super &     \\ \hline
   $I$ &  \cm & \cm & \cm & \cm  & \cm  &     &     & \cm &     \\ \hline
   $F_2$ &  \cm & \cm & \cm & \cm  & \cm  &     &     & \cm &     \\ \hline
  \end{tabular}
\end{center}
\caption{We mark an entry with  $\checkmark$ when a measure satisfies the property i) $\sim$ix).
``Super'' means that a quantity do not satisfy the multiplicativity but the super-multiplicativity.
$*_1$: The properties ii) $\sim$ iii) for $D_{tr}$ and $JS$ are defined based on $1-D_{tr}$ and $1-JS$, instead of themselves.
$*_2$: The QJS divergence satisfies the convexity, instead of the concavity vi) $\sim$ vii).
}
\label{table:comparison}
\end{table}

\begin{description}
\item[i)]
$0 \leq \ca{F}(\rho,\rho') \leq 1$

\item[ii)]
$\ca{F}(\rho,\rho')=1$ if and only if $\rho=\rho'$

\item[iii)]
$\ca{F}(\rho,\rho')=0 $ if and only if $\rho\rho'=0$

\item[iv)]
$\ca{F}(\rho,\rho')=\ca{F}(\rho',\rho)$

\item[v)]
$\ca{F}(U \rho \dg{U} , U \rho' \dg{U})=\ca{F}(\rho,\rho')$ for any unitary operator $U$

\item[vi)]
$\ca{F}\pa{\sum_i p_i \rho_i   , \rho'  }\geq\sum_i p_i \ca{F}\pa{ \rho_i   , \rho'  }$ \ \ \ 
for any $p_i\geq0$ s.t. $\sum_i p_i =1$ \ \ \ 
({\it separable concavity})

\item[vii)]
$\ca{F}\pa{\sum_i p_i \rho_i   ,  \sum_j p_j \rho'_j  }\geq\sum_i p_i \ca{F}\pa{ \rho_i   , \rho'_i  }$  \ \ \ 
for any $p_i\geq0$ s.t. $\sum_i p_i =1$
({\it joint concavity})

\item[viii)]
$\ca{F}\pa{\rho_1 \otimes \rho_2  ,  \rho'_1 \otimes \rho'_2 } = \ca{F}(\rho_1,\rho'_1) \ca{F}(\rho_2,\rho'_2)   $  
({\it multiplicativity})

\item[viii) (Super)]
$\ca{F}\pa{\rho_1 \otimes \rho_2  ,  \rho'_1 \otimes \rho'_2 } \geq \ca{F}(\rho_1,\rho'_1) \ca{F}(\rho_2,\rho'_2)   $  
({\it super-multiplicativity})

\item[ix)]
$\ca{F} \pa{ \ca{E}(\rho)  , \ca{E}(\rho')  } \geq \ca{F}(\rho,\rho')$ for any completely positive trace preserving (CPTP) map $\ca{E}$.

\end{description}

\providecommand{\href}[2]{#2}\begingroup\raggedright

\end{document}